\documentstyle[psfig,epsfig]{mn}

\newcommand{\HI}{H\,{\sc i}}
\newcommand{\HII}{H\,{\sc ii}}
\newcommand{\Ha}{H$\alpha$}
\newcommand{\kms}{~km\,s$^{-1}$}
\newcommand{\kkms}{km\,s$^{-1}$}
\newcommand{\vhel}{$v_{\rm hel}$}
\newcommand{\vopt}{$v_{\rm opt}$}

\newcommand{\vgsr}{$v_{\rm GSR}$}

\newcommand{\vsys}{$v_{\rm sys}$}
\newcommand{\vlg}{$v_{\rm LG}$}
\newcommand{\vmax}{$v_{\rm max}$}

\newcommand{\vrot}{$v_{\rm rot}$}
\newcommand{\Rmax}{$R_{\rm max}$}
\newcommand{\FHI}{$F_{\rm HI}$}
\newcommand{\LB}{$L_{\rm B}$}

\newcommand{\NHI}{$N_{\rm HI}$}
\newcommand{\MHI}{$M_{\rm HI}$}
\newcommand{\Mdyn}{$M_{\rm dyn}$}
\newcommand{\Mstellar}{$M_{\rm *}$}
\newcommand{\RHI}{$R_{\rm HI}$}
\newcommand{\DHI}{$D_{\rm HI}$}
\newcommand{\Dopt}{$D_{\rm opt}$}

\newcommand{\MMsun}{M$_{\odot}$}
\newcommand{\Msun}{~M$_{\odot}$}
\newcommand{\LLsun}{L$_{\odot}$}
\newcommand{\Lsun}{~L$_{\odot}$}
\newcommand{\Ho}{H$_{\rm o}$}
\newcommand{\AB}{$A_{\rm B}$}

\newcommand\tiri{{\sc TiRiFiC}}

\title[The Local Volume HI Survey (LVHIS)]
      {The Local Volume HI Survey (LVHIS)\thanks{The observations were 
         obtained with the Australia Telescope which is funded by the 
         Commonwealth of Australia for operations as a National Facility 
         managed by CSIRO. - E-mail: Baerbel.Koribalski@csiro.au}}

\author[B.S.~Koribalski et al.]
       {B\"arbel S. Koribalski$^1$, Jing~Wang$^{1,6}$, P. Kamphuis$^{1,7}$,
        T. Westmeier$^2$, L. Staveley-Smith$^2$, \newauthor
        S.-H. Oh$^{2,8}$, \'A.R. L\'opez-S\'anchez$^3$, O. I. Wong$^2$, 
        J. Ott$^4$, W.J.G. de Blok$^5$, and L. Shao$^1$  \\
        $^1$Australia Telescope National Facility, 
	    CSIRO Astronomy \& Space Science, 
            P.O. Box 76, Epping, NSW 1710, Australia \\
        $^2$International Centre for Radio Astronomy Research (ICRAR),
            University of Western Australia, 
            35 Stirling Highway, Crawley, WA 6009, Australia \\
        $^3$Australian Astronomical Observatory, 105 Delhi Road, North Ryde,
            NSW 2113, Australia \\
        $^4$National Radio Astronomy Observatory, P.O. Box O,
            1003 Lopezville Road, Socorro, NM 87801, USA \\
        $^5$Netherlands Institute for Radio Astronomy, Postbus 2, 7990 AA
            Dwingeloo, The Netherlands \\
        $^6$Kavli Institute for Astronomy and Astrophysics, Peking University, 
            Beijing 100871, China \\
        $^7$National Centre for Radio Astrophysics, TIFR, Ganeshkhind,  
            Pune 411007, India \\
        $^8$Korea Astronomy and Space Science Institute (KASI), 
            Daedeokdae-ro 776, Yuseong-gu, Daejeon 34055, Republic of Korea \\
}

\date{Received date; accepted date}

\begin{document}

\maketitle

\begin{abstract}
The {\em `Local Volume HI Survey'} (LVHIS) comprises deep \HI\ spectral line
and 20-cm radio 
continuum observations of 82 nearby, gas-rich galaxies, supplemented by 
multi-wavelength images. Our sample consists of all galaxies with Local Group 
velocities \vlg\ $< 550$\kms\ or distances $D < 10$~Mpc that are detected in 
the \HI\ Parkes All Sky Survey (HIPASS). Using full synthesis observations in 
at least three configurations of the Australia Telescope Compact Array (ATCA), 
we obtain detailed \HI\ maps for a complete sample of gas-rich galaxies with 
$\delta \la -30\degr$. Here we present a comprehensive LVHIS galaxy atlas, 
including the overall gas distribution, mean velocity field, velocity 
dispersion and position-velocity diagrams, together with a homogeneous set of 
measured and derived galaxy properties. Our primary goal is to investigate 
the \HI\ morphologies, kinematics and environment at high resolution and 
sensitivity. LVHIS galaxies represent a wide range of morphologies and sizes; 
our measured \HI\ masses range from $\sim$10$^7$ to 10$^{10}$\Msun, based on 
independent distance estimates. The LVHIS galaxy atlas (incl. FITS files) is 
available on-line.
\end{abstract}

 
\section{Introduction} 
The {\em `Local Volume'} (LV), defined here as the sphere of radius 10~Mpc 
centered on the Local Group (LG), includes more than 500 known galaxies, many 
of which congregate in well-known groups. Most prominent in the southern 
hemisphere are the relatively loose Sculptor Group and the more compact 
Centaurus\,A Group. Together, their gas-rich members comprise about half of 
the {\em `Local Volume HI Survey'} (LVHIS) galaxy sample presented here. With 
accurate distances available for the majority of the LV galaxies (Karachentsev 
et al. 2013, hereafter K13, and references therein), it is now possible to 
study their morphologies, dynamics and star formation with respect to their 
surroundings.

Our long-term aim is to obtain high-resolution \HI\ maps of all LV galaxies 
and measure their properties in a homogeneous and unbiased way. Wang et 
al. (2016), for example, present \HI\ diameters (\DHI) of over 500 nearby 
galaxies measured out to an \HI\ mass (\MHI) density of 1\Msun\,pc$^{-2}$ and 
discuss the tightness of the \MHI-\DHI\ relation. Large \HI\ projects with 
radio interferometers, which have targeted LV galaxies, are listed in Table~1.
The {\em `Faint Irregular Galaxies GMRT Survey'} 
(FIGGS) by Begum et al. (2008), which contains \HI\ results for $\sim$60 
dwarf irregular LV galaxies, is the largest \HI\ study of nearby galaxies 
with the Giant Meterwave Radio Telescope (GMRT). The most prominent Very 
Large Array (VLA) \HI\ projects are {\em `The HI Nearby Galaxy Survey'} 
(THINGS; Walter et al. 2008), {\em Little THINGS} (Hunter et al. 2012), and 
{\em VLA-ANGST} (Ott et al. 2012). \HI\ galaxy surveys with the Westerbork 
Synthesis Radio Telescope (WSRT) include the {\em `Westerbork HI Survey of 
Irregular and Spiral Galaxies'} (WHISP; van der Hulst et al. 2001) and 
{\em HALOGAS} (Heald et al. 2011). Interferometric \HI\ surveys of galaxies 
beyond the Local Volume include the {\em VLA Imaging of Virgo in Atomic Gas} 
(VIVA; Chung et al. 2009), ATLAS-3D, targeting 166 early-type galaxies 
(Serra et al. 2012), and BlueDisks (Wang et al. 2013).

The LVHIS project comprises the largest number of nearby galaxies studied 
with the Australia Telescope Compact Array (ATCA); an overview and \HI\ galaxy 
atlas are presented in this paper. Previous publications based on LVHIS 
include Koribalski (2008, 2010, 2015, 2017), Bonne (2008), Koribalski \& 
L\'opez-S\'anchez (2009), van Eymeren et al. (2008, 2009c, 2010), Kirby et 
al. (2012), L\'opez-S\'anchez et al. (2008, 2012, 2015), Johnson et al. 
(2015), Kamphuis et al. (2015), Wang et al. (2016, 2017), and Oh et al. 
(2018).  \\

\begin{table*} 
\caption{Major \HI\ surveys of Local Volume galaxies and related surveys 
         at other wavelengths.}
\begin{flushleft}
\begin{tabular}{llrlcl}
\hline
 Survey Name  & Telescope \& & No. of   & Galaxy & Distance & Reference \\
              & wavelength   & Galaxies &  types & [Mpc]    & \\
\hline
LVHIS (south) & ATCA \HI &  82~~~~~&       & $<$10  & this paper \\
LVHIS (north) & WSRT \HI &  23~~~~~&       & $<$10  & PI: E. J\"utte \\
WHISP         & WSRT \HI & 375 (0) & dIrr+S&   & van der Hulst et al. (2001) \\
HALOGAS       & WSRT \HI &  24 (0) &spirals& 3 -- 11& Heald et al. (2011) \\
FIGGS         & GMRT \HI &  60 (8) & dIrr  &        & Begum et al. (2008) \\
THINGS        & VLA \HI  &  34 (3) &       & 3 -- 15& Walter et al. (2008) \\
Little THINGS & VLA \HI  &  41 (0) & dwarfs& $<$10  & Hunter et al. (2012) \\
VLA-ANGST     & VLA \HI  &  35 (0) & dIrr+S& $<$4   & Ott et al. (2012) \\
\hline
              & B, R     & 72  (8) & dIrr  & $<$10 &Parodi et al. (2002, 2003)\\
LSI           & AAT NIR  & 57 (24) &       & $<$10  & Kirby et al. (2008a,b) \\
LSI           & AAT NIR  & 40 (32) &       & $<$10  & Young et al. (2014) \\
ANGST         & HST optical&69 (9) &dwarfs & 1 --  4& Dalcanton et al. (2009)\\
11HUGS        & \Ha, R      & 400 (29)  &  & 1 -- 11& Lee et al. (2009) \\
LVL           & Spitzer MIR & 258 (27)  &  & $<$11  & Dale et al. (2009) \\
              & GALEX UV    & 459 (54)  &  & $<$11  & Lee et al. (2011) \\
              & \Ha         & 436 (49)  &  & $<$11  & Kennicutt et al. (2008)\\
SINGS         & Spitzer MIR &  75 (5)   &  & $<$30  & Kennicutt et al. (2003)\\
KINGFISH      & Herschel FIR&  61 (4)   &  & $<$30  & Kennicutt et al. (2011)\\
HERACLES      & IRAM 30-m CO&  48 (0)   &  & $<$30  & Leroy et al. (2009) \\
SINGG         & CTIO \Ha, R &  93 (13)  &  &        & Meurer et al. (2006) \\
\hline
\end{tabular}
\end{flushleft}
Notes: Col.~(3) gives the number of galaxies observed in the specified survey
  and in brackets the overlap with LVHIS.
\end{table*}

\begin{figure*} 
  \mbox{\psfig{file=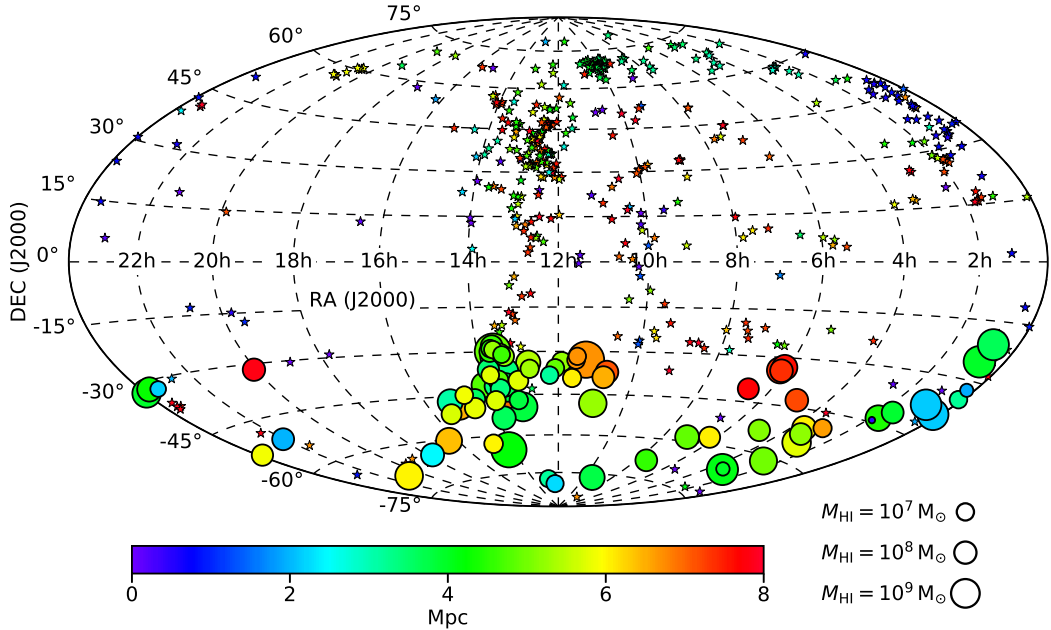,width=14cm}}
\caption{Aitoff distribution of Local Volume (LV) galaxies, highlighting the
  southern LVHIS galaxies (Dec $\la -30$\degr) presented in this paper. The
  \HI\ masses of LVHIS galaxies are indicated by their symbol sizes. Other 
  LV galaxies from the K13 sample are indicated by small stars. The symbol 
  colours indicate the galaxy distances.}
\end{figure*}

The first catalog of LV galaxies was presented by Kraan-Korteweg \& Tammann
(1979) and contained 179 galaxies with Local Group velocities, \vlg, less than 
500\kms. More recently, the LV sample defined by Karachentsev et al. (2004) 
included 451 galaxies with \vlg\ $<$ 550\kms\ or independent distance $D < 
10$~Mpc. A further expansion of the volume (\vlg\ $<$ 600\kms\ or $D < 11$~Mpc)
led to the most recent nearby galaxy catalog (K13; available 
at {\em www.sao.ru/lv/lvgdb}) which contains 
869 galaxies. Of these, only 261 galaxies (30\%) lie in the southern sky. 
Fig.~1 shows the locations of the LVHIS galaxies in the southern sky. \\


The \HI\ sizes of known LV galaxies cover more than two orders of magnitude, 
ranging from low-mass dwarf galaxies with diameters of less than 500~pc (e.g., 
Leo~T, Ryan-Weber et al. 2008) to grand-design spirals with \HI\ diameters of 
nearly 100~kpc, e.g., Circinus (For et al. 2012) and M\,83 (Koribalski 2015,
2017). Consequently, their \HI\ masses span more than four orders of magnitudes,
ranging from a few times 10$^5$\Msun\ to 10$^{10}$\Msun\ (Wang et al. 2016). 
The majority of LV galaxies are dwarf galaxies; their morphologies --- 
usually defined in the optical regime --- range from the typically gas-poor 
elliptical (dE) and spheroidal (dSph) dwarfs to gas-rich irregular (dIrr), 
Magellanic (dM) and blue compact dwarfs (BCD). In between sit the class of 
dwarf transitional galaxies (dSph/dIrr), many of which show \HI\ gas offset 
from their stellar disc (e.g., Phoenix, St-Germain et al. 1999). \\


Our primary goal is to obtain detailed \HI\ gas distributions of LV galaxies, 
analyse their structure and gas kinematics, measure their overall \HI\ extent, 
and search for companions. Furthermore, we investigate the influence of the 
galaxy environment on the shape of the outer \HI\ disc, where gas accretion 
as well as effects of ram pressure stripping and tidal interactions may be
detected. As an example, we refer to the multi-wavelength study of the gas 
dynamics and star formation in the nearby galaxy pair NGC~1512/1510 (HIPASS 
J0404--43) by Koribalski \& L\'opez-S\'anchez (2009). The \HI\ disc of the 
barred spiral galaxy NGC~1512 is spectacular and among the largest in the 
Local Volume (Koribalski 2017). Its gas distribution and kinematics show the 
effects of mild interaction with the BCD companion NGC~1510. In regions 
of high \HI\ column density star formation is prominent, giving rise to a 
well-defined spiral pattern in the outer disc. Furthermore, three tidal dwarf 
galaxy candidates with \HI\ masses around 10$^7$\Msun\ have been discovered 
within the extrapolated spiral/tidal arm of NGC~1512. Another example is the
multi-wavelength study of the dIrr galaxies NGC~5408 and IC\,4662 by van 
Eymeren et al. (2010), who find \HI\ discs extending well beyond their stellar 
extent. \\

Our paper is structured as follows: in Section~2 we introduce the LVHIS galaxy 
sample, followed by a description of the ATCA observations and data reduction 
in Section~3. Our results are presented in Sections~4 \& 5 with the latter 
containing short paragraphs for all LVHIS galaxies and associated galaxies. 
This is followed by our summary and outlook in Section~6. In the on-line 
Appendix we present the \HI\ moment maps and position-velocity ($pv$) diagrams 
for the majority of LVHIS galaxies.

\section{The LVHIS Galaxy Sample}

The {\em `Local Volume HI Survey'} (LVHIS)\footnote{LVHIS project webpage:
{\em www.atnf.csiro.au/research/LVHIS}} comprises deep interferometric \HI\ 
spectral line and 20-cm radio continuum observations of a complete sample of 
nearby, gas-rich galaxies and their surroundings. 
Following Karachentsev et al. (2004), we selected galaxies with \vlg\ $<$ 
550\kms\ or, when available, independently determined distances of $D < 10$
Mpc. To enable high-resolution \HI\ studies of LV galaxies with the Australia 
Telescope Compact Array (ATCA), we also require the selected galaxies to have 
declinations $\delta \la -30$\degr\ and be detected in the \HI\ Parkes All Sky 
Survey (HIPASS). This ensures that our target galaxies are bright enough for a 
detailed study of their \HI\ gas distribution and dynamics with a reasonably 
symmetric synthesized beam. Applying these criteria leads to an ATCA sample of 
82 LV galaxies. Fig.~2 shows the HIPASS \vlg\ of all LVHIS galaxies against 
their best available distances.

\begin{table*} 
\caption{Optical properties of LVHIS galaxies}
{\tiny
\begin{flushleft}
\begin{tabular}{llcccccccccccc}
\hline
(1) & (2) & (3) & (4) & (5) & (6) & (7) & (8) & (9) & (10) & (11) & (12)
    & \multicolumn{2}{c}{(13)} \\
HIPASS  Name      & Galaxy Name & $D$  &  type  & $A_B$ & $B_T$ & log \LB 
    & $R_T$ & ($B-R$)& \Dopt & $i$ & $PA$ &\multicolumn{2}{c}{group \& no. of}\\
    & & [Mpc] & & [mag] & [mag] & [\LLsun] & [mag] & [mag] & [arcsec] & [deg] 
    & [deg] & \multicolumn{2}{c}{close neighbours} \\
\hline
HIPASS J0008--34  & ESO349-G031	& 3.21 & IBm    & 0.043 & 15.54 &  6.99
  & 14.99 & 0.55 & 150 & --- & --- & Sculptor & 3 \\ 
                  & ESO294-G010	& 1.92 & dS0/Im & 0.021 & 15.66 &  6.49
  & 14.44 & 1.22 &  78 & 40 &  6 & Sculptor & 1 \\
HIPASS J0015--32  & ESO410-G005	& 1.92 & dS0-a  & 0.049 & 14.90 &  6.81
  & 14.01 & 0.89 & 102 & 35 & 54 & Sculptor & 2 \\ 
HIPASS J0015--39  & NGC~55	& 2.13 & SBm    & 0.048 & ~8.58 &  9.42 
  &  7.84 & 0.74 &2220 & 74 &108 & Sculptor & 2   \\ 
HIPASS J0047--20  & NGC~247     & 3.65 & SABd   & 0.065 & ~9.60 &  9.49
  &  8.80 & 0.80 &1800 & 73 &174 & Sculptor & 4 \\ 
  & \\
HIPASS J0047--25  & NGC~253	& 3.94 & SABc   & 0.068 & ~8.18 & 10.13
  &  6.66 & 1.52 &1920 & 76 & 52 & Sculptor & 5 \\ 
HIPASS J0054--37  & NGC~300	& 2.15 & Sd     & 0.046 & ~8.69 &  9.39
  &  7.46 & 1.23 &1800 & 48 &111 & Sculptor & 2 \\ 
HIPASS J0135--41  & NGC~625     & 3.89 & SBm    & 0.059 & 11.50 &  8.78
  & 10.61 & 0.89 & 390 & 72 & 92 & Sculptor & 2 \\ 
HIPASS J0145--43  & ESO245-G005 & 4.43 & IBm    & 0.059 & 12.74 &  8.40
  & 11.74 & 1.00 & 312 & 16 &122 & Sculptor & 2 \\ 
HIPASS J0150--44  & ESO245-G007 & 0.42 & Im     & 0.058 & 13.08 &  6.26
  & 12.01 & 1.07 & 300 & 26 & 90 & LG & 2 \\  
  & \\
HIPASS J0237--61  & ESO115-G021 & 4.99 & SBdm   & 0.094 & 13.26 &  8.31
  & 12.06 & 1.20 & 510 & 81 & 44 & & 0 \\ 
HIPASS J0256--54  & ESO154-G023 & 5.76 & SBm    & 0.060 & 12.71 &  8.61 
  & 12.01 & 0.70 & 540 & 80 & 39 & & 1 \\ 
HIPASS J0258--49  & ESO199-G007 & 6.6~ & Sd     & 0.078 & 16.44 &  7.27
  & 16.00 & 0.44 &  60 & 66 &  4 & & 1 \\ 
HIPASS J0317--66  & NGC~1313    & 4.07 & SBd    & 0.395 & ~9.38 &  9.81
  &  8.82 & 0.56 & 720 & 34 &  39  & dw-sp pair & 1 \\ 
HIPASS J0320--52  & NGC~1311    & 5.22 & SBm    & 0.078 & 13.23 &  8.35
  &(12.43) & (0.85)  & 210 & 73 & 40 & & 3 \\ 
  & \\
HIPASS J0321--66  & AM0319--662	& 3.98 & dIrr   & 0.303 & (17.6)& (6.46)
  &(16.70) & 0.90 &  (72)  & ---  &  --- & dw-sp pair & 1 \\ 
HIPASS J0333--50  & IC\,1959    & 6.05 & SBm    & 0.040 & 13.22 &  8.47
  &(12.41) & 0.81  & 228 & 78 &147 & & 1 \\ 
HIPASS J0454--53  & NGC~1705    & 5.11 & S0     & 0.029 & 12.82 &  8.48
  & 12.08 & 0.74 & 102 & 45 & 50 & & 0 \\ 
HIPASS J0457--42  & ESO252-IG001$^1$& 7.2~ &dIrr& 0.046 & 15.28 &  7.80
  & 15.04 & 0.24 &  90 & 75  & 56 & & 1 \\  
HIPASS J0605--33  & ESO364-G?029& 7.6~ & IBm    & 0.163 & 13.60 &  8.57
  & 13.23 & 0.37 & 210 & 44 & 52 & NGC~2188 & 2 \\ 
  & \\
HIPASS J0607--34  & AM0605--341	& 7.4~ & SBdm   & 0.132 &(14.28)& (8.25)
  &(14.36) & --0.08 & (48) & (76)  & (110) & NGC~2188 & 2 \\ 
HIPASS J0610--34  & NGC~2188   	& 7.4~ & SBm    & 0.118 & 12.10 &  9.13
  & 11.39 & 0.71 & 360 & 77 &175 & NGC~2188 & 2 \\ 
HIPASS J0615--57  & ESO121-G020	& 6.05 & Im     & 0.151 & 15.85 &  7.46
  & 15.49 & 0.36 &  78 & 32 & 49  & dw-dw pair& 1 \\ 
HIPASS J0639--40  & ESO308-G022	& 7.7~ & dIrr   & 0.327 & 16.23 &  7.59
  & 16.75 & --0.52 &  78 &  0 & ---  & & 0 \\ 
HIPASS J0705--58  & AM0704--582	& 4.90 & SBm    & 0.435 &(14.44)& (7.96)
  & (13.70) & (0.74)  & 55 & 45 & 174  & & 0 \\  
  & \\
HIPASS J0731--68  & ESO059-G001	& 4.57 & IBm    & 0.535 & 13.74 &  8.22
  & 12.80 & 0.94 & 126 & 18 & 84 & & 0 \\ 
HIPASS J0926--76  & NGC~2915   	& 3.78 & I0     & 0.997 & 12.93 &  8.56
  & 11.90 & 1.03 & 120 & 60 &130 & & 0  \\ 
HIPASS J1043--37  & ESO376-G016	& 7.1~ & dIrr   & 0.212 & 15.44 &  7.79
  & 15.24 & 0.20 &  60 & 37 &129  & & 1 \\ 
HIPASS J1047--38  & ESO318-G013	& 6.5~ & SBd    & 0.278 & 15.02 &  7.91
  & 13.96 & 1.06 & 192 & 81 & 75 &  & 1 \\ 
HIPASS J1057--48  & ESO215-G?009& 5.25 & dIrr   & 0.801 & 16.03 &  7.53
  & 15.02 & 1.01 & 120 & 60 & 72 &  & 0 \\ 
  & \\
HIPASS J1118--32  & NGC~3621    & 6.70 & Sd     & 0.292 & ~9.44 & 10.17
  &  8.07 & 1.37 &1200 & 60 &159 & NGC~3621 & 2 \\ 
HIPASS J1131--31  & new        	& 6.7~ & dIrr   & 0.253 & ---   & ---
  & --- & --- & --- & --- & --- & NGC~3621 & 2 \\ 
HIPASS J1132--32  & new        	& 6.7~ & dIrr   & 0.226 &(17.04)& (7.11)
  &(16.17) & (0.87) &  31 & 55 & 113 & NGC~3621 & 3 \\ 
HIPASS J1137--39  & ESO320-G014	& 6.08 & dIrr   & 0.519 & 15.85 &  7.62
  & 15.17 & 0.68 &  66 & 35 & 86 & & 0 \\ 
HIPASS J1154--33  & ESO379-G007	& 5.22 & dIrr   & 0.270 & 16.60 &  7.08
  & 15.50 & 1.10 &  78 & 32 & 90  &  & 3 \\ 
  & \\
HIPASS J1204--35  & ESO379-G024	& 4.9~ & dIrr   & 0.270 & 16.58 &  7.04
  & 16.37 & 0.21 &  72 & 41 & 30 &  & 2 \\ 
HIPASS J1214--38  & ESO321-G014	& 3.18 & IBm    & 0.342 & 15.21 &  7.24
  & 14.11 & 1.10 & 126 & 65 & 20 &  & 1 \\ 
HIPASS J1219--79  & IC\,3104   	& 2.27 & IBm    & 1.486 & 13.65 &  8.03
  & 12.04 & 1.61 & 180 & 55 & 45 &  & 2 \\ 
HIPASS J1244--35  & ESO381-G018	& 5.32 & dIrr   & 0.228 & 15.79 &  7.41
  & 14.84 & 0.95 &  72 & 54 & 83 & wide pair  & 1 \\ 
HIPASS J1246--33  & ESO381-G020	& 5.44 & IBm    & 0.238 & 14.24 &  8.05
  & 13.50 & 0.74 & 270 & 66 &138 & wide pair & 1 \\ 
  & \\
HIPASS J1247--77  & new        	& 3.16 & Im     & 2.747 &(17.49)& (7.30)
  &(16.25) & 1.24 &  ---  & ---  & ---  &  & 2 \\ 
HIPASS J1305--40  & CEN06      	& 5.78 & dIrr   & 0.372 &(15.92)& (7.48)
  &(15.04) & 0.88 &  (60)  & ---  & ---  & (Cen\,A) & 4 \\ 
HIPASS J1305--49  & NGC~4945   	& 3.80 & SBcd   & 0.640 & ~9.31 &  9.87
  &  7.55 & 1.76 &1560 & 77 & 43  & Cen\,A & 3 \\ 
HIPASS J1310--46A & ESO269-G058	& 3.80 & I0     & 0.394 & 12.50 &  8.50
  & 11.31 & 1.19 & 180 & 48 & 69  & Cen\,A & 8 \\ 
HIPASS J1321--31  & new      & 5.22 & dSph/dIrr & 0.222 &(17.1) & (6.86)
  & --- & --- & (78) & --- & --- & (M\,83) & 10 \\  
  & \\
HIPASS J1321--36  & NGC~5102   	& 3.40 & S0     & 0.199 & ~9.74 &  9.43
  &  8.47 & 1.27 & 720 & 65 & 48 & Cen\,A & 4 \\ 
HIPASS J1324--30  & AM1321--304 & 4.63 & dIrr   & 0.250 &(16.25)& (7.11)
  & --- & --- & (51) & ---  & ---  & M\,83 & 9 \\ 
HIPASS J1324--42  & NGC~5128   	& 3.77 & S0     & 0.416 & ~7.60 & 10.45
  &  6.29 & 1.31 &1800 & 40 & 35 & {\bf Cen\,A} & 9 \\ 
HIPASS J1326--30A & IC\,4247   	& 4.97 & S?     & 0.235 & 14.41 &  7.90
  & 13.74 & 0.67 &  90 & 66 &158 & M\,83 & 14 \\ 
HIPASS J1327--41  & ESO324-G024	& 3.73 & Im     & 0.409 & 12.91 &  8.32
  & 12.13 & 0.78 & 270 & 41 & 50 & Cen\,A & 9 \\ 
  & \\
HIPASS J1334--45  & ESO270-G017 & 6.95 & SBm    & 0.404 & 11.69 &  9.35 
  & 10.03 & 1.66 &1020 & 83 &118 &  Cen\,A & 9 \\
HIPASS J1336--29  & UGCA~365   	& 5.25 & Im     & 0.192 & 15.49 &  7.50
  & 14.81 & 0.68 & 102 & 66 & 31 & M\,83 & 10 \\ 
HIPASS J1337--29  & NGC~5236   	& 4.92 & Sc     & 0.241 & ~8.22 & 10.37
  &  7.34 & 0.88 &1080 &  0 & ---& {\bf M\,83} & 9 \\ 
HIPASS J1337--39  & new        	& 4.83 & Im     & 0.271 &(16.08)& (7.22)
  &(16.27) & (--0.19) & (36) & --- & --- & Cen\,A & 7 \\ 
HIPASS J1337--42  & NGC~5237   	& 3.40 & I0     & 0.358 & 13.26 &  8.08
  & 12.34 & 0.92 & 114 & 33 &128 & Cen\,A & 8 \\ 
  & \\
HIPASS J1337--28  & ESO444-G084 & 4.61 & Im     & 0.249 & 15.01 &  7.60
  & 15.05 & --0.04 &  96 & 36 &(126) & M\,83 & 8 \\ 
HIPASS J1339--31  & NGC~5253   	& 3.56 & Im     & 0.202 & 11.17 &  8.90
  &(10.11) & (1.06)  & 360 & 71 & 45 & Cen\,A & 10 \\ 
HIPASS J1340--28  & IC\,4316   	& 4.41 & IBm    & 0.198 &(14.97)& (7.56)
  & --- & --- &  96 & 36 &(56) & M\,83 & 9 \\ 
HIPASS J1341--29  & NGC~5264   	& 4.53 & IBm    & 0.184 & 12.58 &  8.53
  & 11.49 & 1.09 & 210 & 31 &(65) & M\,83 & 9 \\ 
HIPASS J1345--41  & ESO325-G?011& 3.40 & IBm    & 0.319 & 14.02 &  7.76
  & 13.49 & 0.53 & 240 & 60 &118 & Cen\,A & 6 \\
  & \\
HIPASS J1348--37  & new         & 5.75 & dIrr   & 0.281 &(17.00)& (7.01)
  &(16.45) & (0.55) & --- & --- & --- & Cen\,A & 5 \\ 
HIPASS J1348--53  & ESO174-G?001& 3.6~ & Im?    & 1.817 &(14.44)& (8.24)
  &(14.20) & (0.24) & 180 & 71 &170  & Cen\,A & 0 \\
HIPASS J1349--36  & ESO383-G087 & 3.45 & SBdm   & 0.260 & 11.00 &  8.96
  & 10.11 & 0.89 & 360 & 24 &(93) & Cen\,A & 3 \\ 
HIPASS J1351--47  & new         & 5.73 & dIrr   & 0.523 &(16.51)& (7.30)
  &(15.83) & (0.68) & --- &--- & --- & Cen\,A & 2 \\ 
HIPASS J1403--41  & NGC~5408    & 4.81 & IBm    & 0.248 & 12.59 &  8.61
  & 11.96 & 0.63 & 156 & 52 & 62 & Cen\,A & 3 \\ 
  & \\
HIPASS J1413--65  & Circinus	& 4.2~ & Sb     & 5.279 &  8.90 &  9.86
  & --- & --- & 714 & 65 & 210 & & 1 \\
HIPASS J1428--46  & UKS1424--460& 3.58 & IBm    & 0.472 &  ---  &  ---
  & --- & --- & (21)&(69)& ---  & Cen\,A & 2 \\ 
HIPASS J1434--49  & ESO222-G010 & 5.8~ & dIrr   & 0.978 &(14.82)& (8.17)
  &(14.57) & (0.25) &  60 & 60 & 8 &  & 3 \\ 
HIPASS J1441--62  & new         & 6.0~ & dIrr   & 4.909 &  ---  &  ---
  & --- & --- & --- & ---  & --- & 1 \\ 
HIPASS J1443--44  & ESO272-G025 & 5.9~ & dIrr   & 0.595 & 14.79 &  8.04 
  & 14.00 & 0.79 &  84 & 50 & 62 &  & 3 \\
  &  \\
HIPASS J1501--48  & ESO223-G009 & 6.49 & Im     & 0.942 &(12.28)& (9.27)
  &(12.53) & (--0.25) & 240 & 41 &135 &  & 4 \\ 
HIPASS J1514--46  & ESO274-G001 & 3.09 & Sd     & 0.914 & 12.00 &  8.85
  & 10.97 & 1.03 & 900 & 80 & 38 &  & 2 \\ 
HIPASS J1526--51  & new         & 5.7~ & dIrr   & 2.299 &  ---  & ---
  & --- & --- & --- & --- & --- &  & 2 \\
HIPASS J1620--60  & ESO137-G018 & 6.40 & Sc     & 0.888 & 12.23 &  9.26 
  & 11.03 & 1.20 & 240 & 68 & 29 & & 0 \\ 
HIPASS J1747--64  & IC\,4662    & 2.44 & IBm    & 0.254 & 12.33 &  8.12
  & 11.01 & 1.32 & 180 & 58 &104 & & 1 \\
  & \\
HIPASS J2003--31  & ESO461-G036 & 7.83 & dIrr   & 1.100 & 17.06 &  7.58 
  &(16.35) & (0.71)  &  72 & 60 & 22 &  & 0 \\ 
HIPASS J2052--69  & IC\,5052    & 6.03 & SBd    & 0.184 & 11.58 &  9.18 
  & 10.91 & 0.67 & 390 & 79 &142 & & 0 \\ 
HIPASS J2202--51  & IC\,5152    & 1.97 & Im     & 0.91  & 11.03 &  8.72 
  & 10.19 & 0.84 & 360 & 48 &100 & Sculptor & 1 \\ 
HIPASS J2326--32  & UGCA~438    & 2.18 & IBm    & 0.053 & 13.89 &  7.32 
  & 13.19 & 0.70 & 102 & 28 &(138) & Sculptor & 1 \\ 
HIPASS J2343--31  & UGCA~442    & 4.27 & SBm    & 0.061 & 13.46 &  8.08 
  & 13.06 & 0.40 & 300 & 78 & 48 & Sculptor & 3 \\ 
  & \\
HIPASS J2352--52  & ESO149-G003 & 5.9~ & IBm    & 0.050 & 15.05 &  7.72
  &(14.73) & (0.32)   & 180 & 80 &148 & Sculptor & 0 \\
HIPASS J2357--32  & NGC~7793    & 3.91 & Sd     & 0.070 & ~9.72 &  9.50 
  &  8.71 & 1.01 & 840 & 55 & 98 & Sculptor & 2 \\ 
\hline
\end{tabular}
\end{flushleft}
}
\flushleft{\small 
{\bf Notes:} --- Col.~(3): Tully-Fisher (TF), Hubble (\Ho), and membership 
 (mem) distances are here given to one decimal accuracy, while TRGB and 
 cepheid distances are given to two decimal points.
 $^1$The optical properties listed are for ESO252-IG001 NED01.}
\end{table*}

\subsection{LVHIS --- Optical Galaxy Properties}
The optical properties of LVHIS galaxies, as obtained from the literature, are 
given in Table~2. Once multi-wavelength images from the SkyMapper Southern Sky 
Survey and the Large Synoptic Survey Telescope (LSST) are available, it will 
be possible to obtain a homogeneous set of optical properties to complement 
the \HI\ properties of all LVHIS galaxies. The Table~2 columns are:
Col.~(1+2) HIPASS and optical galaxy name; Col.~(3) best available galaxy 
distance, $D$, with the reference given in Section~5 (also available from K13
and references therein); Col.~(4) morphological type (from RC3, de Vaucouleurs 
et al. 1991, when available); Col.~(5) $B$-band extinction, \AB, from Schlafly 
\& Finkbeiner (2011); Col.~(6) $B$-band magnitude from Lauberts \& Valentijn 
(1989), using the Cousins $B_{\rm T}$ (typical uncertainties are $\pm$0.09 
mag), when available; Col.~(7) log \LB\ as calculated from \LB\ = $D^2 \times 
10^{10-0.4(B_{\rm T}-A_{\rm B}-M_{\rm B,\odot})}$ \Lsun\ (assuming an absolute 
$B$ magnitude for the Sun of $M_{\rm B,\odot}$ = 5.45 mag, Blanton et al. 
2003); Col.~(8) $R$-band magnitude (as above); Col.~(9) $B-R$ colour; 
Cols.~(10--12) $B$-band diameter at 25.5 mag\,arcsec$^{-2}$, inclination
angle, $i$, and position angle, $PA$, from Lauberts (1982); Col.~(13) galaxy 
subgroup and number of close neighbours (within 300\arcmin\ and \vsys\ 
$< 800$\kms). When magnitudes are not available in the Cousins filters
(Lauberts \& Valentijn 1989), we use SuperCosmos $B_{\rm J}$ and $R$ 
magnitudes from Doyle et al. (2005) for AM0605--34, HIPASS J1337--29, 
ESO174-G?001, ESO222-G010 and ESO223-G009, and HST F606W/814W magnitudes 
for HIPASS J1247--77, CEN06, HIPASS J1348--37, and HIPASS J1351--47; these 
are less reliable due to calibration issues and are given in brackets. 

For most of our sample, we use distance estimates from HST observations of the 
tip of the red giant branch (TRGB), which typically have uncertainties of 10\%.
Occasionally we use distances determined from the luminosity of cepheids, 
surface brightness fluctuations (SBF), the Tully-Fisher (TF) relation or based 
on group membership. For details see the descriptions of individual galaxies 
in Section~5. 

Histograms of LVHIS galaxy properties are shown in Fig.~\ref{fig:histograms},
highlighting the wide ranges of, e.g., stellar colours, stellar masses 
(\Mstellar) and \HI\ mass to light ratios (\MHI\,/\,\LB). We note that Table~2 
provides absolute $B$-band magnitudes and optical diameters (\Dopt; see 
Col.~10 in Table~2) for most of the galaxies with \MHI\ $> 10^7$\Msun\ and 
$R$-band magnitudes for most of galaxies with \MHI\ $> 10^8$~\Msun. 


\begin{figure} 
  \mbox{\epsfig{file=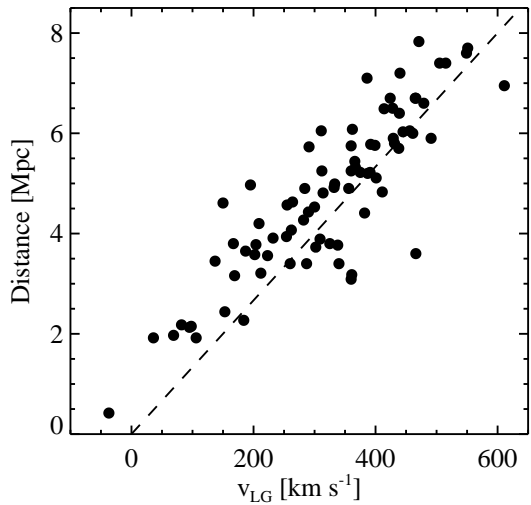,width=8.5cm}} 
\caption{Distance versus Local Group velocity \vlg\ for all LVHIS galaxies as 
   listed in Tables~2 and 4, respectively. The dashed line corresponds to $D$ 
   = \vlg\,/\Ho, where \Ho\ is the Hubble constant (here 75\kms\,Mpc$^{-1}$).}
\end{figure}

\begin{figure*} 
\mbox{\epsfig{file=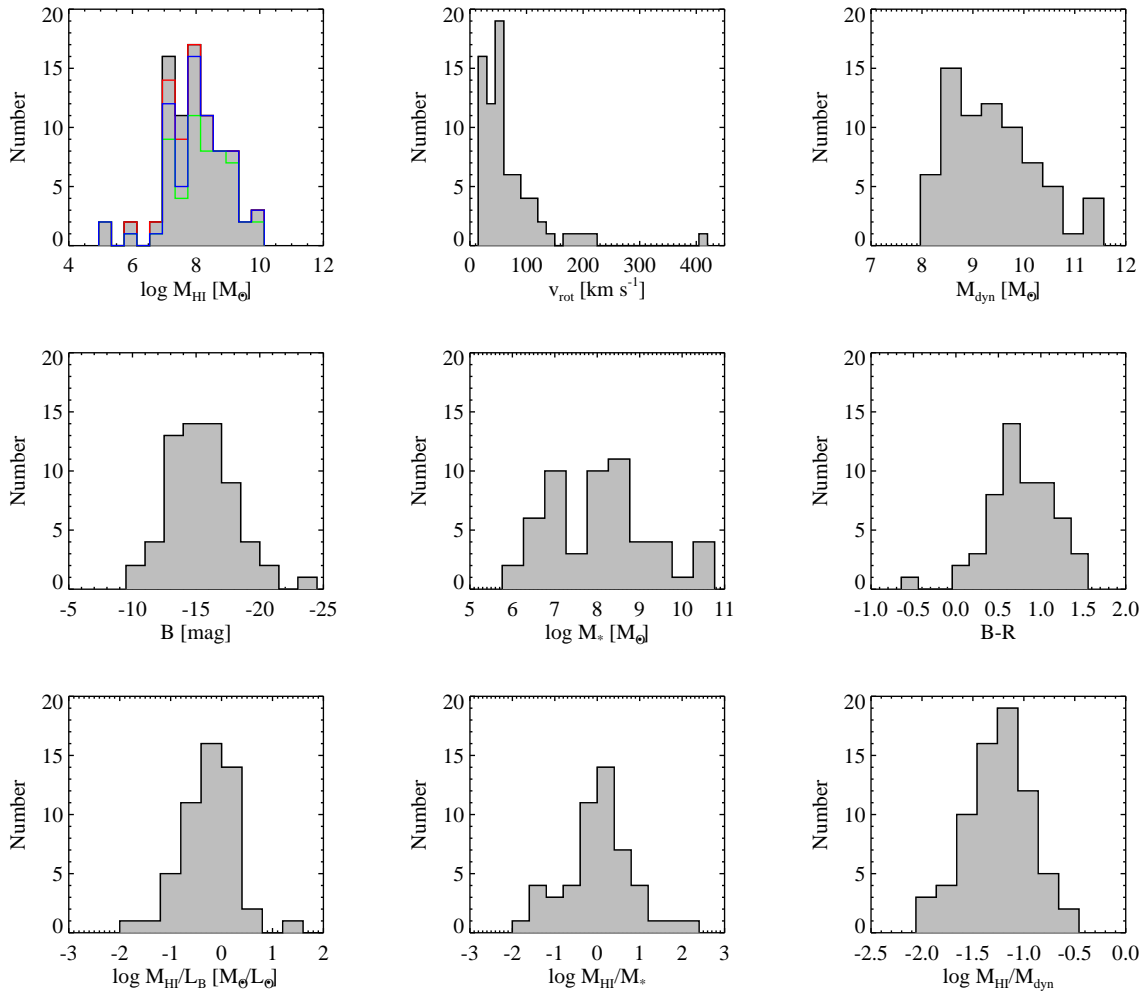,width=16cm}} 
\caption{Histograms of LVHIS galaxy properties as given Tables~2, 6 and 8. 
    The stellar mass, \Mstellar, is estimated from the $B$-band magnitudes 
    and $B-R$ colours based on the formula by Bell et al. (2003). 
    The \MHI\ histogram (top left) shows all LVHIS galaxies (black), those 
    with available $B$-band magnitudes (red), optical diameters (blue) and 
    both $B$- and $R$-band magnitudes (green). }
\label{fig:histograms}
\end{figure*}

\begin{table*} 
\caption{Multi-wavelength coverage for LVHIS galaxies}
{\tiny
\begin{flushleft}
\begin{tabular}{lllllll}
\hline
 HIPASS  Name    & Galaxy Name  &  NIR         & \Ha     & Other 
& Other \HI\ surveys \\
		 &              &                \\
\hline
HIPASS J0008--34 & ESO349-G031  & H-band (K08) & K07, B09  \\
                 & ESO294-G010  & H-band (K08) & B09     & ANGST \\
HIPASS J0015--32 & ESO410-G005  & H-band (Y14) & B09-N   & ANGST \\
HIPASS J0015--39 & NGC~55       &              & F96     & ANGST, 11HUGS \\
HIPASS J0047--20 & NGC~247      &              & LR99    & ANGST, 11HUGS \\
 & \\
HIPASS J0047--25 & NGC~253      &              & (SINGG) & ANGST, 11HUGS \\
HIPASS J0054--37 & NGC~300      &              & LR99    & ANGST, 11HUGS \\
HIPASS J0135--41 & NGC~625      & H-band (K08) & SINGG   & 11HUGS \\
HIPASS J0145--43 & ESO245-G005  & H-band (K08) & SINGG   & 11HUGS \\
HIPASS J0150--44 & ESO245-G007  & \\
 & \\
HIPASS J0237--61 & ESO115-G021  & H-band (K08) &         & 11HUGS \\
HIPASS J0256--54 & ESO154-G023  & H-band (K08) & SINGG   & 11HUGS \\
HIPASS J0258--49 & ESO199-G007  & H-band (Y14) &         & \\
HIPASS J0317--66 & NGC~1313     & H-band (K08) & (SINGG) & 11HUGS \\
HIPASS J0320--52 & NGC~1311     & H-band (K08) & SINGG   & 11HUGS \\
 & \\
HIPASS J0321--66 & AM0319--662  & H-band (K08) \\
HIPASS J0333--50 & IC\,1959     & H-band (K08) & SINGG   & 11HUGS \\
HIPASS J0454--53 & NGC~1705     & H-band (K08) & SINGG   & 11HUGS, SINGS  \\
HIPASS J0457--42 & ESO252-IG001 &              & SINGG   & \\
HIPASS J0605--33 & ESO364-G?029 & H-band (K08) \\
 & \\
HIPASS J0607--34 & AM0605--341  & H-band (Y14) \\
HIPASS J0610--34 & NGC~2188     & H-band (Y14) & RD03 \\
HIPASS J0615--57 & ESO121-G020  & H-band (K08) \\
HIPASS J0639--40 & ESO308-G022  & H-band (K08) \\
HIPASS J0705--58 & AM0704--582  & H-band (K08) \\
 & \\
HIPASS J0731--68 & ESO059-G001  & H-band (K08) \\
HIPASS J0926--76 & NGC~2915     & H-band (K08) &  & 11HUGS, SINGS, KINGFISH \\
HIPASS J1043--37 & ESO376-G016  & \\
HIPASS J1047--38 & ESO318-G013  & H-band (Y14) \\
HIPASS J1057--48 & ESO215-G?009 & \\
 & \\
HIPASS J1118--32 & NGC~3621     &   & & 11HUGS, SINGS, KINGFISH & THINGS \\
HIPASS J1131--31 & new          & \\
HIPASS J1132--32 & new          & \\
HIPASS J1137--39 & ESO320-G014  & H-band (Y14) \\
HIPASS J1154--33 & ESO379-G007  & H-band (Y14) & B09 & & FIGGS \\
 & \\
HIPASS J1204--35 & ESO379-G024  & H-band (Y14) \\
HIPASS J1214--38 & ESO321-G014  & H-band (Y14) & B09 & ANGST & FIGGS \\
HIPASS J1219--79 & IC\,3104     & \\
HIPASS J1244--35 & ESO381-G018  & \\
HIPASS J1246--33 & ESO381-G020  &              & B09 & 11HUGS \\
 & \\
HIPASS J1247--77 & new          & \\
HIPASS J1305--40 & CEN06        & H-band (Y14) & C09 \\
HIPASS J1305--49 & NGC~4945     & \\
HIPASS J1310--46A& ESO269-G058  & H-band (Y14) \\
HIPASS J1321--31 & new          & H-band (Y14) -- N & SINGG -- N & & FIGGS \\
 & \\
HIPASS J1321--36 & NGC~5102     & \\
HIPASS J1324--30 & AM1321--304  & H-band (Y14) & B09 & & FIGGS \\
HIPASS J1324--42 & NGC~5128     &              &       & 11HUGS \\
HIPASS J1326--30A& IC\,4247     & H-band (Y14) &       & 11HUGS \\
HIPASS J1327--41 & ESO324-G024  & H-band (Y14) & C09 & 11HUGS \\
 & \\
HIPASS J1334--45 & ESO270-G017  & \\
HIPASS J1336--29 & UGCA~365     & H-band (Y14) & C09 & & FIGGS \\
HIPASS J1337--29 & NGC~5236     &              & SINGG & 11HUGS & THINGS \\
HIPASS J1337--39 & new          & H-band (Y14) -- M & G07 \\
HIPASS J1337--42 & NGC~5237     & H-band (Y14) & C09 \\
 & \\
HIPASS J1337--28 & ESO444-G084  & H-band (Y14) & SINGG & 11HUGS \\
HIPASS J1339--31 & NGC~5253     & H-band (Y14) & SINGG & 11HUGS  \\
HIPASS J1340--28 & IC\,4316     & H-band (Y14) & C09   &        & FIGGS \\
HIPASS J1341--29 & NGC~5264     & H-band (Y14) & C09   & 11HUGS \\
HIPASS J1345--41 & ESO325-G?011 & H-band (Y14) & C09 \\
 & \\
HIPASS J1348--37 & new          & H-band (Y14) -- N \\
HIPASS J1348--53 & ESO174-G?001 & H-band (K08) -- N \\
HIPASS J1349--36 & ESO383-G087  & H-band (Y14) & C09 \\
HIPASS J1351--47 & new          & H-band (Y14) -- N \\
HIPASS J1403--41 & NGC~5408     &              & C09 & SINGS, KINGFISH \\
 & \\
HIPASS J1413--65 & Circinus     &              & E98 \\
HIPASS J1428--46 & UKS1424--460 & H-band (Y14) -- N & K07, C09 & & FIGGS \\
HIPASS J1434--49 & ESO222-G010  & H-band (Y14) & K07, C09 \\
HIPASS J1441--62 & new          & \\
HIPASS J1443--44 & ESO272-G025  & H-band (Y14) & K07, C09 \\
 & \\
HIPASS J1501--48 & ESO223-G009  & H-band (Y14) & C09 \\
HIPASS J1514--46 & ESO274-G001  & H-band (Y14) & RD03, C09 \\
HIPASS J1526--51 & new          & \\
HIPASS J1620--60 & ESO137-G018  &              & K07 \\
HIPASS J1747--64 & IC\,4662     & H-band (K08) & K07 & 11HUGS \\
 & \\
HIPASS J2003--31 & ESO461-G036  & H-band (K08) & & & FIGGS \\
HIPASS J2052--69 & IC\,5052     & H-band (K08) & SINGG, RD03, K07 & 11HUGS \\
HIPASS J2202--51 & IC\,5152     & H-band (K08) & K07 & ANGST, 11HUGS \\
HIPASS J2326--32 & UGCA~438     & H-band (K08) & K07 & ANGST, 11HUGS & FIGGS \\
HIPASS J2343--31 & UGCA~442     & H-band (K08) & (SINGG) & 11HUGS \\
 & \\
HIPASS J2352--52 & ESO149-G003  & H-band (Y14) & (SINGG), K07 & 11HUGS \\
HIPASS J2357--32 & NGC~7793     & H-band (K08) & (SINGG), SINGS, LR99 
                        & 11HUGS, SINGS, KINGFISH & THINGS \\
\hline
\end{tabular}
\end{flushleft}
}
\flushleft
{\bf Notes:} F96 = Ferguson et al. (1996), E98 = Elmouttie et al. (1998a), 
 LR99 - Larsen \& Richtler (1999), RD03 = Rossa \& Dettmar (2003), 
 K08 = Kirby et al. (2008b), Y14 = Young et al. (2014), G07 = 
 Grossi et al. (2007), K07 = Kaisin et al. (2007), B09 = Bouchard et al. 
 (2009), C09 = C\^ot\'e et al. (2009). --- Brackets indicate that the data have
 not yet been publicly released. N = not detected; M = marginal detection.
 --- The named surveys are described in Table~2.
\end{table*}

In Table~3 we give an (incomplete) overview of complimentary datasets available
for LVHIS galaxies. For example, Kirby et al. (2008a,b) and Young et al. (2014)
obtained deep AAT $H$-band images of LVHIS dwarf galaxies to better understand 
their stellar populations. \Ha\ surveys of LV galaxies were carried out by ---
among others --- Larsen \& Richtler (1999), Rossa \& Dettmar (2003), 
Karachentsev et al. (2005), Kaisin \& Karachentsev (2006), Meurer et al. 
(2006), Kaisin et al. (2007), Bouchard et al. (2009), and C\'ote et al. (2009).

\begin{table*} 
\caption{HIPASS properties of the LVHIS galaxies}
{\tiny
\begin{flushleft}
\begin{tabular}{llcrrcccrcccc}
\hline
 ~~~~~(1) & ~~~~~(2) & (3) & (4)~ & (5)~ & (6) & (7) & (8) & (9) & (10) & (11)\\
 HIPASS  Name     & Galaxy Name  & \vlg & \FHI & e\FHI & log \MHI & \vhel 
  & w50 & w20 & HIPASS & notes \\
 & & [\kkms] & \multicolumn{2}{r}{[Jy\kms]} & [\MMsun] & 
   \multicolumn{3}{c}{[\kkms]} & catalog \\
\hline
HIPASS J0008--34  & ESO349-G031  & 212 &   5.8  &    1.6 & 7.15 & 221 
  & 30 & 79 & BGC \\
                  & ESO294-G010  &(106)&   ---  &    --- & ---  & --- 
  & ---& --- & --- & (ATCA) \\ 
HIPASS J0015--32  & ESO410-G005  & (36)&   ---  &    --- & ---  & --- 
  & ---& --- & --- & (ATCA) \\
HIPASS J0015--39  & NGC~55       &  95 &1990.2  &  145.1 & 9.33 & 129 
  & 169 & 197 & BGC & e \\
HIPASS J0047--20  & NGC~247      & 187 & 608.2  &   42.1 & 9.28 & 156
  & 198 & 224 & BGC & e \\
  & \\
HIPASS J0047--25  & NGC~253      & 254 & 692.9  &   42.2 & 9.40 & 243
  & 407 & 431 & BGC & e \\
HIPASS J0054--37  & NGC~300      &  98 &1972.6  &  156.1 & 9.33 & 146
  & 147 & 166 & BGC & e \\
HIPASS J0135--41  & NGC~625      & 309 &  30.9  &    3.6 & 8.04 & 396 
  & 75 &  99 & BGC \\
HIPASS J0145--43  & ESO245-G005  & 290 &  81.0  &    9.1 & 8.57 & 391 
  & 60 &  85 & BGC \\
                  & ESO245-G007  &(--37)&  ---  &   ---  & ---  & (--23)
  & ---&  ---& --- & (ATCA) \\ 
  & \\
HIPASS J0237--61  & ESO115-G021  & 333 &  97.6  &    8.2 & 8.76 & 515
  & 121 & 145 & BGC \\
HIPASS J0256--54  & ESO154-G023  & 399 & 139.2  &   11.6 & 9.01 & 574
  & 122 & 143 & BGC \\
HIPASS J0258--49  & ESO199-G007  & 479 &   2.1  &    1.5 & 7.33 & 631
  &  55 &  74 & HICAT \\
HIPASS J0317--66  & NGC~1313     & 262 & 462.7  &   32.6 & 9.26 & 470
  & 168 & 196 & BGC & e  \\
HIPASS J0320--52  & NGC~1311     & 387 &  14.6  &    3.2 & 7.97 & 568 
  &  80 & 105 & BGC \\
  & \\
HIPASS J0321--66  & AM0319--662  & (532) &  ---   &  ---   & ---  & --- 
  & --- & --- & --- & (ATCA) \\  
HIPASS J0333--50  & IC\,1959     & 456 &  27.2  &    3.2 & 8.37 & 640
  & 128 & 155 & BGC \\
HIPASS J0454--53  & NGC~1705     & 401 &  15.4  &    2.6 & 7.98 & 632
  & 108 & 165 & BGC \\
HIPASS J0457--42  & ESO252-IG001 & 440 &  10.9  &    1.9 & 8.13 & 657
  &  58 &  99 & BGC \\
HIPASS J0605--33  & ESO364-G?029 & 549 &  17.6  &    2.5 & 8.38 & 786
  &  75 &  94 & BGC \\
  & \\
HIPASS J0607--34  & AM0605--341  & 515 &   9.0  &    2.1 & 8.07 & 766
  & 123 & 211  & HICAT \\
HIPASS J0610--34  & NGC~2188     & 505 &  32.5  &    3.9 & 8.62 & 747
  & 111 & 149 & BGC \\
HIPASS J0615--57  & ESO121-G020  & 311 &  14.1  &    2.9 & 8.09 & 577
  &  65 &  96 & BGC & c \\
HIPASS J0639--40  & ESO308-G022  & 551 &   3.8  &    1.5 & 7.73 & 822
  &  52 &  74 & HICAT \\
HIPASS J0705--58  & AM0704--582  & 284 &  34.8  &    4.4 & 8.29 & 564
  &   68 &  84 & BGC \\
  & \\
HIPASS J0731--68  & ESO059-G001  & 255 &  17.7  &    2.5 & 7.94 & 530
  &  82 & 104 & BGC \\
HIPASS J0926--76  & NGC~2915     & 204 & 108.4  &   13.9 & 8.56 & 468
  & 148 & 164 & BGC & e  \\
HIPASS J1043--37  & ESO376-G016  & 386 &  10.3  &    1.9 & 8.09 & 668
  &  33 &  53 & BGC \\
HIPASS J1047--38  & ESO318-G013  & 428 &   8.6  &    3.0 & 7.93 & 711
  &  42 &  71 & BGC \\
HIPASS J1057--48  & ESO215-G?009 & 312 & 104.4  &   11.5 & 8.83 & 598
  &  67 &  83 & BGC \\
  & \\
HIPASS J1118--32  & NGC~3621     & 466 & 884.3  &   56.2 & 9.97 & 730
  & 271 & 293 & BGC & e  \\
HIPASS J1131--31  & new          & 465 &   2.5  &    1.8 & 7.42 & 717
  &  29  &  ---  & ---   \\
HIPASS J1132--32  & new          & 424 &   2.8  &    1.8 & 7.47 & 699 
  &  59 & 105 & HICAT & uncertain \\
HIPASS J1137--39  & ESO320-G014  & 362 &   2.5  &    1.4 & 7.34 & 654 
  &  40 &  61 & HICAT \\
HIPASS J1154--33  & ESO379-G007  & 391 &   5.2  &    1.7 & 7.52 & 641 
  &  25 &  45 & BGC \\
  & \\
HIPASS J1204--35  & ESO379-G024  & 356 &   3.4  &    1.4 & 7.28 & 631 
  &  39 &  58 & HICAT \\
HIPASS J1214--38  & ESO321-G014  & 361 &   6.4  &    1.6 & 7.18 & 610 
  &  30 &  49 & BGC \\
HIPASS J1219--79  & IC\,3104     & 184 &  10.3  &    2.5 & 7.10 & 429
  &  40 &  63 & BGC \\
HIPASS J1244--35  & ESO381-G018  & 367 &   3.3  &    1.3 & 7.34 & 625
  &  40 &  62 & HICAT \\
HIPASS J1246--33  & ESO381-G020  & 366 &  30.9  &    3.7 & 8.34 & 589
  &  83 & 100 & BGC \\
  & \\
HIPASS J1247--77  & new          & 169 &   4.7  &    1.4 & 7.04 & 413
  &  32 &  46 & BGC \\
HIPASS J1305--40  & CEN06        & 392 &   5.1  &    1.9 & 7.60 & 617
  &  33 &  46 & BGC \\
HIPASS J1305--49  & NGC~4945     & 325 & 319.1  &   20.7 & 9.04 & 563
  & 361 & 386 & BGC & e   \\
HIPASS J1310--46A & ESO269-G058  & 167 &   7.2  &    2.3 & 7.39 & 400
  &  62 &  84 & B99 \\
HIPASS J1321--31  & new          & 375 &   5.9  &    1.6 & 7.58 & 571
  &  31 &  47 & BGC \\
  & \\
HIPASS J1321--36  & NGC~5102     & 260 &  80.1  &    6.0 & 8.34 & 468
  & 200 & 222 & BGC   \\
HIPASS J1324--30  & AM1321--304  & 264 &   3.9  &    2.5 & 7.30 & 500
  &  34 &  66 & B99   \\
HIPASS J1324--42  & NGC~5128     & 338 &  91.8  &   13.2 & 8.48 & 556
  & 477 & 542 & BGC & er   \\ 
HIPASS J1326--30A & IC\,4247     & 195 &   3.4  &    0.8 & 7.30 & 420
  &  33 & 49        & \multicolumn{2}{c}{B99, HIDEEP} \\ 
HIPASS J1327--41  & ESO324-G024  & 302 &  47.5  &    5.5 & 8.19 & 516
  &  81 & 107 & BGC   \\
  & \\
HIPASS J1334--45  & ESO270-G017  & 611 & 199.4  & 15.1   & 9.36 & 828
  & 141 & 158 & BGC \\ 
HIPASS J1336--29  & UGCA~365     & 360 &   1.2  &  1.2   & 7.04 & 572
  &  30 &  43 & HICAT \\ 
HIPASS J1337--29  & NGC~5236     & 332 &1630.3  &   95.8 & 9.97 & 513
  & 259 & 287 & BGC & ec  \\
HIPASS J1337--39  & new          & 287 &   6.6  &    1.8 & 7.56 & 492
  &  37 &  53 & BGC \\
HIPASS J1337--42  & NGC~5237     & 150 &  12.1  &    2.6 & 7.52 & 361
  &  77 &  96 & BGC \\
  & \\
HIPASS J1337--28  & ESO444-G084  & 411 &  21.1  &    3.2 & 8.02 & 587
  &  56 &  75 & BGC  \\ 
HIPASS J1339--31A & NGC~5253     & 223 &  44.4  &    4.7 & 8.12 & 407
  &  67 & 104 & BGC \\
HIPASS J1340--28  & IC\,4316     & 382 &   2.1  &    0.2 & 6.98 & 581
  &  ?  &  50 & HIDEEP  \\ 
HIPASS J1341--29  & NGC~5264     & 300 &  12.8  &    2.4 & 7.79 & 478
  &  35 &  55 & BGC   \\
HIPASS J1345--41  & ESO325-G?011 & 340 &  26.6  &    3.7 & 7.86 & 545
  &  59 &  75 & BGC \\
  & \\
HIPASS J1348--37  & new          & 360 &   2.5  &    1.5 & 7.29 & 581 
  &  39 &  60 & HICAT \\
HIPASS J1348--53  & ESO174-G?001 & 466 &  55.1  &    5.9 & 8.23 & 688
  &  71 & 103 & BGC \\
HIPASS J1349--36  & ESO383-G087  & 137 &  27.7  &    4.2 & 7.89 & 326
  &  33 &  52 & BGC \\
HIPASS J1351--47  & new          & 291 &   3.5  &    1.3 & 7.43 & 530
  &  39 &  59 & HICAT \\
HIPASS J1403--41  & NGC~5408     & 314 &  61.5  &    6.7 & 8.53 & 506 
  &  62 & 112 & BGC \\
  & \\
HIPASS J1413--65  & Circinus     & 209 &1450.5  &   97.9 & 9.78 & 434
  & 242 & 284 & BGC & e  \\
HIPASS J1428--46  & UKS1424--460 & 202 &  17.3  &    2.6 & 7.72 & 390
  &  48 &  66 & BGC \\
HIPASS J1434--49  & ESO222-G010  & 431 &   7.0  &    2.0 & 7.74 & 622
  &  38 &  62 & BGC \\
HIPASS J1441--62  & new          & 461 &   4.7  &    2.8 & 7.60 & 672
  &  52 &  68 & BGC \\
HIPASS J1443--44  & ESO272-G025  & 429 &   1.7  &    1.3 & 7.15 & 627
  & 42 &  71 & HICAT \\ 
  & \\
HIPASS J1501--48  & ESO223-G009  & 414 & 101.3  &   11.3 & 9.00 & 588
  &  61 &  89 & BGC \\
HIPASS J1514--46  & ESO274-G001  & 360 & 120.2  &   10.3 & 8.43 & 522 
  & 167 & 181 & BGC \\
HIPASS J1526--51  & new          & 438 &   6.0  &    2.6 & 7.66 & 605
  &  39 &  60 & BGC \\
HIPASS J1620--60  & ESO137-G018  & 439 &  37.4  &    4.9 & 8.56 & 605
  & 139 & 155 & BGC \\
HIPASS J1747--64  & IC\,4662     & 153 & 130.0  &   12.0 & 8.26 & 302
  &  86 & 133 & BGC \\
  & \\
HIPASS J2003--31  & ESO461-G036  & 471 &   7.5  &    1.8 & 8.04 & 427
  &  72 & 105 & HICAT \\
HIPASS J2052--69  & IC\,5052     & 445 & 101.7  &    7.6 & 8.94 & 584 
  & 174 & 203 & BGC \\
HIPASS J2202--51  & IC\,5152     &  69 &  97.2  &    9.5 & 7.95 & 122
  &  84 & 100 & BGC \\
HIPASS J2326--32  & UGCA~438     & (82)&  ---   &   ---  & ---  & ---
  & --- & --- & --- & (ATCA) \\ 
HIPASS J2343--31  & UGCA~442     & 282 &  50.1  &    5.3 & 8.33 & 267 
  &  94 & 112 & BGC \\
  & \\
HIPASS J2352--52  & ESO149-G003  & 491 &   6.9  &    1.6 & 7.75 & 576 
  &  39 &  70 & BGC \\
HIPASS J2357--32  & NGC~7793     & 232 & 278.5  &   20.4 & 9.00 & 227 
  & 172 & 191 & BGC & e \\
\hline
\end{tabular}
\end{flushleft}
}
\flushleft
Notes: Col.~(10) HIPASS catalogs: B99 (Banks et al. 1999), HIDEEP (Minchin 
  et al. 2003), BGC (Koribalski et al. 2004), HICAT (Meyer et al. 2004); 
  Col.~(11) e = extended, c = confused, r = severe baseline ripple.
%
\end{table*}

\subsection{LVHIS --- HIPASS Galaxy Properties}
HIPASS covers two thirds of the sky (up to DEC = +25\degr) and has an r.m.s. 
noise of $\sim$13~mJy\,beam$^{-1}$ per 13.2\kms\ channel width. The HIPASS
velocity resolution is 18\kms. For details of the observations, calibration 
and imaging techniques see Barnes et al. (2001). A typical 3$\sigma$ \HI\ 
flux (\FHI) detection limit for galaxies with a velocity width of 50\kms\ is 
2~Jy\kms, and the respective \HI\ column density limit is $N_{\rm HI} \sim 
10^{18}$~cm$^{-2}$ (for \HI\ gas filling the 15\farcm5 gridded beam). Efforts
are currently under way to create an improved, significantly deeper version 
of the survey (HIPASS~2); for details and first results see Calabretta et al. 
(2014) and Westmeier et al. (2017). While most of the galaxies selected here 
are listed in the 
HIPASS Bright Galaxy Catalog (BGC; Koribalski et al. 2004) and/or in the 
southern HIPASS catalog (HICAT; Meyer et al. 2004), a few of the fainter LV 
galaxies missed out either because of the velocity cutoff used for HICAT 
(\vgsr\ $>$ 300\kms; to avoid inclusion of HVCs and Galactic \HI\ emission) 
or confusion with neighbouring large galaxies. The HIPASS properties of LVHIS
galaxies are useful for comparison with our integrated ATCA \HI\ spectra,
in particular to estimate the amount of diffuse \HI\ emission filtered out
by the interferometer. \\

Table~4 gives the HIPASS properties of the LVHIS galaxies; the columns are as 
follows: Col.~(1+2) HIPASS and optical galaxy name; Col.~(3) the Local Group 
velocity, \vlg, as calculated in Koribalski et al. (2004); Col.~(4+5) the \HI\ 
flux density, \FHI, and its uncertainty, e\FHI; Col.~(6) logarithm of the \HI\ 
mass, \MHI, as calculated from \FHI, as listed here, using the distance $D$ 
given in Table~2; Col.~(7) \HI\ systemic velocity in the optical, heliocentric 
velocity frame, \vhel; Cols.~(8+9) \HI\ velocity widths determined at 50\% and 
20\% of the \HI\ peak flux; Col.~(10) HIPASS catalog from which the data were 
extracted: B99 = Banks et al. (1999), HIDEEP = Minchin et al. (2003), BGC = 
Koribalski et al. (2004), and HICAT = Meyer et al. (2004); and Col.~(11) notes 
on the HIPASS sources: e = extended, c = confused, r = baseline ripple.

\section{Observations and Data Reduction} 

Using the Australia Telescope Compact Array (ATCA) we obtained sensitive \HI\ 
spectral line and 20-cm radio continuum data for a complete sample of 82 
gas-rich galaxies in the Local Volume. Each LVHIS galaxy was typically observed
for a full synthesis (12 hrs) in three different ATCA configurations, providing 
good sensitivity to both small- and large-scale structures. The three ATCA 
configurations combined provide 45 antenna baselines with lengths ranging 
from 30~m to 6~km and a total on-source integration time of typically 
$\sim$30 hrs per target galaxy. 

The observations, which are summarised in Table~5, were carried out with the 
original ATCA correlator using two intermediate frequency (IF) bands. The 
first band (IF\,1) was typically centred on 1418 MHz with a bandwidth of 8~MHz,
divided into 512 channels. This covers an \HI\ velocity range from about 
--200\kms\ to +1200\kms\ with a velocity resolution of 4\kms. The ATCA primary 
beam is 33\farcm6 at 1418~MHz, i.e. the sensitivity of a single-pointing ATCA 
observation drops to 50\% at a distance of 16\farcm8 from the pointing centre. 
Large galaxies like NGC~55, NGC~247, NGC~300, M\,83, NGC~3621 and Circinus 
were mosaicked to ensure that the extended \HI\ emission of the outer disc 
was not missed.
The second band (IF\,2) was centred on 1384~MHz with a bandwidth of 128 MHz, 
divided into 32 channels to observe the 20-cm radio continuum emission. The 
results will be presented in a companion paper (Shao et al. 2017). 

Data reduction was carried out with the {\sc miriad} software package (Sault,
Teuben \& Wright 1995) using standard procedures. The IF\,1 data were split 
into a narrow band 20-cm radio continuum and an \HI\ line data set using a 
first order fit to the line-free channels. For this paper the \HI\ channel 
maps were made using `natural' (na) weighting of the {\em uv}-data in the 
velocity range covered by the \HI\ emission using steps of 4\kms. We use 
the shortest 30 baselines of the combined data sets to study the large-scale 
\HI\ emission in and around the selected galaxies. Together these provide 
good $uv$-coverage, angular resolution $\ga40''$ (ie., $\sim$1~kpc at $D$ = 
5 Mpc) and sensitivity to structures up to $\sim12'$. An on-source integration 
time of 30 hrs results in a theoretical r.m.s. sensitivity of 
$\sim$1.5 mJy\,beam$^{-1}$ per 4\kms\ channel, which is reached in most of 
our data cubes. 

In addition we made \HI\ cubes with low velocity resolution (20\kms), covering 
the whole observed velocity range. These cubes were then searched for 
extragalactic \HI\ from companions and more distant galaxies (up to 
$\sim$1200\kms) within the primary beam area. The rms of these cubes is 
typically 0.8 mJy\,beam$^{-1}$.

\HI\ moment maps were obtained by applying a $\sim$3$\sigma$ cutoff to the \HI\
data cubes. No masking was applied here, but may in future be used to better 
study the outer \HI\ edges and haloes of the LVHIS galaxies. Our 3$\sigma$ \HI\
column density (\NHI) sensitivity (for $\sigma$ = 1.5 mJy\,beam$^{-1}$) is 
$\sim$2.8 $\times 10^{19}$ cm$^{-2}$ over 20\kms\ for gas filling a 60\arcsec\ 
synthesized beam.
All ATCA \HI\ properties for LVHIS galaxies are obtained from primary-beam 
corrected cubes and moment maps.

\begin{table} 
\caption{ATCA \HI\ observations of LVHIS galaxies}
{\tiny
\begin{flushleft}
\begin{tabular}{llccc}
\hline
 HIPASS  Name     & Galaxy Name  & & ATCA & \\
		  &              &\multicolumn{3}{c}{configurations}\\
\hline
HIPASS J0008--34  & ESO349-G031  & EW352 & 750B & 1.5A \\
                  & ESO294-G010  &       & 750D &      \\ 
HIPASS J0015--32  & ESO410-G005  & EW367 & 750B & 1.5C \\
HIPASS J0015--39**& NGC~55   & EW352/367 & \multicolumn{2}{c}{\HI\ mosaic} \\ 
HIPASS J0047--20**& NGC~247  & EW352/367 & \multicolumn{2}{c}{\HI\ mosaic} \\ 
\\
HIPASS J0047--25**& NGC~253  & 375/EW367 & 750A/B & 1.5A/B/G \\ 
HIPASS J0054--37**& NGC~300  & EW352/367 & \multicolumn{2}{c}{\HI\ mosaic} \\ 
HIPASS J0135--41* & NGC~625      & EW352 & 750D & 1.5A \\
HIPASS J0145--43  & ESO245-G005  & EW352 & 750D & 1.5C \\
HIPASS J0150--44  & ESO245-G007  & 375/750D & \multicolumn{2}{c}{\HI\ mosaic}\\ 
\\
HIPASS J0237--61* & ESO115-G021  & EW352 & 750D & 1.5C \\
HIPASS J0256--54* & ESO154-G023  & EW367 & 750D & 1.5C \\
HIPASS J0258--49  & ESO199-G007  & EW352 & 750B & 1.5C \\
HIPASS J0317--66* & NGC~1313     &   375 &750A/C& 1.5C/D \\
HIPASS J0320--52  & NGC~1311     & EW352 & 750B & 1.5C \\
\\
HIPASS J0321--66  & AM0319--662  & EW367 & 750D & 1.5C \\
HIPASS J0333--50  & IC\,1959     & EW352 & 750B & 1.5C \\
HIPASS J0454--53  & NGC~1705     & EW352 & 750D & 1.5B \\
HIPASS J0457--42  & ESO252-IG001 & EW352 & 750D & 1.5C \\
HIPASS J0605--33  & ESO364-G?029 &   375 & 750C,e & 1.5B \\
\\
HIPASS J0607--34  & AM0605--341  & EW352 & 750D & 1.5C \\
HIPASS J0610--34  & NGC~2188     & EW352 & 750D & 1.5C \\
HIPASS J0615--57  & ESO121-G020  & EW352 & 750D & 1.5B \\
HIPASS J0639--40  & ESO308-G022  & EW367 & 750B & 1.5C \\
HIPASS J0705--58  & AM0704--582  & EW352 & 750C & 1.5A \\
\\
HIPASS J0731--68  & ESO059-G001  & EW352 & 750C & 1.5B \\
HIPASS J0926--76  & NGC~2915     & ~~375 & 750C & 1.5D \\
HIPASS J1043--37  & ESO376-G016  & EW352 & 750D & 1.5B \\
HIPASS J1047--38  & ESO318-G013  & EW352 & 750C & 1.5B \\
HIPASS J1057--48  & ESO215-G?009 & EW352 & 750A & 1.5C, 6A \\ 
\\
HIPASS J1118--32**& NGC~3621     & ~~375 & 750A & 1.5A \\ 
HIPASS J1131--31  & new          & EW352 & 750C & 1.5B \\
HIPASS J1132--32  & new          & EW352 & 750C & 1.5B \\
HIPASS J1137--39  & ESO320-G014  & EW352 & 750C & 1.5B \\
HIPASS J1154--33  & ESO379-G007  & EW367 & 750A & 1.5B \\
\\
HIPASS J1204--35  & ESO379-G024  & EW352 & 750A & 1.5B \\
HIPASS J1214--38  & ESO321-G014  & EW352 & 750A & 1.5B \\
HIPASS J1219--79  & IC\,3104     & EW367 & 750D & 1.5B \\
HIPASS J1244--35  & ESO381-G018  & EW352 & 750A & 1.5B \\
HIPASS J1246--33  & ESO381-G020  & EW352 & 750A & 1.5B \\
\\
HIPASS J1247--77  & new          & EW367 & 750D & 1.5B \\
HIPASS J1305--40  & CEN06        & EW367 & 750D & 1.5B \\
HIPASS J1305--49**& NGC~4945     & EW367 & 750A & ---  \\ 
HIPASS J1310--46A & ESO269-G058  & EW352 & 750A & 1.5D \\
HIPASS J1321--31  & new          & EW367 & 750D & 1.5D \\
\\
HIPASS J1321--36* & NGC~5102     & EW367 & 750B & 1.5D \\
HIPASS J1324--30  & AM1321--304  & EW367 & 750A & 1.5B \\  
HIPASS J1324--42**& NGC~5128     &  ---  & 750A & 1.5A, 6A \\
HIPASS J1326--30A & IC\,4247     & EW367 & 750A & 1.5B \\ 
HIPASS J1327--41  & ESO324-G024  &   375 & 750D & 1.5D \\
\\
HIPASS J1334--45  & ESO270-G017  & EW367 & 750A & 6A, 6C \\
HIPASS J1336--29  & UGCA~365     & EW367 & 750A & 1.5A \\ 
HIPASS J1337--39  & new          & EW367 & 750D & 1.5D \\
HIPASS J1337--42  & NGC~5237     & ~~375 & 750D & 1.5C,D \\
\\
HIPASS J1337--28  & ESO444-G084  & ~~375 & 750D &   6D \\
HIPASS J1339--31* & NGC~5253     & EW367 & 750A & 1.5A \\ 
HIPASS J1340--28  & IC\,4316     & EW367 & 750A & 1.5B \\ 
HIPASS J1341--29  & NGC~5264     & ~~375 & 750D & 1.5A \\
HIPASS J1345--41  & ESO325-G?011 & EW367 & 750A & 1.5B \\
\\
HIPASS J1348--37  & new          & EW352 & 750A & 1.5B \\
HIPASS J1348--53  & ESO174-G?001 & EW352 & 750B & 1.5B \\
HIPASS J1349--36  & ESO383-G087  & ~~375 & 750D & 1.5A \\
HIPASS J1351--47  & new          & EW367 & 750B & 1.5B \\ 
HIPASS J1403--41  & NGC~5408     & ~~375 & 750D & 1.5A \\
\\
HIPASS J1413--65* & Circinus     & ~~375 & \multicolumn{2}{c}{\HI\ mosaic}  \\ 
HIPASS J1428--46  & UKS1424--460 & EW367 & 750A & 1.5D \\
HIPASS J1434--49  & ESO222-G010  & EW367 & 750A & 1.5D \\
HIPASS J1441--62  & new          & EW367 & 750A & 1.5D \\
HIPASS J1443--44  & ESO272-G025  & EW352 & 750A & 1.5D \\
\\
HIPASS J1501--48  & ESO223-G009  & EW367 & 750A & 1.5D \\
HIPASS J1514--46**& ESO274-G001  & EW367 & 750A & 1.5D \\ 
HIPASS J1526--51  & new          & EW367 & 750A & 1.5D \\
HIPASS J1620--60  & ESO137-G018  & EW367 & 750A & 1.5D \\
HIPASS J1747--64  & IC\,4662     & EW367 & 750A & 1.5D \\
\\
HIPASS J2003--31  & ESO461-G036  & EW367 & 750B & 1.5A \\
HIPASS J2052--69* & IC\,5052     & EW367 & 750B & 1.5B \\ 
HIPASS J2202--51* & IC\,5152     & EW367 & 750A & 1.5B \\ 
HIPASS J2326--32  & UGCA~438     & EW367 & 750A & 1.5D \\ 
HIPASS J2343--31* & UGCA~442     & EW367 & 750B & 1.5C \\ 
\\
HIPASS J2352--52  & ESO149-G003  & EW352 & 750B & 1.5C \\
HIPASS J2357--32* & NGC~7793  & EW352/367 & \multicolumn{2}{c}{\HI\ mosaic} \\ 
\hline
\end{tabular}
\end{flushleft}
Notes: Galaxies marked with a single (double) star in Col.~(1) have optical 
  diameters $> 5'$ ($10'$), and additional \HI\ mosaics in the very compact 
  ATCA H75 array (and in some cases the EW214 array) have been completed.}
\flushleft
\end{table}

The Table~5 columns are: Col.~(1+2) HIPASS and optical galaxy name; Col.~(3--5)
ATCA configuration names. --- Galaxies marked with a single (double) star in 
Col.~(1) have optical diameters $> 5'$ ($10'$), and additional \HI\ mosaics in 
the very compact ATCA H75 array (and in some cases the EW214 array) have been 
completed.

\section{Results} 

Here we present the ATCA \HI\ atlas of LVHIS galaxies, consisting of \HI\ 
moment maps and position-velocity diagrams which are provided in the on-line 
Appendix. Colour figures are provided on our LVHIS webpages together with 
the opportunity to download FITS files of the ATCA \HI\ data products (\HI\ 
data cubes and moment maps). Most of the maps presented here were made with 
`natural' weighting to maximise sensitivity to diffuse, extended \HI\ emission 
in the outer discs of galaxies. The \HI\ galaxy properties, as measured from 
the primary-beam corrected ATCA integrated \HI\ intensity maps (mom0 maps), 
are listed in Table~6. 

The Table~6 columns are: Col.~(1+2) HIPASS and optical galaxy name; Col.~(3) 
total \HI\ flux density, \FHI; Col.~(4) total \HI\ mass, log \MHI; Cols.~(5--7)
\HI\ radius, \RHI, inclination angle, $i$, and position angle, $PA$, as 
measured at 1\Msun\,pc$^{-2}$, where cos($i$) is the ratio of the fitted \HI\ 
minor to major axes (see Wang et al. 2016); 
Col.~(8) $F^*_{\rm HI}$ enclosed within that diameter; Col.~(9) \HI\ mass to 
optical light ratio, \MHI/\LB, calculated as $1.5 \times 10^{-7}$ \FHI\ 
$10^{0.4(m_{\rm B}-A_{\rm B})}$ \Msun/\Lsun, ie. independent of distance; 
and Col.~(10) ratio of the \DHI\ to \Dopt.

\begin{figure}  
 \mbox{\psfig{file=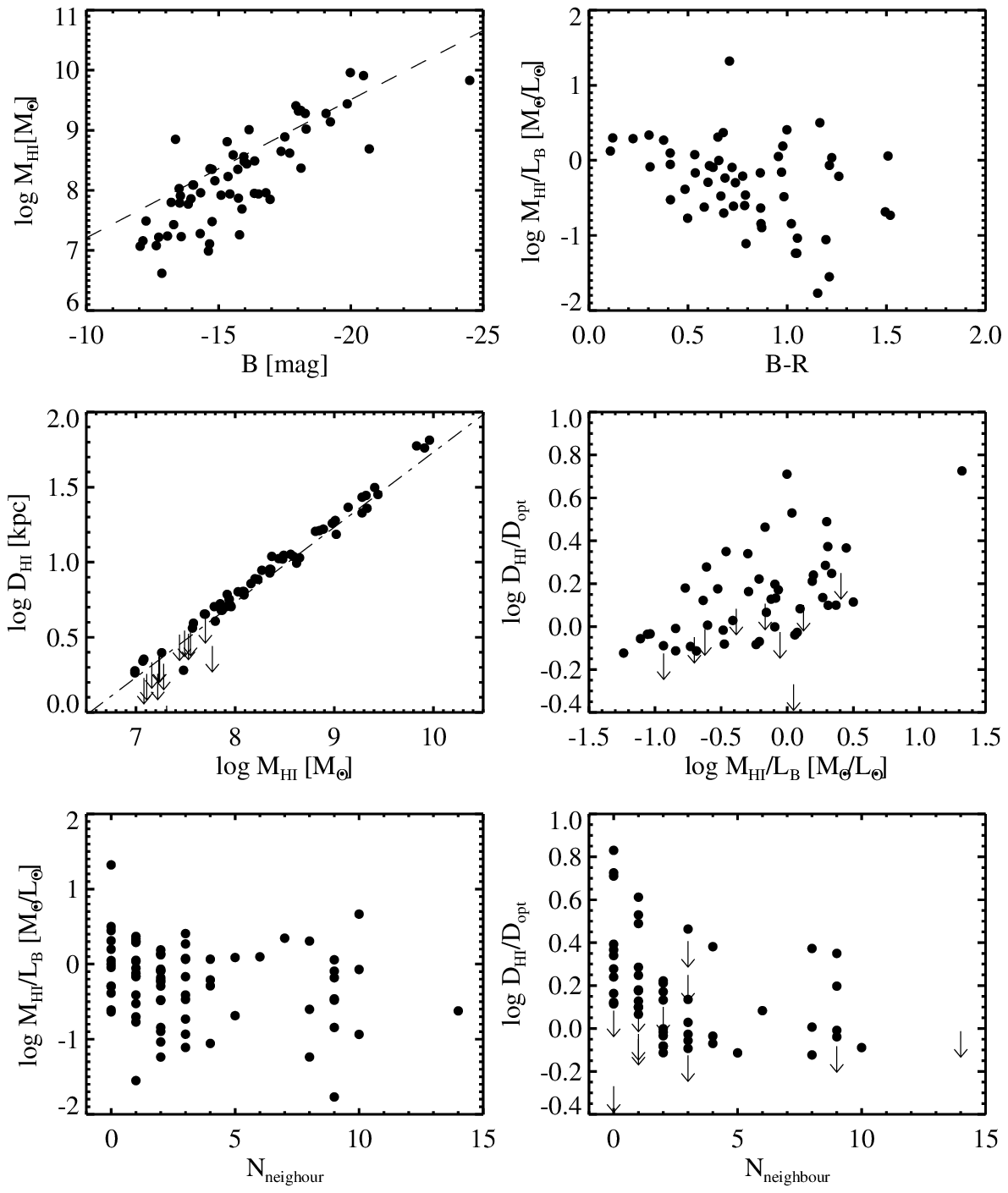,width=8.5cm}}
\caption{Correlations between optical and \HI\ properties for the LVHIS sample. 
  The dashed line in the top left panel shows the average relation from D\'enes
  et al. (2014); the dashed line in the middle left panel shows the average 
  relation from Wang et al. (2016). Arrows indicate the upper limits for the 
  unresolved galaxies.} 
\label{fig:correlations}
\end{figure}

\begin{figure}   
 \mbox{\psfig{file=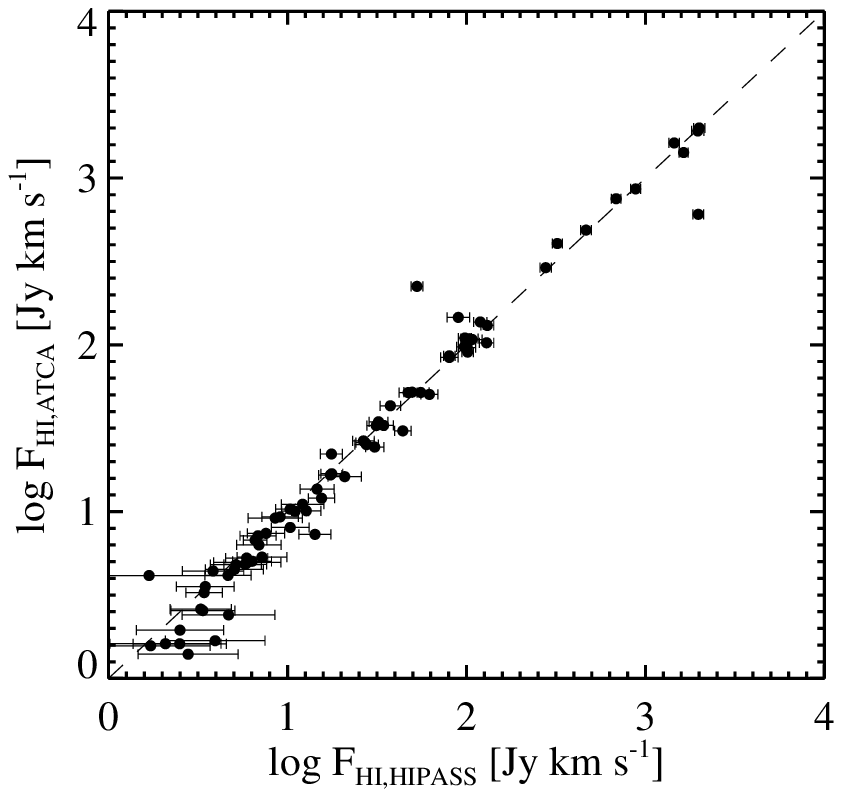,width=8cm}}
\caption{Comparison of the \HI\ flux densities of LVHIS galaxies as measured 
       in HIPASS (Table~4) and with the ATCA in this study (Table~6). }
\label{fig:FHI_atca_hipass}
\end{figure}

Correlations between these ATCA \HI\ properties and between \HI\ and optical 
properties are shown in Fig.~\ref{fig:correlations}. The \MHI\ versus $B$-band 
magnitudes and \MHI\ versus \DHI\ relations are consistent with the literature 
(e.g., D\'enes et al. 2014, Wang et al. 2016). There is a weak anti-correlation
between \MHI/\LB\ and $B-R$ colour, and a weak correlation between \DHI/\Dopt\ 
and \MHI/\LB. Both correlations show large scatter, which may be partly caused 
by the inhomogeneous optical dataset used in this study. 
Fig.~\ref{fig:FHI_atca_hipass} shows the measured HIPASS \FHI\ compared to 
the ATCA \FHI, listed in Tables~4 \& 6, respectively, for all LVHIS galaxies.\\

Table~7 gives the ATCA \HI\ centre positions for (a) galaxies newly detected 
in HIPASS, ie without a previously known optical or infrared counterpart, and
(b) newly detected companions to the LVHIS sample galaxies. The accurate \HI\
positions allow the secure identification of the stellar counterparts as shown 
in the ATCA \HI\ moment maps. 
The Table~7 columns are: Col.~(1) Galaxy name; Col.~(2) notes; Col.~(3--5) 
ATCA \HI\ centre position, deconvolved major and minor axes diameters, and 
$PA$, as obtained from a Gaussian fit to the \HI\ distribution (mom0). More 
accurate estimates will be obtained from the ATCA \HI\ high resolution maps 
made with robust/uniform weighting (an example is given in Fig.~11). \\

\begin{table*} 
\caption{ATCA \HI\ properties of LVHIS galaxies}
{\tiny
\begin{flushleft}
\begin{tabular}{llrrrrrrcccc}
\hline
 (1) & (2) & (3) & (4) & (5) & (6) & (7) & (8) & (9) & (10) & (11) & (12) \\
HIPASS Name & Galaxy Name & \FHI &log \MHI & \RHI & $i$ & $PA$ & \FHI$^*$ 
  & \MHI\,/\,\LB & \DHI\,/\,\Dopt & Figure & notes \\
 & & [Jy\kms] & [\MMsun] & [$"$] & [deg] & [deg] & [Jy\kms]  
  & [\MMsun/\LLsun] & ratio & No. & \\
\hline
HIPASS J0008--34 & ESO349-G031 &   4.8 & 7.07 &  70& 38& 32& 3.5& 1.2& 0.94
 & A1  & dSph/dIrr, (2) \\
                 & ESO294-G010 &   0.1 & 4.94 &---&---&---&---& ---& ---
 & --- & dSph/dIrr, (1) \\  
HIPASS J0015--32 & ESO410-G005 &   0.9 & 5.89 &  18&35& 54&   0.1 &0.1& 0.36
 & A2  & dSph/dIrr, (2) \\
HIPASS J0015--39 & NGC~55      &2025.4 & 9.34 &1106&69&103&1796.0 &0.8& 1.00
 & A3 & W2013 \\ 
HIPASS J0047--20 & NGC~247     & 662.5 & 9.32 &1544&73&173& 661.2 &0.9& 1.72
 & A4 \\
 & \\
HIPASS J0047--25 & NGC~253     & 746.8 & 9.44 & 739&73& 45& 713.6 &0.2& 0.77
 & A5 & \HI\ absorption \\ 
HIPASS J0054--37 & NGC~300     &1758.3 & 9.28 &1335&46&147&1486.1 &0.8& 1.48
 & A6 & W2011 \\
HIPASS J0135--41 & NGC~625     &  24.6 & 7.94 &150 &64 &106 &17.5 &0.1 & 0.77 
 & A7 \\
HIPASS J0145--43 & ESO245-G005 &  83.1 & 8.58 &253 &42 &106 &77.8 &1.5 &1.63 
 & A8  \\ 
HIPASS J0150--44 & ESO245-G007 &   2.3 & 4.98 &--- &---&--- & --- &0.1 & ---
 & Fig.~6 &  dSph/dIrr \\
 & \\
HIPASS J0237--61 & ESO115-G021 & 110.7 & 8.81 &331 &76 & 43 & 93.6&3.2&1.30
 & A9  \\
HIPASS J0256--54 & ESO154-G023 & 130.3 & 9.01 &339 &73 & 29 &110.5&2.3&1.26
 & A10 \\ 
HIPASS J0258--49 & ESO199-G007 &   1.6 & 7.22 & 39 &66 &  4 &  0.6&0.9&1.31
 & A11 & (2)\\
HIPASS J0317--66 & NGC~1313    & 491.3 & 9.28 &540 &51 & 14 &479.4&0.3&1.50
 & A12 \\ 
HIPASS J0320--52 & NGC~1311    &  13.6 & 7.94 &112 &73 & 40 &  9.7&0.4&1.07 
 & A14 \\
 & \\
HIPASS J0321--66 & AM0319-662  &   0.3 & 6.05 &--- & ---& ---& --- &0.3& ---
 & A13 & dSph/dIrr, (1) \\
HIPASS J0333--50 & IC\,1959    &  26.0 & 8.35 &152& 69 &150 &22.3 &0.8 &1.34 
 & A15 \\
HIPASS J0454--53 & NGC~1705    &  12.2 & 7.88 & 96 &45 & 50 & 8.7 &0.2 &1.90 
 & A16 \\
HIPASS J0457--42 & ESO252-IG001&   9.9 & 8.08 & 86 &46 & 68 & 8.7 &1.9 &1.93 
 & A17 \\
HIPASS J0605--33 & ESO364-G?029&  22.3 & 8.48 &142 &44 & 52 &19.6 &0.8 &1.36 
 & A18 \\
 & \\
HIPASS J0607--34 & AM0605-341  &   9.3 & 8.08 & 89 &47 & 84 & 7.7 &0.7 &3.71 
 & A19 \\
HIPASS J0610--34 & NGC~2188    &  34.3 & 8.65 &149 &54 &  2 &30.9 &0.3 &0.83 
 & A20 \\
HIPASS J0615--57 & ESO121-G020 &   7.3 & 7.80 & 69 &32 & 56 & 7.0 &2.2 &1.77 
 & A21 \\
HIPASS J0639--40 & ESO308-G022 &   4.4 & 7.79 & 67 &37 & 12 & 3.4 &1.6 &1.74 
 & A22 \\
HIPASS J0705--58 & AM0704-582  &  33.0 & 8.27 &186 &46 & 92 &30.2 &2.0 &3.38 
 & A23 \\
 & \\
HIPASS J0731--68 & ESO059-G001 &  16.9 & 7.92 &137 &48 &143 &14.0 &0.5 &2.19 
 & A24 \\
HIPASS J0926--76 & NGC~2915    & 108.7 & 8.56 &308 &56 &102 &50.7 &1.0 &5.13 
 & A25 \\ 
HIPASS J1043--37 & ESO376-G016 &  10.2 & 8.08 & 92 &36 &129 & 8.9 &2.0 &3.08 
 & A26 \\
HIPASS J1047--38 & ESO318-G013 &   9.1 & 7.96 &--- &---&--- & --- &1.1 &---
 & A27 & (1) \\
HIPASS J1057--48 & ESO215-G?009& 110.1 & 8.85 &318 &24 &139 &105.4&21.2& 5.32
 & A28 \\ 
 & \\
HIPASS J1118--32 & NGC~3621    & 856.8 & 9.96 &999 &75 &167 &684.8 &0.6 &1.67 
 & A29 \\ 
HIPASS J1131--31 & new         &   1.1 & 7.07 & ---&---& ---& --- & --- & ---
 & A30 & (1) \\
HIPASS J1132--32 & new         &   1.4 & 7.17 & ---&---& ---& --- & 0.9 & ---
 & A31 & (1) \\
HIPASS J1137--39 & ESO320-G014 &   2.0 & 7.24 & 49 &35 & 86 & 1.3 &0.4 &1.51
 & A32 & (2) \\
HIPASS J1154--33 & ESO379-G007 &   4.8 & 7.49 & 75 &32 & 90 & 3.7 &2.5 &1.94 
 & A33 \\
 & \\
HIPASS J1204--35 & ESO379-G024 &   2.6 & 7.17 & 52 &40 & 30 & 1.5 &1.3 &1.45
 & A34 & (2) \\
HIPASS J1214--38 & ESO321-G014 &   5.1 & 7.08 & 73 &65 & 20 & 3.2 &0.7 &1.16 
 & A35 \\
HIPASS J1219--79 & IC\,3104    &   8.1 & 6.99 & 83 &38 & 39 & 7.0 &0.1 &0.93 
 & A36 \\
HIPASS J1244--35 & ESO381-G018 &   2.6 & 7.24 & 53 &53 & 83 & 1.4 &0.7 &1.50 
 & A37 & (2)  \\
HIPASS J1246--33 & ESO381-G020 &  32.8 & 8.36 &169 &60 &136 &29.4 &2.0 &1.26 
 & A38 \\
 & \\
HIPASS J1247--77 & new         &   4.2 & 6.99 & 61 &48 & 55 & 3.4 &0.5 & --- 
 & A39 \\
HIPASS J1305--40 & CEN06       &   4.5 & 7.55 & 69 &57 & 81 & 2.4 &1.2 & ---
 & A40 & (2) \\
HIPASS J1305--49 & NGC~4945    & 405.3 & 9.14 &629 &70 & 48 &364.3&0.2 &0.81 
 & A41 & \HI\ absorption \\ 
HIPASS J1310--46A& ESO269-G058 &   5.4 & 7.26 & 67 &48 & 69 & 4.3 &0.1 &0.75 
 & A42 \\
HIPASS J1321--31 & new         &   5.2 & 7.52 & 73 &40 & 19 & 2.1 &4.6 &4.91 
 & A43 & dSph/dIrr, (2) \\
 & \\
HIPASS J1321--36 & NGC~5102    &  85.0 & 8.36 &331 &70 & 46 &53.6 &0.1 &0.92 
 & A44 \\
HIPASS J1324--30 & AM1321--304 &   1.7 & 6.93 & 38 &45 &106 & 0.6 &0.7 & ---
 & A45 & (1) \\
HIPASS J1324--42 & NGC~5128    & (144) &(8.68)&(814)&---&---&---&(0.02)& --- 
 & --- & \HI\ absorption \\ 
HIPASS J1326--30 & IC\,4247    &   3.3 & 7.28 & 54 &66 &158 & 1.2 &0.2 &1.21 
 & A46 & (2) \\
HIPASS J1327--41 & ESO324-G024 &  52.0 & 8.23 &212 &59 & 54 &44.1 &0.8 &1.58 
 & A47 \\
 & \\
HIPASS J1334--45 & ESO270-G017 & 224.7 & 9.41 &466 &79 &108 &199.4&1.1 &0.91 
 & A48 \\
HIPASS J1336--29 & UGCA~365    &   2.9 & 7.27 &--- &---&--- & --- &0.6 & ---
 & A49 & (1) \\ 
HIPASS J1337--29 & NGC~5236    &1428.5 & 9.91 &1208&51 &179 &941.9&0.3 &2.62 
 & A50 \\
HIPASS J1337--39 & new         &   6.7 & 7.57 & 77 &45 &  8 & 5.7 &2.2 &3.89 
 & A51 & dSph/dIrr \\
HIPASS J1337--42 & NGC~5237    &  11.0 & 7.48 & 57 &59 & 26 & 2.0 &0.2 &1.02 
 & A52 & dSph/dIrr \\
 & \\
HIPASS J1337--28 & ESO444-G084 &  16.3 & 7.91 &113 &55 & 79 &14.9 &2.0 &2.36 
 & A53 \\
HIPASS J1339--31 & NGC~5253    &  30.2 & 7.96 &146 &42 & 21 &26.8 &0.1 &0.81 
 & A54 \\
HIPASS J1340--28 & IC\,4316    &   2.6 & 7.08 & 53 &35 & 52 & 1.4 &0.3 &1.12 
 & A55 & (2) \\
HIPASS J1341--29 & NGC~5264    &  10.2 & 7.69 &102 &30 & 54 & 8.2 &0.1 &0.98 
 & A56 \\
HIPASS J1345--41 & ESO325-G?011&  26.5 & 7.86 &145 &50 &126 &23.7 &1.2 &1.21 
 & A57 \\
 & \\
HIPASS J1348--37 & new         &   1.6 & 7.10 &--- &---&--- & --- &1.2 & ---
 & A58 & dSph/dIrr, (1) \\
HIPASS J1348--53 & ESO174-G?001&  52.3 & 8.20 &222 &45 &160 &41.6 &0.9 &2.47 
 & A59 \\
HIPASS J1349--36 & ESO383-G087 &  25.4 & 7.85 &158 &50 & 19 &20.9 &0.1 &0.88 
 & A60 \\
HIPASS J1351--47 & new         &   3.5 & 7.43 & 67 &50 & 19 & 2.3 &1.4 & --- 
 & A61 & (2) \\
HIPASS J1403--41 & NGC~5408    &  49.9 & 8.43 &226 &57 &120 &42.6 &0.7 &2.91 
 & A62 \\
 & \\
HIPASS J1413--65 & Circinus    &1624.5 & 9.83 &1460&47 & 25&1318.0&0.9 &6.96 
 & A63 \\
HIPASS J1428--46 & UKS1424-460 &  16.7 & 7.70 &129 &57 &125 &14.3 & --- & ---
 & A64 \\
HIPASS J1434--49 & ESO222-G010 &   6.3 & 7.70 & 85 &60 &  8 & 4.3 &0.3 &2.87 
 & A65 & (2) \\
HIPASS J1441--62 & new         &   2.4 & 7.31 & 42 &58 &111 & 0.7 & --- & ---
 & A66 & (2) \\
HIPASS J1443--44 & ESO272-G025 &   1.6 & 7.12 & 42 &49 & 62 & 0.8 &0.1 &1.00 
 & A67 & (2) \\
 & \\
HIPASS J1501--48 & ESO223-G009 &  97.0 & 8.98 &288 &42 &165 &72.8 &0.5 &2.41 
 & A68 \\
HIPASS J1514--46 & ESO274-G001 & 138.4 & 8.49 &371 &77 & 35 &120.8&0.6 &0.82 
 & A69 \\
HIPASS J1526--51 & new         &   5.0 & 7.58 & 71 &34 &137 & 2.5 & --- & --- 
 & A70 \\
HIPASS J1620--60 & ESO137-G018 &  43.5 & 8.62 &158 &55 & 29 &41.6 &0.2 &1.32 
 & A71 \\
HIPASS J1747--64 & IC\,4662    & 103.5 & 8.16 &304 &35 & 79 &86.5 &1.1 &3.38 
 & A72 \\
 & \\
HIPASS J2003--31 & ESO461-G036 &   7.5 & 8.03 & 83 &60 & 22 & 4.4 &2.8 &2.33 
 & A73 \\
HIPASS J2052--69 & IC\,5052    &  90.0 & 8.89 &284 &71 &121 &69.5 &0.5 &1.46 
 & A74 \\
HIPASS J2202--51 & IC\,5152    &  98.4 & 7.95 &272 &46 & 94 &89.5 &0.2 &1.51 
 & A75 \\
HIPASS J2326--32 & UGCA~438    &   3.7 & 6.62 & 54 &28 & 89 & 1.0 &0.2 &1.06 
 & A76 & (2) \\
HIPASS J2343--31 & UGCA~442    &  52.2 & 8.35 &204 &67 & 43 &47.2 &1.9 &1.37 
 & A77 \\
 & \\
HIPASS J2352--52 & ESO149-G003 &   7.1 & 7.77 & 55 &80 &148 & 1.1 &1.1 &0.62 
 & A78 & (2) \\
HIPASS J2357--32 & NGC~7793    & 292.6 & 9.02 &991 &55 & 98 &290.1&0.4 &1.21 
 & A79 \\
\hline
\end{tabular}
\end{flushleft}
}
\flushleft
{\bf Notes:} (1) the \HI\ distribution is unresolved; (2) the \HI\ distribution
  is poorly resolved (\DHI\ $<2~{\rm B_{maj}}$), ie. \DHI\ and \DHI\,/\,\Dopt\ 
  are upper limits, and $i$ and $PA$ may differ significantly from the measured
  values; (3) \FHI\ is a lower limit due to significant \HI\ absorption. 
\end{table*}

\begin{table*} 
\caption{ATCA \HI\ properties for galaxies newly discovered in HIPASS and here}
\begin{flushleft}
\begin{tabular}{lccccr}
\hline
 (1) & (2) & (3) & (4) & (5) & (6)      \\
Galaxy Name       & notes         & $\alpha,\delta$ (J2000)  & \FHI 
   & \HI\ dimensions & $PA$   \\
   & & [hms, dms] & [Jy\kms] & (Gaussian fit) & [degr] \\
\hline
HIPASS J1131--31  & behind a star & 11:31:34.6, --31:40:28.3 & 1.13
   & unresolved \\ 
HIPASS J1132--32  &               & 11:33:10.6, --32:57:45.2 & 1.41 
   & unresolved \\ 
HIPASS J1247--77  &               & 12:47:32.4, --77:34:53.9 & 4.28
   & $61\arcsec \times 37\arcsec$ & 48 \\ 
HIPASS J1321--31  &               & 13:21:09.4, --31:32:01.2 & 5.24 
   & $143\arcsec \times 111\arcsec$ & 53 \\  
HIPASS J1337--39  &               & 13:37:25.0, --39:53:46.7 & 6.81 
   & $79\arcsec \times 50\arcsec$ & --12 \\  
HIPASS J1348--37  &               & 13:48:34.1, --37:58:08.0 & 1.54
   & $49\arcsec \times 34\arcsec$ & --5 \\ 
HIPASS J1351--47  &               & 13:51:21.2, --46:59:53.0 & 3.79
   & $81\arcsec \times 51\arcsec$ & 6   \\ 
HIPASS J1441--62  &               & 14:41:42.2, --62:46:04.2 & 2.61
   & $58\arcsec \times 29\arcsec$ & 78  \\ 
HIPASS J1526--51  &               & 15:26:22.4, --51:10:30.2 & 5.36
   & $97\arcsec \times 70\arcsec$ & --53 \\
\hline
 ATCA J023658--611838 & companion to ESO115-G021       
   & 02:36:58.8, --61:18:38.5 & 0.11 
   & unresolved \\ 
 ATCA J025640--543537 & background galaxy to ESO154-G023 
   & 02:56:40.3, --54:35:38.8 & 0.31 
   & unresolved \\  
 ATCA J045659--424758 & HIPASS J0457--42$^1$
   & 04:56:59.1, --42:47:58.3 & 9.96
   & $84\arcsec \times 41\arcsec$ & 75 \\
 ATCA J060511--332534 & near ESO364-G?029
   & 06:05:10.8, --33:25:34.4 & 0.99
   & $69\arcsec \times 34\arcsec$ & 20 \\
 ATCA J061608--574552 & companion to ESO121-G020       
   & 06:16:08.9, --57:45:52.3 & 2.10
   & $33\arcsec \times 27\arcsec$ & 36 \\ 
 ATCA J124850--774930 & companion to HIPASS J1247--77 
   & 12:48:50.3, --77:49:30.6 & 0.33 
   & unresolved \\ 
\hline
\end{tabular}
\end{flushleft}
Note: $^1$ \HI\ detected galaxy in ESO252-IG001 NED01
\end{table*}

\begin{table} 
\caption{Derived properties of the LVHIS galaxies}
{\tiny
\begin{flushleft}
\begin{tabular}{llccc}
\hline
 ~~~~~(1)         & ~~~~~(2)    & (3)     & (4)       &     (5)        \\
 HIPASS  Name     & Galaxy Name &   \vrot & log \Mdyn & log \MHI/\Mdyn \\
                  &             & [\kkms] & [\MMsun]  &                \\
\hline
HIPASS J0008--34  & ESO349-G031 &      51 &     8.82 &  --1.75 \\
                  & ESO294-G010 &     --- &     --- &     --- \\
HIPASS J0015--32  & ESO410-G005 &     --- &     --- &     --- \\
HIPASS J0015--39  &      NGC~55 &      97 &    10.39 &  --1.06 \\
HIPASS J0047--20  &     NGC~247 &     109 &    10.87 &  --1.56 \\
 \\
HIPASS J0047--25  &     NGC~253 &     217 &    11.19 &  --1.75 \\
HIPASS J0054--37  &     NGC~300 &     104 &    10.54 &  --1.26 \\
HIPASS J0135--41  &     NGC~625 &      46 &     9.14 &  --1.20 \\ 
HIPASS J0145--43  & ESO245-G005 &      52 &     9.52 &  --0.94 \\
HIPASS J0150--44  & ESO245-G007 &     --- &     ---  &     --- \\
 \\
HIPASS J0237--61  & ESO115-G021 &      66 &     9.91 &  --1.10 \\
HIPASS J0256--54  & ESO154-G023 &      66 &     9.98 &  --0.98 \\
HIPASS J0258--49  & ESO199-G007 &      32 &     8.46 &  --1.25 \\
HIPASS J0317--66  &    NGC~1313 &     116 &    10.52 &  --1.24 \\
HIPASS J0320--52  &    NGC~1311 &      47 &     9.15 &  --1.21 \\
 \\
HIPASS J0321--66  &  AM0319-662 &     --- &     --- &     --- \\
HIPASS J0333--50  &    IC\,1959 &      74 &     9.76 &  --1.41 \\
HIPASS J0454--53  &    NGC~1705 &     105 &     9.79 &  --1.91 \\
HIPASS J0457--42  &ESO252-IG001 &      58 &     9.36 &  --1.28 \\
HIPASS J0605--33  &ESO364-G?029 &      56 &     9.58 &  --1.10 \\
 \\
HIPASS J0607--34  &  AM0605-341 &     133 &    10.12 &  --2.04 \\
HIPASS J0610--34  &    NGC~2188 &      82 &     9.92 &  --1.28 \\
HIPASS J0615--57  & ESO121-G020 &      75 &     9.43 &  --1.63 \\
HIPASS J0639--40  & ESO308-G022 &      48 &     9.13 &  --1.34 \\
HIPASS J0705--58  &  AM0704-582 &      47 &     9.36 &  --1.09 \\
 \\
HIPASS J0731--68  & ESO059-G001 &      59 &     9.39 &  --1.47 \\
HIPASS J0926--76  &    NGC~2915 &      89 &    10.02 &  --1.45 \\
HIPASS J1043--37  & ESO376-G016 &      31 &     8.86 &  --0.78 \\
HIPASS J1047--38  & ESO318-G013 &     --- &     ---  &     --- \\ 
HIPASS J1057--48  &ESO215-G?009 &      82 &    10.10 &  --1.25 \\
 \\
HIPASS J1118--32  &    NGC~3621 &     143 &    11.19 &  --1.23 \\
HIPASS J1131--31  &         new &     --- &     ---  &     --- \\
HIPASS J1132--32  &         new &     --- &     ---  &     --- \\ 
HIPASS J1137--39  & ESO320-G014 &      39 &     8.71 &  --1.47 \\
HIPASS J1154--33  & ESO379-G007 &      27 &     8.52 &  --1.03 \\
 \\
HIPASS J1204--35  & ESO379-G024 &      33 &     8.48 &  --1.32 \\
HIPASS J1214--38  & ESO321-G014 &      18 &     7.94 &  --0.85 \\
HIPASS J1219--79  &    IC\,3104 &      38 &     8.49 &  --1.50 \\
HIPASS J1244--35  & ESO381-G018 &      29 &     8.42 &  --1.18 \\
HIPASS J1246--33  & ESO381-G020 &      48 &     9.38 &  --1.02 \\
 \\
HIPASS J1247--77  &         new &      20 &     7.94 &  --0.95 \\
HIPASS J1305--40  &       CEN06 &      18 &     8.15 &  --0.61 \\
HIPASS J1305--49  &    NGC~4945 &     197 &    11.02 &  --1.88 \\
HIPASS J1310--46  & ESO269-G058 &      46 &     8.78 &  --1.51 \\
HIPASS J1321--31  &         new &      24 &     8.39 &  --0.87 \\
 \\
HIPASS J1321--36  &    NGC~5102 &     110 &    10.18 &  --1.82 \\
HIPASS J1324--30  &  AM1321-304 &      35 &     8.39 &  --1.46 \\
HIPASS J1324--42  &    NGC~5128 &     --- &     --- &     --- \\
HIPASS J1326--30  &    IC\,4247 &      18 &     7.99 &  --0.71 \\
HIPASS J1327--41  & ESO324-G024 &      53 &     9.40 &  --1.17 \\
 \\
HIPASS J1334--45  & ESO270-G017 &      72 &    10.28 &  --0.87 \\
HIPASS J1336--29  &    UGCA~365 &     --- &     ---  &     --- \\
HIPASS J1337--29  &    NGC~5236 &     174 &    11.31 &  --1.40 \\
HIPASS J1337--39  &         new &      26 &     8.45 &  --0.89 \\
HIPASS J1337--42  &    NGC~5237 &      47 &     8.67 &  --1.20 \\
 \\
HIPASS J1337--28  & ESO444-G084 &      36 &     8.88 &  --0.97 \\
HIPASS J1339--31  &    NGC~5253 &      66 &     9.40 &  --1.45 \\
HIPASS J1340--28  &    IC\,4316 &      30 &     8.36 &  --1.29 \\
HIPASS J1341--29  &    NGC~5264 &      39 &     8.90 &  --1.20 \\
HIPASS J1345--41  &ESO325-G?011 &      39 &     8.91 &  --1.05 \\
 \\
HIPASS J1348--37  &         new &     --- &     --- &     --- \\
HIPASS J1348--53  &ESO174-G?001 &      62 &     9.53 &  --1.33 \\
HIPASS J1349--36  & ESO383-G087 &      23 &     8.53 &  --0.68 \\
HIPASS J1351--47  &         new &      28 &     8.53 &  --1.10 \\
HIPASS J1403--41  &    NGC~5408 &      57 &     9.60 &  --1.17 \\
 \\
HIPASS J1413--65  &    Circinus &     183 &    11.36 &  --1.53 \\
HIPASS J1428--46  & UKS1424-460 &      30 &     8.66 &  --0.96 \\
HIPASS J1434--49  & ESO222-G010 &      27 &     8.59 &  --0.89 \\
HIPASS J1441--62  &         new &      31 &     8.42 &  --1.11 \\
HIPASS J1443--44  & ESO272-G025 &      36 &     8.57 &  --1.45 \\
 \\
HIPASS J1501--48  & ESO223-G009 &      55 &     9.79 &  --0.81 \\
HIPASS J1514--46  & ESO274-G001 &      85 &     9.96 &  --1.47 \\
HIPASS J1526--51  &         new &      39 &     8.85 &  --1.26 \\
HIPASS J1620--60  & ESO137-G018 &      85 &     9.91 &  --1.29 \\
HIPASS J1747--64  &    IC\,4662 &     102 &     9.94 &  --1.77 \\
 \\
HIPASS J2003--31  & ESO461-G036 &      51 &     9.28 &  --1.25 \\
HIPASS J2052--69  &    IC\,5052 &      99 &    10.27 &  --1.39 \\
HIPASS J2202--51  &    IC\,5152 &      58 &     9.31 &  --1.36 \\
HIPASS J2326--32  &    UGCA~438 &     --- &     ---  &     --- \\
HIPASS J2343--31  &    UGCA~442 &      52 &     9.42 &  --1.07 \\
 \\
HIPASS J2352--52  & ESO149-G003 &      27 &     8.44 &  --0.67 \\
HIPASS J2357--32  &    NGC~7793 &     107 &    10.69 &  --1.67 \\
\hline
\end{tabular}
\end{flushleft}
}
\flushleft
\end{table}

The Table~8 columns are: Cols.~(1+2) HIPASS and optical galaxy name; Col.~(3) 
rotational velocity, calculated from the HIPASS 20\% profile width (given in
Table~4) corrected for instrumental broadening ($\sim$0.5 $\times$ velocity 
resolution = 9\kms), turbulence ($\sim$7\kms) and the galaxy inclination 
angle (from Table~6): \vrot\ = 0.5 ($w_{\rm 20}$ -- 16\kms)/sin($i$); Col.~(4) 
dynamical mass \Mdyn\ = $2.31 \times 10^5$ \vrot$^2$ $R_{\rm kpc}$, where 
$R_{\rm kpc}$ is \RHI\ in kpc; and Col.~(5) the \HI\ to dynamical mass ratio, 
\MHI/\Mdyn. \\

\begin{table*} 
\caption{ATCA \HI\ kinematic properties of LVHIS galaxies.}
\begin{flushleft}
\begin{tabular}{llcccccccc}
\hline
 (1) & (2) & (3) & (4) & (5) & (6) & (7) & (8) & (9) & (10) \\
HIPASS Name & Galaxy Name & \vrot & $R_{\rm max}$ & $i$ & $PA$ & log \Mdyn 
 & \MHI\,/\Mdyn & No. of & Notes \\
 & &[\kkms] & [kpc] & [degr] & [degr] & [\MMsun] & & beams & \\
\hline
HIPASS J0008--34 & ESO349-G031 & ~19.3 & ~2.0 & 42~~ &  298~~ & ~8.24 & 0.066 
  & 9  & C2000 \\ 
HIPASS J0015--39 & NGC~55      & ~69.7 & 20.1 &85--67& 110--93& 10.35 & 0.096
  & 8+ & warped, W2013 \\
HIPASS J0047--25 & NGC~253     & 200.0 & 17.2 &77--80&  229~~ & 11.20 & 0.017 
  & 8+ & starburst \\
HIPASS J0054--37 & NGC~300     & ~82.7 & 20.8 &40--50&290--332& 10.52 & 0.058
  & 8+ & warped, W2011 \\
HIPASS J0145--43 & ESO245-G005 & ~51~~ & ~6.4 & 36~~ & 70--98 & ~9.59 & 0.100  
  & $\sim$12 & warped, K2012 \\ 
                 &             & ~47.7 & ~6.7 & 54~~ &  ~88~~ & ~9.55 & 0.110  
  & $\sim$36 & C2000 \\ 
HIPASS J0237--61 & ESO115-G021 & ~63.0 & 10.4 &90--80&$\sim$45& ~9.97 & 0.068 
  & 8+ &  \\
HIPASS J0256--54 & ESO154-G023 & ~63.6 & 15.7 & 79.7 &218.3 & 10.17   & 0.070 
  & 8+ & warped  \\
HIPASS J0317--66 & NGC~1313    & 220.0 & 10.3 & 20.0 &$\sim$10&11.06  & 0.017 
  & 8+ \\
HIPASS J0320--52 & NGC~1311    & ~42.4 & ~3.6 & 71.4 &~40.5 & ~9.18   & 0.062 
   & $<$8 \\
HIPASS J0333--50 & IC\,1959    & ~65.9 & ~5.3 & 78.8 &149.0 & ~9.73   & 0.042 
   & $<$8 \\
HIPASS J0605--33 & ESO364-G?029& ~40.0 & ~4.4 & 70.5 &~60.0 & ~9.21   & 0.186
   & ? \\ 
HIPASS J0607--34 & AM0605--341 & ~85~~ & ~4.3 & (50) &274~~ & ~9.86   & 0.016  
   & $\sim$5 & K2012 \\ 
HIPASS J0615--57 & ESO121-G020 & ~48.7 & ~3.2 & (40) &265~~ & ~9.24   & 0.036  
   & $\sim$7 & K2012 \\ 
HIPASS J0639--40 & ESO308-G022 & ~40~~ & ~4.1 & (40) &~82~~ & ~9.18   & 0.041  
   & $\sim$5 & K2012 \\ 
HIPASS J0705--58 & AM0704-582  & ~38.5 & ~6.1 & 53.6 &275.9 & ~9.32   & 0.092 
   & 8+ \\
                 &             & ~57~~ & ~5.3 & (35) &276~~ & ~9.60   & 0.047  
   & $\sim$9 & K2012 \\ 
HIPASS J0731--68 & ESO059-G001 & ~61.0 & ~5.4 & 50.2 &323.4 & ~9.67   & 0.019 
   & 8+ \\
                 &             & ~61.8 & ~4.4 & 45~~ &329--319& ~9.59 & 0.021  
   & $\sim$9 & K2012 \\ 
HIPASS J0926--76 & NGC~2915    & ~80.0 & ~7.7 & 52.7 &292.6   & 10.06 & 0.028 
   & 8+  & E2011 \\ 
HIPASS J1057--48 & ESO215-G?009& ~93.0 & ~9.3 & 18.5 &119.7 & 10.27   & 0.040 
   & ? & $i < 20$\degr, uncertain \\ 
                 &             & ~53.8 & ~9.7 & 35~~ &123--116& ~9.81 & 0.109
   & $\sim$9 & K2012 \\ 
HIPASS J1118--32 & NGC 3621    & 130.0 & 39.0 &65--77&345--362& 11.18 & 0.060
   & 8+ & warped \\
HIPASS J1219--79 & IC\,3104    & ~15.3 & ~1.6 & 89.0 &  205.9 & ~7.94 & 0.113 
   & $<$8 \\
HIPASS J1246--33 & ESO381-G020 & ~46.7 & ~5.3 & 55~~ &295--314& ~9.43 & 0.086  
   & $\sim$9  & K2012 \\ 
                 &             & ~50.5 & ~5.8 & 57~~ &  311   & ~9.53 & 0.068
   & 25 & C2000 \\
HIPASS J1305--49 & NGC~4945    & 173.6 & 16.7 & 82.8 &~44.2   & 11.07 & 0.010 
   & 8+ & starburst \\
HIPASS J1321--36 & NGC~5102    & ~94.3 & 10.5 & 75.3 &~42.2   & 10.33 & 0.011 
   & 8+ & warped \\
HIPASS J1324--42 & NGC~5128    & 260.0 &  6.2 &85--105&100--130&10.99 & 0.005 
   & 8+ & warped, S2010 \\
HIPASS J1337--28 & ESO444-G084 & ~63.1 & ~4.2 & 32~~ &104~~   & ~9.59 & 0.021
   & 11 & C2000 \\
HIPASS J1337--29 & NGC~5236    & 150.0 & 54.9 & 37.0 &227--180& 11.46 & 0.028 
   & 8+ & warped \\
HIPASS J1337--42 & NGC~5237    & ~75.2 & ~4.9 & 33.8 &~50.2   & ~9.81 & 0.005 
   & 8+ & dSph/dIrr \\
HIPASS J1345--41 & ESO325-G?011& ~46.0 & ~3.1 & 42~~ &302~~   & ~9.18 & 0.048  
   & $\sim$9  & K2012 \\ 
                 &             & ~43.1 & ~3.4 & 52~~ &310~~   & ~9.16 & 0.050
   & 10 & C2000 \\
HIPASS J1348--53 & ESO174-G?001& ~97.3 & ~4.5 & 22.7 &218.4   & ~9.99 & 0.016
   & 8  & \\
                 &             & ~66~~ & ~6.6 & 40~~ &233--202& ~9.82 & 0.024
   & $\sim$13 & warped, K2012 \\
HIPASS J1413--65 & Circinus    & 161.4 & 47.2 & 62.2 &199.6   & 11.46 & 0.023 
  & 8+ & warped \\
HIPASS J1428--46 & UKS1424--460& ~22.0 & ~3.2 & 74.6 &122.8   & ~8.55 & 0.140 
  & $<$8 \\
HIPASS J1501--48 & ESO223-G009 & ~85.6 & 15.3 & 20.3 & 256.1  & 10.41 & 0.037 
  & 8+ & warped \\
HIPASS J1620--60 & ESO137-G018 & ~71.2 & ~8.6 & 71.5 & ~29.2  & 10.00 & 0.041 
  & 8+ \\
                 &             & ~80.1 & ~5.6 & 50~~ & 33--28 & ~9.92 & 0.050
  & $\sim$8 & K2012 \\
HIPASS J2003--31 & ESO461-G036 & ~51.0 & ~6.8 & 65   &330--350& ~9.61 & 0.026
  & 8+  & warped, K2011  \\
HIPASS J2052--69 & IC\,5052    & ~90.0 & 16.1 &$\sim$70&140--125&10.48& 0.026 
  & 8+ & warped \\
HIPASS J2202--51 & IC\,5152    & ~58.0 & ~3.6 &66--45&275--294& ~9.45 & 0.032 
  & 8+ & warped \\
                 &             & ~59.5 & ~4.0 & 49~~ &271--198& ~9.51 & 0.027
  & $\sim$20 & warped, K2012 \\
HIPASS J2343--31 & UGCA~442    & ~57.8 & ~6.2 & 64~~ &228~~   & ~9.68 & 0.046 
  & 10   & warped, C2000 \\  
HIPASS J2357--32 & NGC 7793    & 105.0 & ~9.1 & 48~~ &290--320& 10.37 & 0.045
  & 8+ & warped \\
\hline
\end{tabular}
\end{flushleft}
\flushleft{References: C2000 (C\^ot\'e et al. 2000), E2011 (Elson et al. 2011b),
   K2011 (Kreckel et al. 2011), K2012 (Kirby et al. 2012), O2017 (Oh et al. 
   2018), S2010 (Struve et al. 2010), W2011 (Westmeier et al. 2011), W2013 
   (Westmeier et al. 2013).}
\end{table*}

\subsection{\HI\ diameter relations}
Wang et al. (2016) compiled \DHI\ and \MHI\ estimates for a large sample 
of 561 nearby galaxies, incl. most of the LVHIS galaxies, and analyse the 
surprisingly tight \MHI--\DHI\ relation. Fig.~\ref{fig:correlations} (middle
left panel) shows the relation for LVHIS galaxies, incl. upper  limits for 
some unresolved LVHIS dwarf galaxies. A galaxy's \HI\ mass roughly scales
as the square of its \HI\ diameter: \MHI\ $\propto$ \DHI$^2$ (see also Broeils 
1992, Broeils \& Rhee 1997). Wang et al. (2016) also determine and compare
average \HI\ radial profiles for some of the galaxy samples in their study. 
The tightness of the \MHI--\DHI\ relation indicates a fundamental common 
mechanism in shaping the structure of \HI\ discs for a wide range of galaxies. 

We can use the \MHI--\DHI\ relation to obtain an approximate estimate of a 
galaxy's \HI\ diameter based on its \HI\ mass alone. This is particularly 
useful for large single-dish \HI\ surveys, like HIPASS, where the majority 
of galaxies are unresolved. We estimate that of the $\sim$5000 catalogued 
HIPASS galaxies (Koribalski et al. 2004, Meyer et al. 2004, Wong et al. 2006) 
$\ga$1000 are larger than 4\arcmin\ (ie eight beams in WALLABY), suitable for 
a detailed \HI\ kinematic analysis once observed in WALLABY (Koribalski 2012).

\subsection{\HI\ kinematic analysis}
A common way of analyzing the \HI\ kinematics of well-resolved disc galaxies 
is to fit tilted-ring models to the \HI\ data, using either the data cube or a 
carefully derived velocity field (Rogstad et al. 1974, de Blok et al. 2008, Oh 
et al. 2011). Kamphuis et al. (2015) recently developed a code for automated 
kinematic modelling of of disc galaxies that pipelines the {\em Tilted Ring 
Fitting Code} (TiRiFiC) by J\'ozsa et al. (2007). The so-called ``Fully 
Automated TiRiFiC'' (FAT) is particularly useful for a kinematic modelling of 
warped galaxies like some of our LVHIS galaxies (e.g., HIPASS J1413--65). 
Ultimately, FAT aims to analyse the \HI\ kinematics of well and marginally 
resolved galaxies from the upcoming SKA pathfinders' large \HI\ galaxy surveys 
like ASKAP WALLABY together with ``2D Bayesian Automated Tilted ring fitter''
(2DBAT; Oh et al. 2018) which is based on a Bayesian method for 2D tilted-ring 
analysis. 

Kamphuis et al. (2015) and Wang et al. (2017) make use of FAT to derive tilted
ring models and rotation curves for 26 and 10 LVHIS galaxies with large \HI\
discs, respectively. An updated list of the derived \HI\ properties for the 
successful fits is given in Table~9. The Kamphuis et al. (2015) sample was also
analysed by Oh et al. (2018) using 2DBAT. The Table~9 columns are: Cols.~(1+2) 
HIPASS and optical galaxy name; Col.~(3) rotational velocity, \vrot, near the 
maximum fitted \HI\ radius; Col.~(4) radius of the fitted \HI\ disc, \Rmax; 
Cols.~(5+6) range of fitted inclination and position angles over the fitted 
\HI\ disc; Col.~(7) dynamical mass \Mdyn\ = $2.31 \times 10^5$ \vrot$^2$ \Rmax;
Col.~(8) the \MHI\,/\,\Mdyn\ ratio; Col.~(9) number of resolution elements 
across the \HI\ disc major axis; and Col.~(10) notes.



\section{Notes on individual galaxies} 
In the following we briefly introduce each of the LVHIS galaxies and discuss 
their stellar morphologies, \HI\ properties and environment. We first discuss 
LVHIS galaxies in the Local Group (\S~5.1), followed by the well-known Sculptor
Group (RA $\sim$ 0~h; \S~5.2) and Cen\,A Group (RA $\sim$ 13~h; \S~5.3), which 
host the majority of LVHIS galaxies (see the group associations in Table~2), 
followed by the remaining galaxies in RA order (\S~5.4).

\begin{figure} 
\begin{tabular}{c}
  \mbox{\psfig{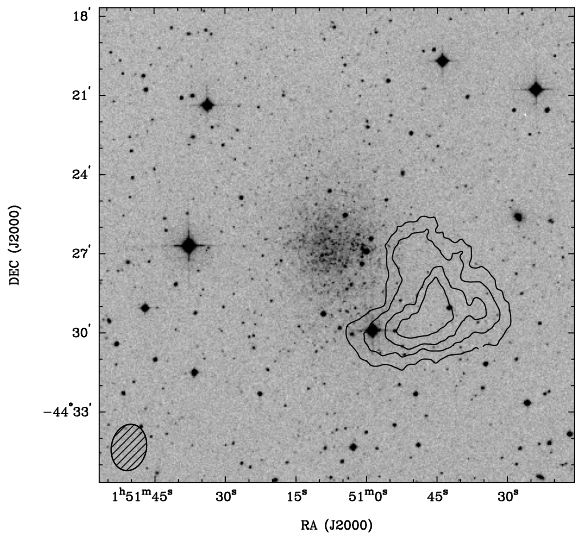}} 
\end{tabular}
\caption{\HI\ distribution (contours) of the galaxy ESO245-G007 
   (HIPASS J0150--44; $D_{\rm TRGB}$ = 0.42~Mpc), also known as the Phoenix 
   Dwarf, overlaid on to a DSS2 $R$-band image. Contour levels are 0.1, 0.2, 
   0.3, and 0.4 Jy\,beam$^{-1}$\kms. The angular resolution is $106\farcs2 
   \times 79\farcs6$ (216 pc $\times$ 162 pc).}
\end{figure}

\subsection{The Local Group} 
A comprehensive overview of galaxies in and around the Local Group is 
provided by McConnachie (2012), who catalogs $\sim$100 galaxies with reliable 
distance estimates of $D \la 3$~Mpc. Of these 75 are Local Group members 
($D < 1$~Mpc), mainly satellite galaxies around the Milky Way and Andromeda 
(M\,31). We note that the Milky Way sub-group contains only two gas-rich 
irregular galaxies, these are the Small and Large Magellanic Clouds, 
and at least 25 dSphs. The M\,31 sub-group is only slightly more diverse. In 
the outer reaches, McConnachie's sample overlaps with nearby galaxy groups, 
such as the Sculptor Group. For recent studies of the \HI\ content of Milky
Way satellites see Westmeier et al. (2015).
Our nearest neighbours, the Magellanic Clouds, have already been studied using 
large-scale Parkes and ATCA \HI\ mosaic observations. Stanimirovic et al. 
(1999) obtain an \HI\ mass of \MHI\ = $4.2 \times 10^8$\Msun\ and \DHI\ =
10.4 kpc for the Small Magellanic Cloud (SMC), while Staveley-Smith et al. 
(2003) derive \MHI\ = $4.8 \times 10^8$\Msun\ and \DHI\ = 18.6 kpc for the 
Large Magellanic Cloud (LMC) (assuming a distance of 50~kpc). \\

{\bf ESO245-G007 (HIPASS J0150--44)}, also known as the Phoenix Dwarf Galaxy, 
is a member of the Local Group ($D_{\rm TRGB}$ =  $420 \pm 10$ kpc). Its 
distance has been well constrained by several authors (e.g., Young et al. 
2007). Our ATCA \HI\ mosaic of ESO245-G007 is shown in Fig.~6, revealing an 
extended \HI\ cloud embracing the stellar core to the south-west. We detect 
\HI\ emission from about --40 to --8\kms, no clear velocity gradient is seen. 
The \HI\ emission associated with the Phoenix galaxy is also clearly detected 
in the re-calibrated HIPASS data. Phoenix is likely a transitional dwarf 
galaxy as previously suggested by Young \& Lo (1997), St-Germain et al. (1999) 
and Young et al. (2007) who carry out detailed studies of its \HI\ emission.
Martin\'ez-Delgado et al. (1999) investigate the stellar content of the Phoenix
Dwarf Galaxy and find a predominantly old population oriented north--south and 
a more compact, young population aligned  east--west; the latter shows an 
asymmetry in its distribution with more blue stars in the south-western part 
of the galaxy. Later stellar velocity measurements of $-13 \pm 9$\kms\ 
(Gallart et al. 2001, Irwin \& Tolstoy 2002) confirmed that the \HI\ cloud is 
indeed associated with the galaxy. Young et al. (2007) use deep VLA \HI\ data 
of Phoenix to investigate mechanisms which may transform gas-rich irregulars 
into gas-poor dwarf spheroidal galaxies. As the \HI\ cloud is associated with 
the most recent star formation in Phoenix, they suggest that the gas expulsion 
may have been caused by winds from supernovae. Using our ATCA maps we measure 
an \HI\ flux density of \FHI\ = 2.3 Jy\kms, corresponding to an \HI\ mass of 
\MHI\ $\sim$ $10^5$\Msun.  For comparison, St-Germain et al. (1999) measure 
\FHI\ = 4.0 Jy\kms\ with the ATCA, while Young et al. (2007) measure \FHI\ = 
$2.95 \pm 0.10$ Jy\kms\ with the VLA (single pointing). \\

\begin{figure*} 
\begin{tabular}{ll}
\mbox{\psfig{file=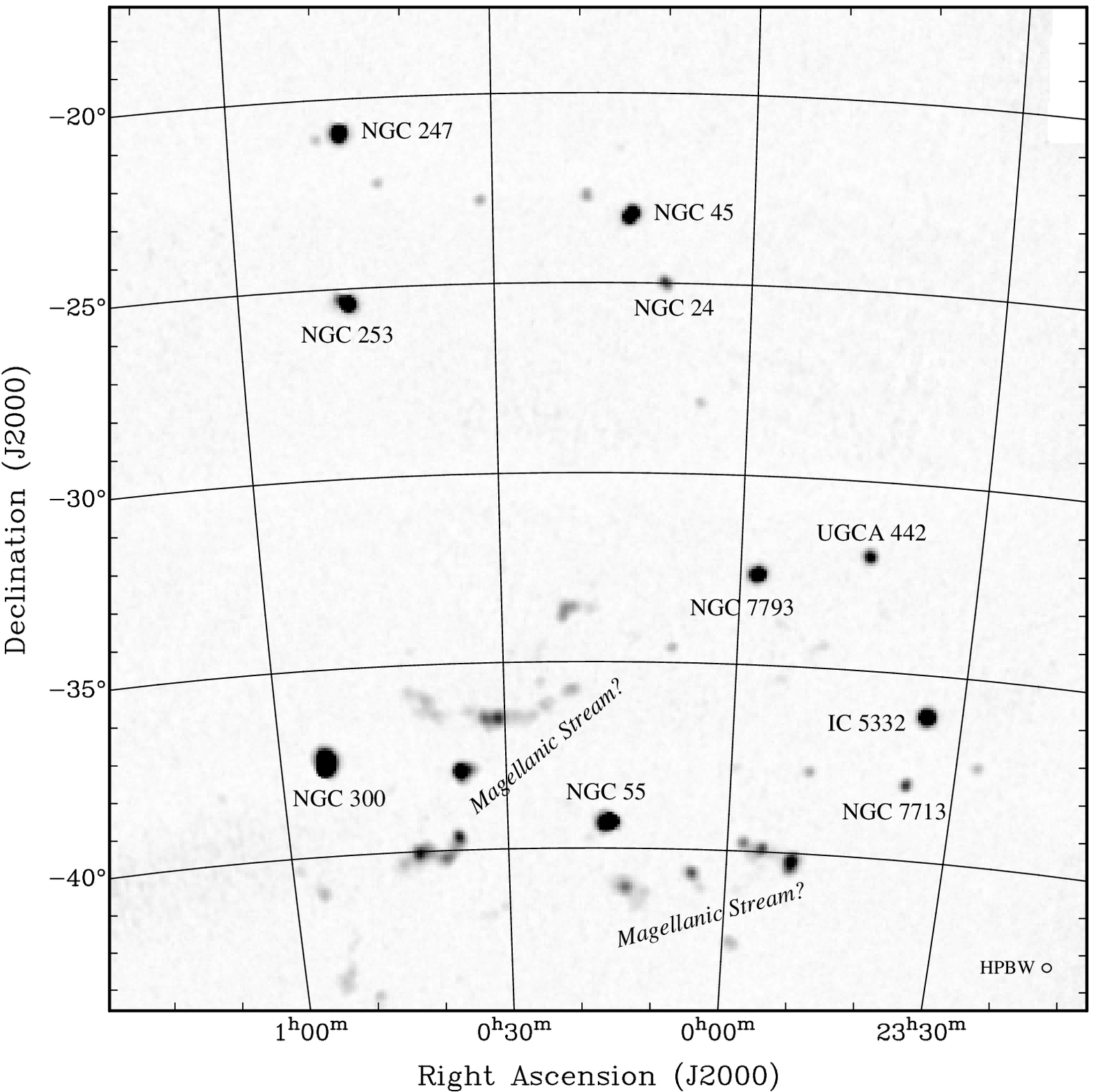,width=11cm,angle=0}} & 
  \begin{tabular}{c}
  \mbox{\psfig{file=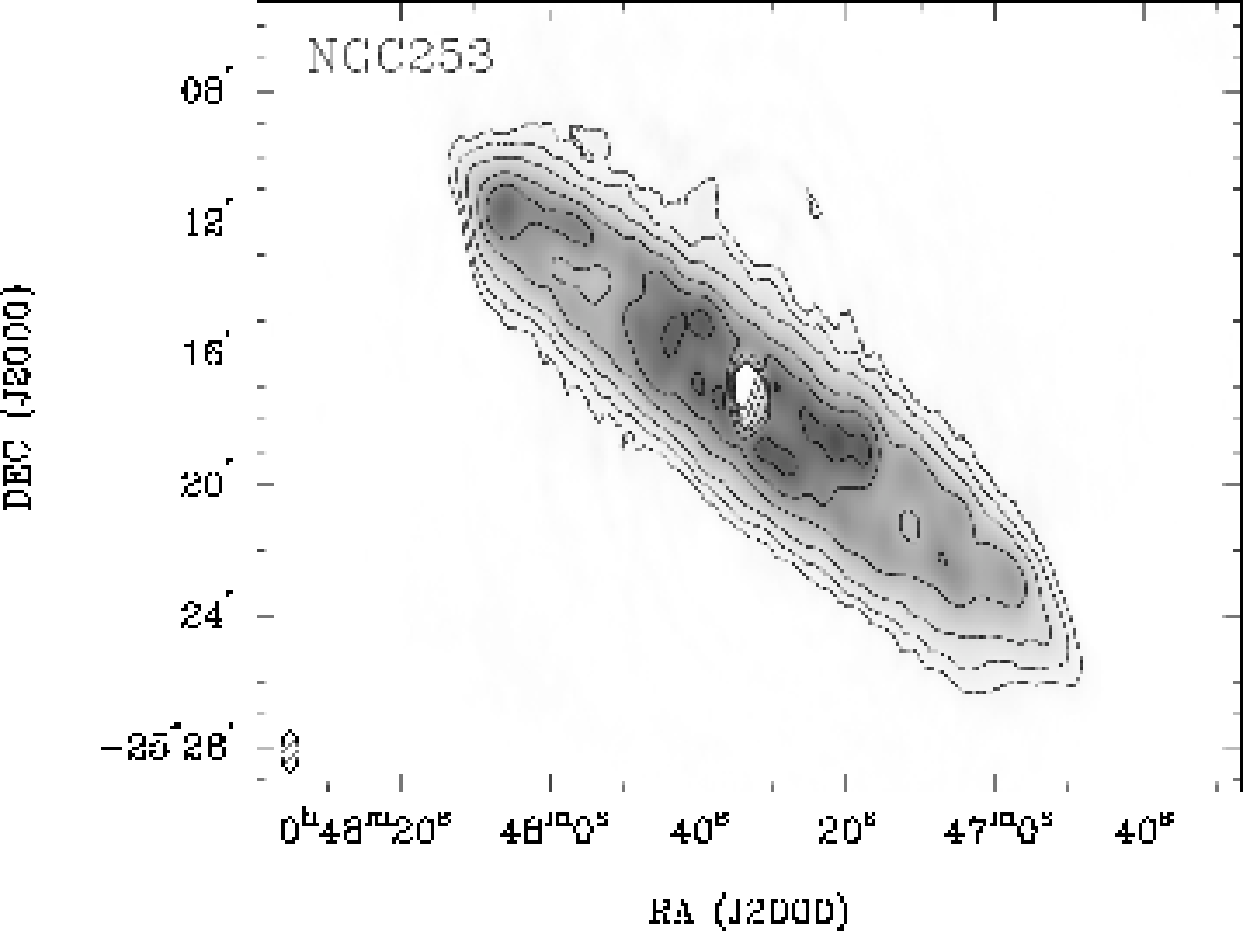,width=4cm,angle=-90}} \\
  \mbox{\psfig{file=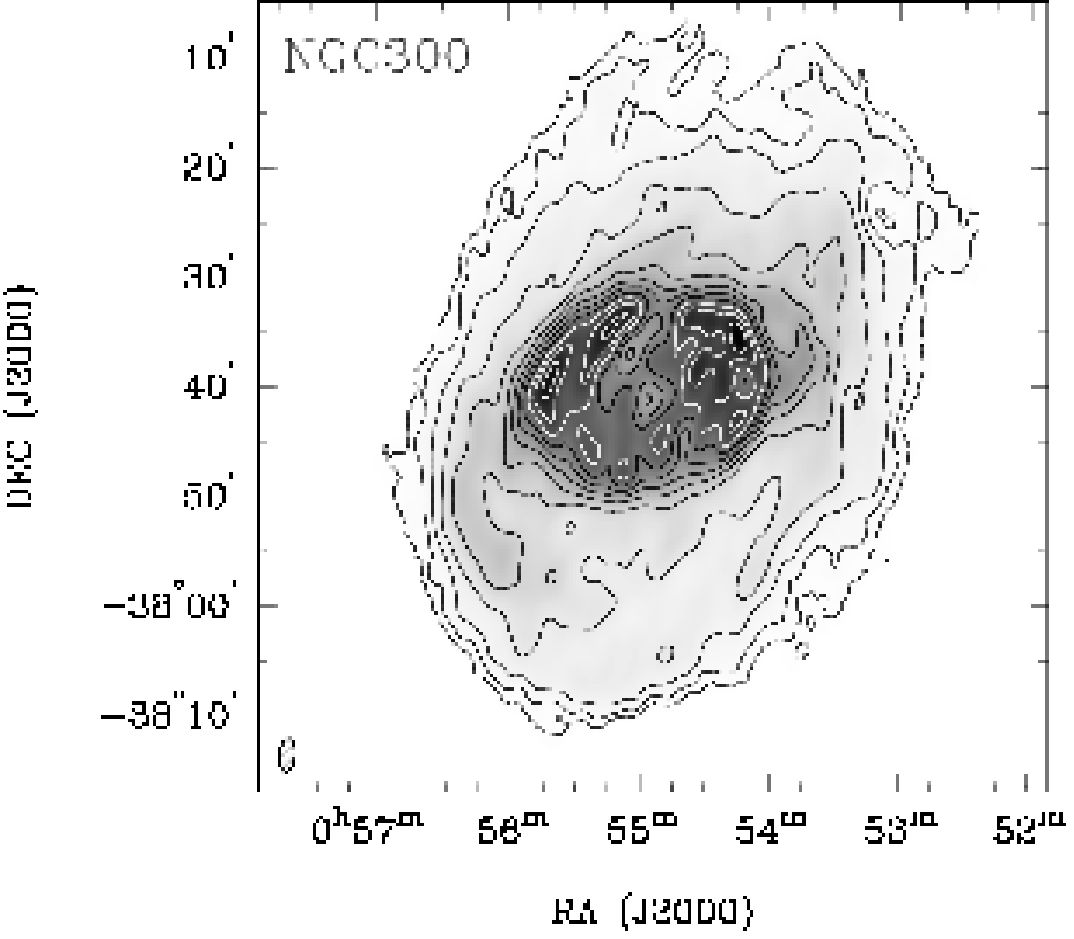,width=4cm,angle=-90}} \\
  \mbox{\psfig{file=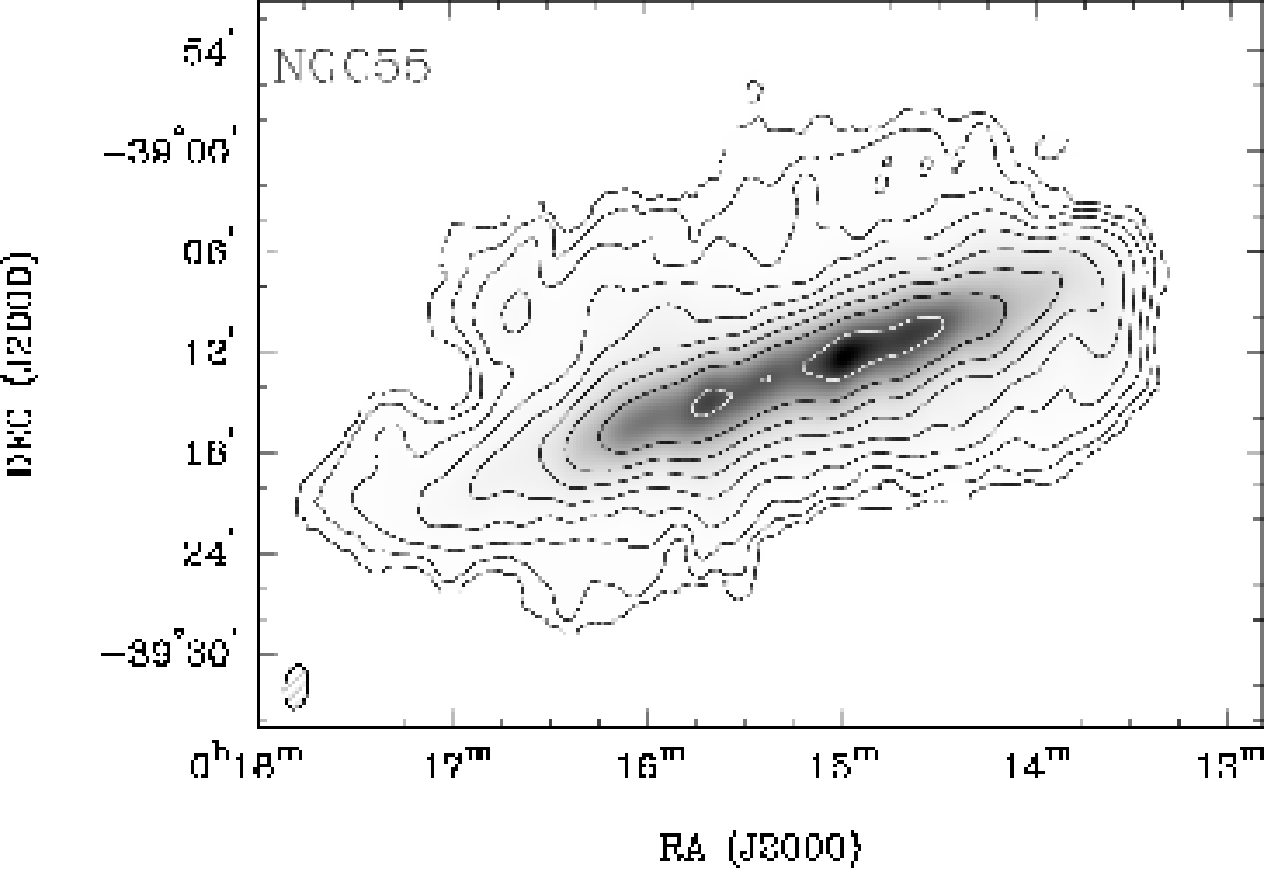,width=3.1cm,angle=-90}} \\
  \end{tabular}
\end{tabular}
\caption{Sculptor Group galaxies. {\bf --- Left:} HIPASS peak flux density map 
  of the central Sculptor Group, marked with the names of the brightest 
  galaxies (see also Westmeier et al. 2017). Extended \HI\ emission in the 
  southern half of the map is associated with the Magellanic Stream. The 
  gridded Parkes beam of 15\farcm5 is displayed in the lower right corner. 
  {\bf --- Right:} ATCA \HI\ distributions of NGC~253 (top), NGC~300 (middle), 
  and NGC~55 (bottom); \HI\ contour levels as given in the Appendix.} 
\end{figure*}

\subsection{The Sculptor Group}

Galaxies in the Sculptor Group, many of which have been observed with the 
ATCA as part of the LVHIS project, are among the closest to the Local Group. 
They span a considerable range in heliocentric radial velocity ($\sim$30 to 
600\kms). Accurate distance measurements are available for the majority of 
group members (see Jerjen et al. 1998; Karachentsev et al. 2000, 2003; Tully 
et al. 2006; Gieren et al. 2008, Dalcanton et al. 2009). The large distance 
spread ($\sim$2 to 5~Mpc) suggests that the Sculptor Group is not a 
gravitationally bound group like the Centaurus~A group, but a loosely bound 
filament of several distinct subgroups which are introduced below. The 
Sculptor Group has over 20 known members; subgroups are identified around
NGC~55/NGC~300 ($D \sim 2$~Mpc), NGC~247/NGC~253 ($D \sim 4$~Mpc), NGC~7793,
NGC~625, and NGC~45 (see Fig.~7). A deep Parkes multibeam \HI\ survey of 
the northern Sculptor Group and the more extended Sculptor filament is 
presented by Westmeier et al. (2017).

\subsubsection{NGC~55 / NGC~300 subgroup}
The large galaxies NGC~55 and NGC~300, together with the dwarf galaxies 
ESO294-G010, ESO410-G005, UGCA~438 and IC\,5152 (all detected in \HI) form 
a small subgroup at the near side of the Sculptor Group at a distance of 
$\sim$2~Mpc. \\

{\bf NGC~55 (HIPASS~J0015--39)} is a Magellanic barred spiral galaxy, viewed 
nearly edge-on, with a stellar diameter of at least half a degree, while 
{\bf NGC~300 (HIPASS J0054--37)} is an equally large, late-type spiral galaxy 
located $\sim$8\degr\ (300~kpc) from NGC~55. Both galaxies and their 
large-scale surroundings were recently mosaicked in \HI\ with the ATCA by 
Westmeier et al. (2011, 2013) who used 32 pointings covering an area of 
$\sim$2\degr $\times$ 2\degr\ to reveal much larger gas envelopes than 
previously known. They also found disturbed outer disc gas in both galaxies 
as well as high-velocity clouds (HVCs) surrounding NGC~55. We refer to 
Westmeier et al. for a detailed analysis and literature overview of both 
Sculptor galaxies. \\

The ATCA \HI\ distribution of the edge-on spiral {\bf NGC~55} is quite 
asymmetric, with the neutral gas more extended towards the east (receding 
side) and north (see Fig.~7). Contours in the west and south are much more 
compressed suggesting the influence of ram pressure stripping. The twisting 
of the \HI\ velocity contours hints at a mild warp of the outer disc 
(Westmeier et al. 2013). Star formation is prominent in the disc, possibly 
responsible for extraplanar \HII\ regions (T\"ullmann et al. 2003). \\

The ATCA \HI\ distribution of {\bf NGC~300} is huge, extending well beyond 
previous measurements (see Fig.~7). The outer \HI\ disc is strongly warped, 
exhibiting a significant twist of the position angle from east-west to nearly 
north-south. The wide-field \HI\ mapping with a compact array was crucial to 
discover the extent of the outer disc, which contains nearly 50\% of the 
\HI\ mass. Westmeier et al. (2011) carry out a detailed analysis of the gas 
kinematics and dark matter distribution, finding a slightly decreasing 
rotation curve ($v_{\rm max} \sim 100$\kms) that extends to a radius 
of $\sim$20 kpc. Significant asymmetries in NGC~300's outer disc hint at 
the possibility of ram-pressure stripping of gas by the intragroup medium. 
We measure \FHI\ = 1758.3 Jy\kms, $\sim$10\% lower than Parkes \HI\ 
measurements (Koribalski et al. 2004, Westmeier et al. 2017).  \\

Of the four dwarf galaxies in this subgroup, three are detected in HIPASS and 
one, ESO294-G010, shows a marginal ATCA \HI\ detection. Because of their low 
velocities, the \HI\ emission of the nearby Sculptor galaxies in some cases 
overlaps with that of Galactic HVCs and \HI\ gas in the Magellanic Stream. \\

{\bf ESO294-G010} is a dwarf galaxy at $D_{\rm TRGB}$ = $1.92 \pm 0.10$~Mpc 
(Karachentsev et al. 2002). Optical images show a rather smooth stellar body 
typical of dwarf spheroidal galaxies, whereas GALEX $UV$ images clearly reveal 
a clumpy inner structure. Some signs of star formation are present (Jerjen et 
al. 1998, Karachentsev et al. 2002). ATCA \HI\ data show a weak source around 
$\sim$106\kms, slightly offset from its optical position as already indicated 
by Bouchard et al. (2005). It remains unclear if the detected \HI\ emission 
belongs to the dwarf galaxy. The optical velocity of $117 \pm 5$\kms\ (Jerjen 
et al. 1998) is also slightly offset. We measure \FHI\ = 0.1 Jy\kms\ and derive 
\MHI\ = $9 \times 10^4$\Msun. ESO294-G010 appears to be similar to dwarf 
galaxies of mixed type like Phoenix, LGS\,3 and KK\,3.  \\

{\bf ESO410-G005 (HIPASS J0015--32)} is a dwarf galaxy at $D_{\rm TRGB}$ = 
$1.92 \pm 0.19$~Mpc (Karachentsev et al. 2000). Its stellar morphology 
resembles that of a dwarf elliptical, while our \HI\ detections at \vhel\ = 
20 -- 52\kms, both with the Parkes telescope and the ATCA, suggest that it 
is most likely a transition-type dwarf galaxy. No \HI\ emission was detected
with the ATCA at velocities around +160\kms\ where Bouchard et al. (2005)
report a detection using the Parkes telescope. Using our ATCA \HI\ data we 
determine a center position of $\alpha,\delta$(J2000) = 00:15:30.9,
--32:10:52, an \HI\ diameter of $1\farcm5 \times 1\farcm0$ (840 pc $\times$
560 pc), $PA$ = 320\degr, a systemic velocity of \vsys\ = $36 \pm 2$\kms\ 
(\vlg\ = 53\kms) and a 50\% (20\%) velocity width of 22.8\kms\ (31.3\kms).
Furthermore, we measure \FHI\ = 0.93 Jy\kms, corresponding to an \HI\ mass of
$8 \times 10^5$\Msun. We calculate a mass-to-light ratio (\MHI/\LB) = 
0.22\Msun/\Lsun. The ATCA \HI\ velocity field shows a gradient perpendicular 
to the stellar disc, similar to that discussed for ESO349-G031. 
High resolution ATCA \HI\ images indicate the \HI\ emission is offset from 
the optical centre, located on both sides of the minor axis. It is unclear 
if the gas is rotating or being accreted/ejected. If the observed velocity 
gradient is due to rotation (\vrot\ = 15\kms) we calculate \Mdyn\ = $2 \times 
10^7$\Msun. \\

{\bf IC\,5152 (HIPASS~J2202--51)} is a dwarf irregular galaxy at a TRGB 
distance of 1.97~Mpc (Tully et al. 2006). Its ATCA \HI\ distribution extends 
well beyond the bright stellar disc (see Fig.~8). We measure \FHI\ = 98.6 
Jy\kms, in agreement with HIPASS (Koribalski et al. 2004), and derive \MHI\ = 
$9.0 \times 10^7$\Msun. The \HI\ velocity field, which shows a twisting $PA$,
was analysed by van Eymeren et al. (2009c) who determine the rotation curve 
up to a radius of 4~kpc.  A comprehensive multi-wavelength description of 
IC\,5152 is given by Kirby et al. (2012), who also carry out some kinematic 
modelling. Using 3D FAT Wang et al. (2017) obtain an \HI\ rotation curve 
indicating \vrot\ = 58\kms\ at \Rmax\ = 3.6 kpc (see Table~9) and \Mdyn\ = 
$2.8 \times 10^9$\Msun. HST images of IC\,5152 reveal several blue star 
complexes and dust patches. Our 20-cm ATCA radio continuum images show two 
bright star forming regions, coincident with ithe MIPS 24$\mu$m emission 
peaks (Shao et al. 2017). \Ha\ emission is detected in and between those 
two regions (Meurer et al. 2006). \\

{\bf UGCA~438 (HIPASS~J2326--32)} is a dwarf irregular galaxy at a TRGB 
distances of $2.18 \pm 0.09$~Mpc (Dalcanton et al. 2009). The results of 
stellar photometry, made difficult by a bright foreground star, are presented 
by Lee \& Byun (1999). Kaisin et al. (2007) spot a single \Ha\ emission 
region, while the GALEX $UV$ emission is clearly extended. Buyle et al. (2006)
report CO non-detections for both UGCA~438 and IC\,5152 and show preliminary 
ATCA \HI\ intensity maps. Here we show --- for comparison with ESO410-G005 ---
high-resolution (30\arcsec) ATCA \HI\ distributions of both galaxies overlaid 
on to DSS2 $B$-band optical images (see Fig.~8). The \HI\ gas associated with 
UGCA~438 is mostly found outside the stellar disc, with \HI\ peaks to the 
north and south plus a minor peak to the east. The gas distribution is highly 
peculiar and gives the appearance of a fragmented \HI\ ring; the \HI\ velocity
field shows no clear signs of rotation. \HI\ emission is detected in the 
velocity range from $\sim$46 to 80\kms. Using our ATCA \HI\ data we measure 
\FHI\ = 3.7 Jy\kms\ which corresponds to an \HI\ mass of only $4.1 \times 
10^6$\Msun. We note that in HIPASS the galaxy UGCA~438 is confused with 
Galactic \HI\ emission. 

\subsubsection{NGC~247 / NGC~253 subgroup}
    
The large spiral galaxies NGC~247 and NGC~253, together with the dwarf galaxies 
ESO540-G030, ESO540-G031 and ESO540-G032 form another small association 
($D \sim 4$~Mpc) within the Sculptor Group. \\

{\bf NGC~247 (HIPASS~J0047--20)} is a late-type spiral galaxy at a cepheid
distance of $D = 3.65 \pm 0.17$~Mpc (Madore et al. 2009). This value agrees 
with its TRGB distance estimate by Karachentsev et al. (2006). Our ATCA \HI\ 
moment maps show a mildly warped, regularly rotating disc extending about a 
factor two beyond the stellar disc. We measure \FHI\ = 662.5 Jy\kms, about    
10\% higher than the HIPASS value (Koribalski et al. 2004).
High-resolution, single-pointing VLA \HI\ maps were obtained by Carignan \& 
Puche (1990b) and Ott et al. (2012), who measure \FHI\ = $528 \pm 18$ Jy\kms\ 
and 382.6 Jy\kms\ (VLA-ANGST project), respectively. Both are missing 
substantial amounts of \HI\ gas due to the lack of very short baselines and 
limited field-of-view. NGC~247's closest neighbour is the dIrr galaxy 
ESO540-G031 (HIPASS J0049--20; UGCA~015). \\ 

{\bf NGC~253 (HIPASS~J0047--25)} is a well-known starburst spiral galaxy and 
the brightest member of the Sculptor Group. It is oriented close to edge-on 
with an optical extent of $32\arcmin \times 8\arcmin$ (see Table~2) and 
$\sim50\arcmin \times 30\arcmin$ at $\sim$28~mag\,arcsec$^{-2}$ (Malin \& 
Hadley 1997). We have adopted a TRGB distance of $3.94 \pm 0.37$~Mpc 
(Karachentsev et al. 2003). Our ATCA data show bright \HI\ emission extending 
barely beyond the stellar disc and \HI\ absorption (visible as the central 
hole) against the prominent starburst region. The mean \HI\ velocity field
highlights the fast rotating disc (\vrot\ = 213\kms, see Table~8) with only
minor deviations from regular rotation in its outskirts. The brightest \HI\
emission is detected in the barred region of NGC~253.

Previous ATCA \HI\ observations of NGC~253 were presented by Koribalski, 
Whiteoak \& Houghton (1995) and Boomsma et al. (2005). Lucero et al. (2015) 
use the Karoo Array Telescope (KAT-7) to map the extended \HI\ structure 
of NGC~253 at low resolution ($\sim$3\farcm5) down to column densities of 
\NHI\ = $1.3 \times 10^{19}$ cm$^{-2}$. They measure \FHI\ = $728 \pm 36$ 
Jy\kms, in agreement with the HIPASS \FHI\ of $692.9 \pm 42.2$ Jy\kms\ 
(Koribalski et al. 2004). Both are lower limits due to significant \HI\ 
absorption against the bright star-forming inner region of NGC~253, which
covers the same velocity range as the \HI\ emission (Koribalski, Whiteoak 
\& Houghton 1995). Using our ATCA \HI\ maps we measure \FHI\ = 746.8 Jy\kms\
and derive \MHI\ = $2.7 \times 10^9$\Msun. Using 3D FAT Wang et al. (2017) 
obtained an \HI\ rotation curve indicating \vrot\ = 200\kms\ at \Rmax\ = 
17.2 kpc (see Table~9) and \Mdyn\ = $1.6 \times 10^{11}$\Msun. \\

In the following we briefly discuss three faint dwarf galaxies (ESO540-G030,
ESO540-G031, and ESO540-G032) near NGC~247; these are not part of the LVHIS 
galaxy atlas presented here. Karachentsev et al. (2003) obtained TRGB distances 
of $3.40 \pm 0.34$ Mpc (ESO540-G030), $3.34 \pm 0.24$ Mpc (ESO540-G031), and 
$3.42 \pm 0.27$ Mpc (ESO540-G032), suggesting that this dwarf grouping lies 
slightly in front of the spiral galaxy NGC~247. \\

{\bf ESO540-G030} is a low surface brightness dwarf galaxy. Jerjen et al. 
(1998, 2000) measured an integrated magnitude of $B_{\rm T}$ = 16.37~mag and 
an SBF distance of $3.19 \pm 0.13$~Mpc. In addition to a large number of red 
stars, Karachentsev et al. (2003) also detect a number of blue stars in the 
central region of ESO540-G030, suggesting that it is a transition dSph/dIrr 
type galaxy. Deep Parkes and ATCA \HI\ observations were presented by Bouchard 
et al. (2005), resulting in a tentative detection. They derive a total \HI\ 
mass of $8.9 \pm 1.9 \times 10^5$\Msun\ and note a positional offset between 
the \HI\ gas and the stellar component. \\

{\bf ESO540-G031} (HIPASS J0049--20) is a dwarf irregular galaxy at a TRGB 
distance of $3.34 \pm 0.24$~Mpc (Karachentsev et al. 2003), located only 
40\arcmin\ east of NGC~247. The HIPASS BGC gives \FHI\ = $3.9 \pm 1.5$ Jy\kms\ 
and \vsys\ = 294\kms\ (Koribalski et al. 2004). \HI\ maps of ESO540-G031 
(UGCA~015, DDO6) obtained with the GMRT (Begum et al. 2008) and the VLA (Ott 
et al. 2012) show \FHI\ = $2.6 \pm 0.3$ and 1.2 Jy\kms, respectively. Using 
our low-resolution ATCA \HI\ mosaic of NGC~247 and surroundings we measure 
\FHI\ $\approx$ 3.2 Jy\kms\ for ESO540-G031 and derive \MHI\ = $8.2 \times 
10^6$\Msun (see also Warren et al. 2007). The \HI\ emission of ESO540-G031 
is offset towards the south-western side of the stellar body. Its relatively 
bright GALEX $UV$ emission suggests significant star formation (SFR = $1.4 
\times 10^{-3}$\Msun\,yr$^{-1}$; Ott et al. 2012). \\

{\bf ESO540-G032} is another transition-type dwarf galaxy. Parkes and ATCA 
\HI\ data by Bouchard et al. (2005) suggest a centre velocity of 228\kms. The 
detection was later confirmed by Da Costa et al. (2008) who also obtain new 
HST-ACS optical data, deriving a TRGB distance of $3.7 \pm 0.2$ Mpc and \MHI\ 
= $9.1 \times 10^5$\Msun. Their ATCA image shows an \HI\ source with a faint 
eastern extension.

\subsubsection{NGC~7793 subgroup}

{\bf NGC~7793 (HIPASS~J2357--32)} is a bright spiral galaxy in the Sculptor 
group at a TRGB distance of $3.91 \pm 0.41$ Mpc (Karachentsev et al. 2003). 
We use the low-resolution ATCA 16-pointing mosaic data of NGC~7793 and its 
surroundings to obtain \HI\ maps of NGC~7793. The ATCA \HI\ distribution 
extends beyond the previously published single-pointing VLA maps (Carignan 
\& Puche 1990a; Walter et al. 2008). We measure \DHI\ = 16\farcm5 (see Table~6)
and \FHI\ = 292.6 Jy\kms, in agreement with the HIPASS BGC \FHI\ of $278.5 
\pm 20.4$ Jy\kms\ (Koribalski et al. 2004). For comparison, Walter et al. 
(2008) measure \FHI\ = 246 Jy\kms\ with the VLA, missing \HI\ gas mainly in 
the extended outer disc. Our mean \HI\ velocity field shows that the position 
angle of NGC~7793's outer disc ($PA$ = 300\degr) is much larger than that of 
its stellar disc ($PA$ = 278\degr), suggesting a significant warp. This is 
also supported by the closed velocity contours. Carignan \& Puche (1990a) also 
notice the $PA$ change. They determine the rotation curve out to a radius of 
7\farcm5 (see also de Blok et al. 2008). Significant uncertainties remain 
about the inclination angle change in the outer disc, affecting the shape of 
the rotation curve. Combining the ATCA and VLA \HI\ data would be of 
significant benefit for studying the outer disc gas distribution and 
kinematics (see also Radburn-Smith et al. 2012, 2014). NGC~7793 has two dwarf 
companions, ESO349-G031 and UGCA~442, located at projected distances of 
$\sim$3\degr.  \\

{\bf ESO349-G031 (HIPASS~J0008--34)}, best known as the Sculptor Dwarf 
Irregular Galaxy (SDIG), has an extremely low stellar luminosity and is most
likely a transition-type dwarf galaxy (C\^ot\'e et al. 2000). For a 
comprehensive 
study of its stellar properties and star formation history see Heisler et al. 
(1997) and Lianou \& Cole (2013), respectively.  Karachentsev et al. (2006) 
determined a TRGB distance of $3.21 \pm 0.26$~Mpc. Its nearest neighbour is 
the spiral galaxy NGC~7793 (see Fig.~7), which lies at a projected distance of 
$\sim$3\degr\ ($\sim$200~kpc). Our ATCA data show a resolved \HI\ source with 
a clear velocity gradient. We measure \FHI\ = 4.8 Jy\kms\ and derive \MHI\ = 
$1.2 \times 10^7$\Msun; our fitted \HI\ position agrees with the optical 
centre. Using higher resolution VLA \HI\ maps C\^ot\'e et al. (2000) find the 
\HI\ kinematical major axis (118\degr) --- also seen in our ATCA \HI\ maps --- 
perdendicular to the optical position angle (30.3\degr), suggesting that the 
gas is rotating about the optical minor axis. This would make SDIG another 
dwarf transitional galaxy. Similar misalignments are seen in Sextans\,A 
(Skillman et al. 1988), NGC~5253 (Kobulnicky \& Skillman 1995;
L\'opez-S\'anchez et al. 2008, 2012), and GR~8 (Carignan et al. 1990). \Ha\ 
images of SDIG show very faint regions of diffuse emission (Meurer et al. 
2006; Kaisin et al. 2007) and two point sources (Bouchard et al. 2009). Higher 
resolution ATCA \HI\ maps will provide more information on the morphology and 
peculiar kinematics of the cold gas. \\

{\bf UGCA~442 (HIPASS~J2343--31)} is a Magellanic barred spiral galaxy oriented 
close to edge-on. Karachentsev et al. (2003) give a TRGB distance of $4.27 \pm 
0.52$~Mpc. Its nearest neighbour is the bright spiral galaxy NGC~7793, just 
over 3\degr\ away (see Fig.~7). Using our ATCA \HI\ data we find an extended, 
regular rotating \HI\ disc apart from minor deviations towards the southern 
(receding) end (see also C\^ot\'e et al. 2000). We measure \FHI\ = 52.2 Jy\kms, 
in agreement with HIPASS (Koribalski et al. 2004), and derive \MHI\ = $2.2 
\times 10^8$\Msun. \Ha\ imaging of UGCA~442 revealed five \HII\ regions within 
the stellar disc, which are discussed by Miller (1996), Lee et al. (2003), 
and Meurer et al. (2006).

\subsubsection{NGC~625 subgroup}

{\bf NGC~625 (HIPASS J0054--37)} is a Magellanic barred spiral galaxy at a 
distance of $D_{\rm TRGB}$ = $3.89 \pm 0.22$~Mpc (Cannon et al. 2003). Its
nearest neighbour is the dwarf irregular galaxy ESO245-G005 (HIPASS J0145--43)
at a projected distance of 170\farcm3. Our ATCA data show \HI\ emission 
extending well beyond the stellar disc, including a peculiar \HI\ feature on
the western side, and a complex velocity field. The latter shows two \HI\
components, likely a rotating disc ($PA \sim 90\degr$) along the major axis 
and an equally prominent kinematic feature along the minor axis.
C\^ot\'e et al. (2000) find that NGC~625's dominant \HI\ velocity gradient 
is along the minor axis and speculate whether a recent merger or accretion 
event could have led to the peculiar kinematics of the \HI\ disc. Using 
higher resolution observations, Cannon et al. (2004) show that the velocity
gradient is due to outflowing \HI\ gas. NGC~625 is also classified as a `blue 
amorphous galaxy' (Marlowe, Meurer \& Heckman 1999), suggesting a recent 
burst of star formation. \\

{\bf ESO245-G005 (HIPASS J0145--43)} is a Magellanic barred irregular galaxy
at a TRGB distance of $4.43 \pm 0.45$ Mpc (Karachentsev et al. 2003). It is 
located $\sim$3\degr\ (200~kpc at a distance of 4~Mpc) from the spiral galaxy 
NGC~625 in the outskirts of the Sculptor Group. Our ATCA \HI\ data show a
large, somewhat warped \HI\ disc extending well beyond the star-forming stellar
body. The \HI\ emission is brightest near three prominent star-forming regions 
shown in the multi-wavelength image by Wang et al. (2017); we observe an \HI\ 
depression towards the galaxy center. 
C\^ot\'e et al. (2000) use ATCA \HI\ maps to carry out a tilted-ring analysis; 
they find the \HI\ kinematical axis of ESO245-G005 to depart from the optical 
major axis. For a detailed discussion of the \HI\ kinematics see Kirby et al. 
(2012). We measure \FHI\ = 83.2 Jy\kms, in agreement with HIPASS 
(Koribalski et al. 2004). SINGG \Ha\ (Meurer et al. 2006) and GALEX $UV$ (Dale 
et al. 2009; Wong et al. 2016) images allow a detailed study of the galaxy's 
local star formation properties with respect to the \HI\ gas density.  \\

{\bf ESO149-G003 (HIPASS~J2352--52)} is a Magellanic barred irregular galaxy 
at a distance of $D_{\rm TF}$ = 5.9~Mpc. It is seen close to edge-on and has a 
remarkable low-surface brightness (LSB) extension, possibly due to accretion of
a dwarf companion. ESO149-G003 is relatively isolated and lies in the southern
outskirts of the Sculptor Group. Ryan-Weber et al. (2004) detect filamentary 
\Ha\ emission in its bright stellar disc. Furthermore they find 
an isolated \HII\ region $\sim$1\farcm5 to the west, which might be associated 
with ESO149-G003. Close inspection of the optical images shows some flaring of 
the outer LSB component and potentially a very small companion (PGC~441599) to 
the south-east. The latter lies within the \HI\ envelope. Our ATCA \HI\ moment 
maps show a slight asymmetry of ESO149-G003's \HI\ distribution, which is 
further emphasized by its unusual two-component velocity field. The peculiar 
\HI\ emission towards the northern (receding) end of the galaxy, which shows
a velocity gradient nearly 90\degr\ different from the rotating disc, is 
offset from the very faint stellar extension. We suggest that ESO149-G003 may 
have undergone a minor merger event, the debris of which are detected in the 
form of \HI\ tidal tails and extended stellar streams. Wang et al. (2016) 
find the galaxy to lie offset from the \DHI--\MHI\ relation, which is not 
surprising given the extended peculiar emission. We measure \FHI\ = 7.2 Jy\kms,
in agreement with HIPASS (Koribalski et al. 2004), and derive an \HI\ mass of 
$5.9 \times 10^7$\Msun.

\subsubsection{NGC~45 subgroup}
    
The NGC~45 subgroup consists of the galaxies NGC~24, NGC~45, NGC~59 (Fouqu\'e 
et al. 1992), and possibly MCG-04-02-003 (all at $\delta \ga -25\degr$). It is 
located at a larger distance than the main Sculptor members, but is often 
considered part of the Sculptor group (see Fig.~7). These galaxies are briefly 
described below, but are not part of the LVHIS galaxy atlas presented here. See
also Westmeier et al. (2017) for deep Parkes \HI\ observations. \\

{\bf NGC~45 (HIPASS~J0014--23)} is a low surface brightness spiral galaxy with 
a TRGB distance measurement of $6.6 \pm 0.2$~Mpc (Jacobs et al. 2009). Its 
\HI\ disc (\MHI\ = $2.0 \times 10^9$\Msun) was studied in detail with the VLA 
by Chemin et al. (2006) who derive a total mass of $3.7 \times 10^{10}$\Msun\ 
within a radius of 16.7~kpc from rotation curve analysis. GALEX $UV$ images 
were presented by Lee at al. (2011). \\
    
{\bf NGC~24 (HIPASS~J0009--24)} is a nearly edge-on spiral galaxy with a TRGB 
distance measurement of 7.7~Mpc. It was studied in \HI\ with the VLA by Chemin 
et al. (2006) who find a rather regular rotating disc and determine \MHI\ = 
$7.5 \times 10^8$\Msun\ and \Mdyn\ = $2.8 \times 10^{10}$\Msun\ within a 
radius of 10.5 kpc.  \\

{\bf NGC~59} is a small, elliptical galaxy of type dS0. Karachentsev et al. 
(2003) list an SBF distance of 5.3~Mpc. It is detected in HIPASS at \vhel\ =
364\kms, but was too faint for inclusion in either the HIPASS BGC (Koribalski
et al. 2004) or HICAT (Meyer et al. 2004). We measure a HIPASS \FHI\ of 2.6 
Jy\kms\ and derived \MHI\ = $1.7 \times 10^7$\Msun. Westmeier et al. (2017)
measure \FHI\ = 2.3 Jy\kms\ in much deeper Parkes \HI\ data. Beaulieu et al. 
(2006) obtain low-resolution ATCA \HI\ data and determine \MHI\ = $1.5 \times 
10^7$\Msun. They show a barely resolved \HI\ source centred on the stellar 
body and derive \MHI\,/\LB\ = 0.07\Msun\,\Lsun, similar to typical dIrr 
galaxies. We suggest that NGC~59 is likely a dwarf transitional galaxy. \\ 
    
{\bf MCG-04-02-003 (HIPASS~J0019--22)} appears to be a blue compact dwarf 
galaxy with an LSB outer disc (radius $\sim$ 110\arcsec) at a Hubble distance 
of 9.5~Mpc (Warren et al. 2006, 2007). Warren et al. obtained deep optical 
images and low-resolution ATCA \HI\ data; they measure \FHI\ = 16.2 Jy\kms, 
in agreement with the HIPASS BGC (Koribalski et al. 2004), and obtain \MHI\ 
= $3.4 \times 10^8$\Msun\ and \MHI/\LB\ = $3.0 \pm 0.3$\Msun/\Lsun. The \HI\
distribution is very extended, reaching well beyond the faint stellar disc.
GALEX $UV$ images show star formation in the galaxy core as well as a ring-like 
structure ($\sim$5\arcmin\ $\times$ 2\arcmin) encompassing the faint stellar
disc. Higher resolution \HI\ observations are needed to study the structure
and kinematics of this galaxy in greater detail.  \\

\begin{figure*}
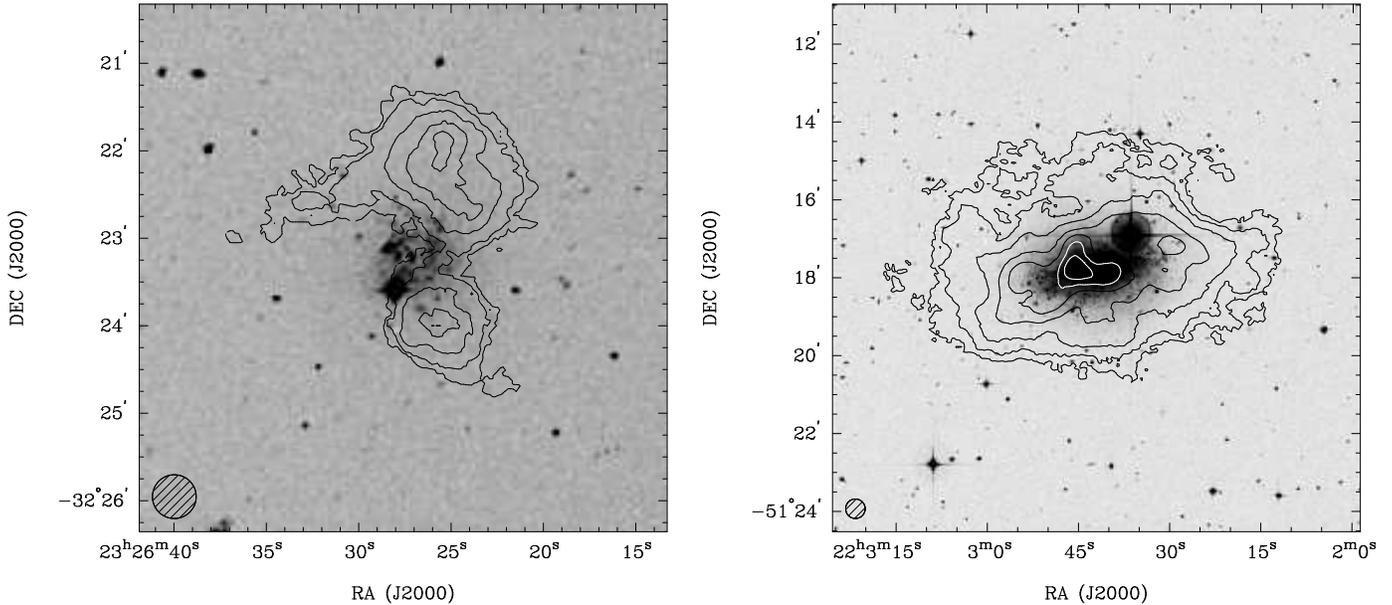
 
  \begin{tabular}{cc}
  \mbox{\psfig{file=ugca438.line.12fwhm.mom0+dss2blue.regrid.ps,width=8cm,angle=-90}} &
  \mbox{\psfig{file=ic5152.line.9fwhm.mom0+dss2blue.regrid.ps,width=8cm,angle=-90}} \\
  \end{tabular}
\caption{High-resolution ATCA \HI\ column density maps of the Sculptor Group 
    dIrr galaxies UGCA~438 (left) and IC\,5152 (right). Contour levels are 
    0.05, 0.1, 0.2, 0.3, and 0.4 Jy\,beam$^{-1}$\kms\ (for UGCA~438) and 0.1, 
    0.2, 0.4 0.8, 1.2, 1.6 and 2.0 Jy\,beam$^{-1}$\kms\ (for IC\,5152). 
    The synthesized beam (30\arcsec) is displayed in the bottom left corner.}
\end{figure*}

\subsection{The Cen\,A Group}
The nearby Cen\,A Group consists of two spatially separated sub-groups, one 
around the spiral galaxy M\,83 (NGC~5236) and the other around the giant 
elliptical galaxy Centaurus\,A (NGC~5128). Karachentsev et al. (2007) estimate 
mean group distances of 4.8 and 3.8~Mpc, respectively, as well as virial masses
of $\sim$0.82 and $8.1 
\times 10^{12}$\Msun. The M\,83 sub-group is significantly smaller, a factor 
10 less massive and more compact than the Cen\,A sub-group. Their velocity 
dispersions differ by a factor two (61\kms\ for the M\,83 and 136\kms\ for the 
Cen\,A sub-group). The gas-rich neighbours of the Cen\,A galaxy, as measured
by the LVHIS project, are nicely visualised in Johnson et al. (2015; their 
Fig.~1). The Cen\,A Group covers $\sim$1000 sq deg on the sky and has at least 
60 confirmed members, of these the majority are dwarf irregular galaxies.
Woodley (2006) estimate a total mass of $9.2 \times 10^{12}$\Msun\ for the 
Cen\,A group. 

\subsubsection{The M\,83 subgroup}

The M\,83 subgroup consists of at least 10 galaxies. Of these six gas-rich 
dwarf galaxies are in the immediate vicinity of M\,83 (see Fig.~10). Thim et 
al. (2003) and Jacobs et al. (2009) discuss the group and determine a group 
distance of $4.5 \pm 0.2$~Mpc. Some of the closest neighbours to M\,83 are 
the dIrr galaxies UGCA\,365 (= ESO444-G078), NGC~5264 (= DDO\,242), IC\,4316, 
ESO444-G084, IC\,4247, AM1321--304, and the peculiar starburst galaxy NGC~5253.
Within the $\sim$10\% uncertainty of their independent distances it appears 
that either NGC~5264, which lies 59\farcm8 (78~kpc) east of M\,83, or 
UGCA~365, located 38\farcm4 (50~kpc) north of M\,83, are closest to M\,83. 
Recently, M\"uller et al. (2015) imaged an area of 60 sq deg around the 
M\,83 subgroup, detecting 16 new dwarf galaxy candidates, which may be group 
members. \\

\begin{figure*} 
\begin{tabular}{cc}
 \mbox{\epsfig{file=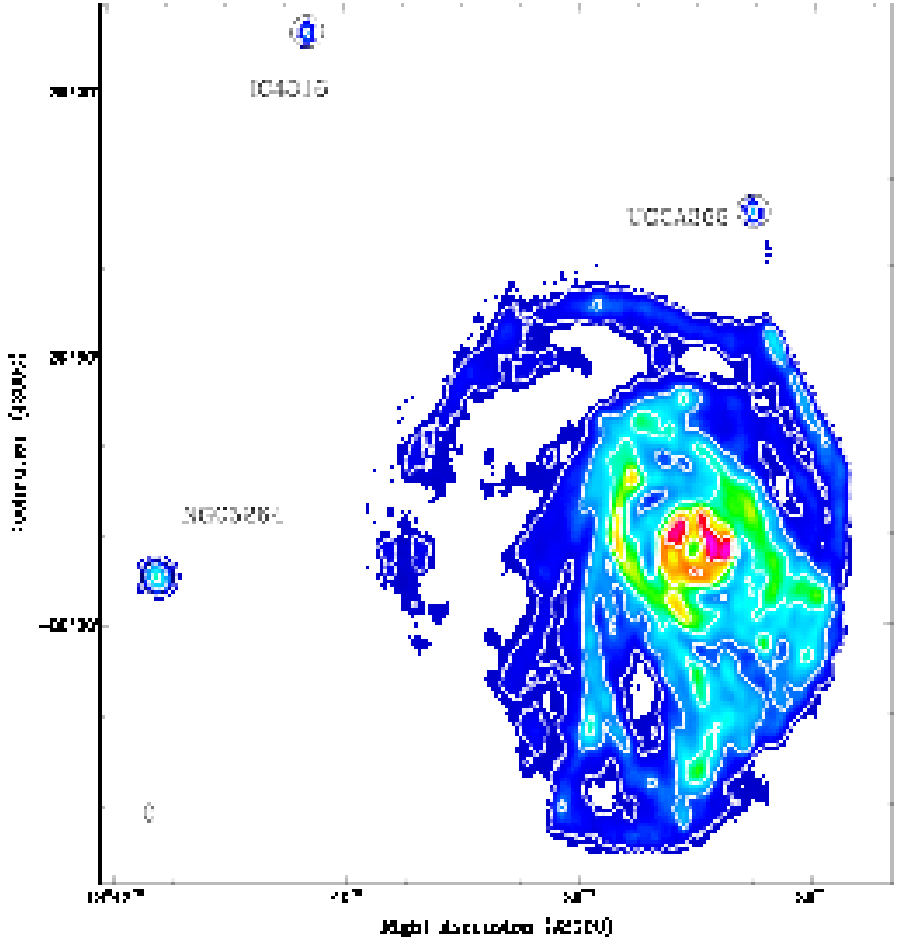,height=9cm,angle=-90}} & 
 \mbox{\epsfig{file=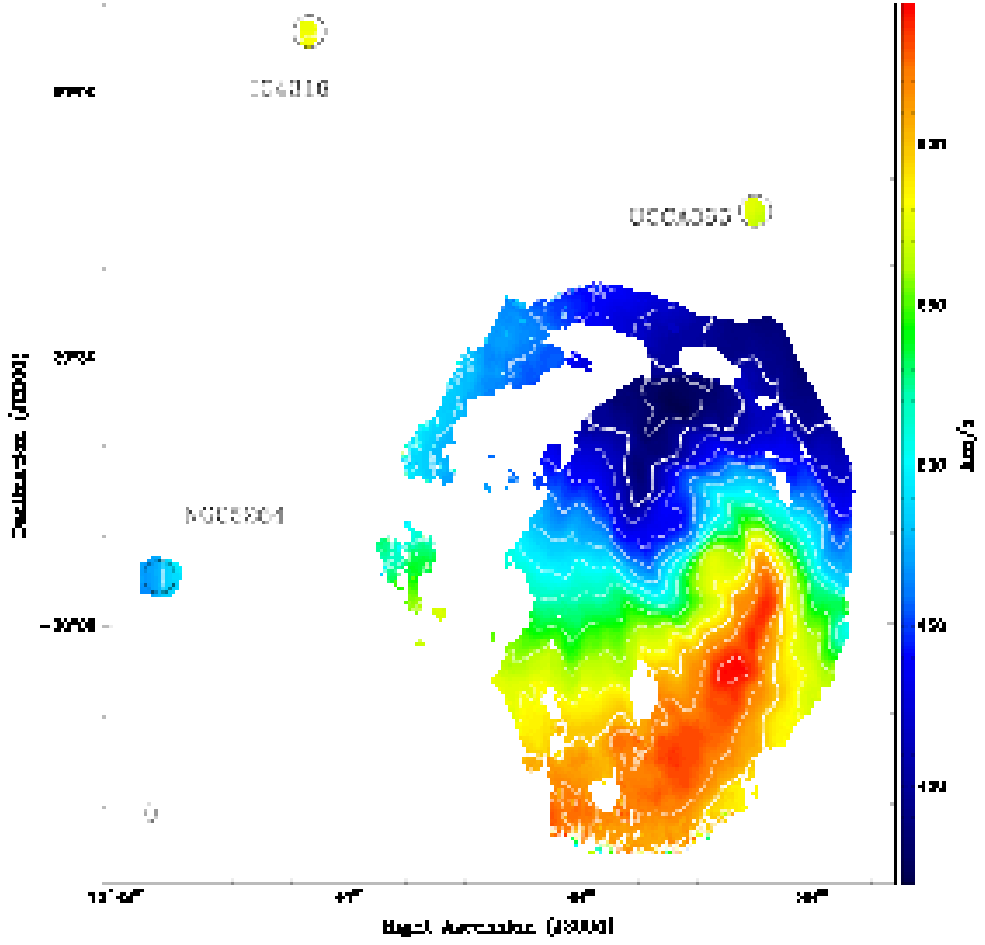,height=9cm,angle=-90}} \\
\end{tabular}
\caption{ATCA \HI\ distribution (left) and mean \HI\ velocity field (right)
   of the giant spiral galaxy M\,83 (HIPASS J1337--29) and its dwarf companions
   NGC~5264, IC\,4316 and UGCA~365. The contour levels are 0.4, 1.5, 3, 5 
   and 7 Jy\,beam$^{-1}$\kms\ (mom0) and 380 to 640\kms\ in steps of 20\kms\ 
   (mom1). The angular resolution is $113\farcs1 \times 64\farcs1$.} 
\end{figure*}

{\bf M\,83 (HIPASS~J1337--29)} is a grand-design spiral galaxy with 
an unusually large \HI\ envelope. It is located in the Cen\,A Group and forms 
the centre of a subgroup consisting of $\sim$10 known galaxies. A large range 
of distance estimates exist for M\,83, the majority of which agree to within 
10\%. We adopt the TRGB distance estimate of $4.92 \pm 0.10$~Mpc by Jacobs et 
al. (2009). For comparison, Thim et al. (2003) estimate $4.5 \pm 0.3$~Mpc from 
ground-based cepheid measurements, while Herrmann et al. (2008) obtain $4.85 
\pm 0.12$~Mpc from the planetary nebula luminosity function. We calculate a 
Hubble distance of 4.4~Mpc from M\,83's Local Group velocity of \vlg\ = 
332\kms\ (Koribalski et al. 2004), assuming \Ho\ = 75\kms\,Mpc$^{-1}$. 
We note that Karachentsev et al. (2002) also estimate a TRGB distance ($4.68 
\pm 0.46$ Mpc) for the stellar stream located 18\farcm5 north of M\,83, likely 
an accreted dSph galaxy. This stream was discovered by Malin \& Hadley 
(1997). 

Our ATCA \HI\ mosaic of M\,83 and its dwarf companions NGC~5264, IC\,4316 and 
UGCA~365, is shown in Fig.~9. The integrated \HI\ flux density of M\,83 as 
measured with the ATCA in the maximum entropy moment map is \FHI\ = 1384 
Jy\kms\ compared to 1440 Jy\kms\ in the CLEANed map, both are $\sim$10\% lower 
than the measured single-dish \HI\ flux density. Walter et al. (2008) recover 
only a quarter of the total \HI\ flux density (361 Jy\kms), due to their 
single pointing VLA observations and low sensitivity to extended, diffuse \HI\ 
emission. The \HI\ distribution of M\,83, as mapped with the ATCA, is most 
remarkable (Koribalski 2015, 2017; Jarrett et al. 2013). It extends well 
beyond the GALEX $XUV$ disc (Thilker et al. 2007), referred to as a giant 
2X-\HI\ disc by Koribalski (2017). No longer does this grand-design spiral 
look regular and undisturbed. Our ATCA \HI\ maps show outer disc streamers, 
irregular enhancements, an asymmetric tidal arm, 
diffuse emission, and a thoroughly twisted velocity field, much in contrast 
to its regular appearance in short-exposure optical images. M\,83's \HI\
distribution is enormous, several times larger than its stellar disc. 
It is also a rather massive galaxy, mildly interacting with the neighbouring
dwarf galaxies. The effect of this interaction on the dwarfs can of course
be rather devastating. It is indeed quite likely that M\,83 has accreted
dwarf galaxies in the past. While the \HI\ distribution of M\,83 has 
previously been studied, we show the first detailed study of the large-scale
emission in M\,83 and its surroundings. 

Huchtmeier \& Bohnenstengel (1981) measured an integrated \HI\ flux density
of about 1632 Jy\kms\ which agrees well with the HIPASS estimate of $1630 
\pm 96$ Jy\kms\ for M\,83 by Koribalski et al. (2004). At a distance of 4.92
Mpc this corresponds to an \HI\ mass of \MHI\ = 9.3 ($\pm$0.5) $\times 
10^9$\Msun. They also derived a total \HI\ extent of 
$76\arcmin$ (EW) $\times 95\arcmin$ (NS) for M\,83 at a column density of 
$N_{\rm HI} = 6 \times 10^{18}$ atoms\,cm$^{-2}$. For comparison we show 
the deep Parkes multibeam \HI\ data of M\,83 and its surroundings in Fig.~10.
Note that the gridded Parkes beam is 15.5\arcmin, and 1 Jy\,beam\kms\
corresponds to $\sim1.4 \times 10^{18}$ atoms\,cm$^{-2}$. We measure a very 
similar \HI\ extent of roughly $80\arcmin \times 88\arcmin$ ($\pm$4\arcmin),
i.e. roughly 100 kpc. The \HI\ distribution is clearly asymmetric
as already noted by Huchtmeier \& Bohnenstengel, the reason for which becomes
immediately obvious in our high-resolution ATCA \HI\ images (see Fig.~9).
The latter reveal an extended arm emerging from the western part of M\,83 
and curving 180\degr\ around to the east. The overall impression of M\,83 in
neutral hydrogen is that of a distorted one-armed spiral, indicating that 
it may have interacted or merged with another, smaller galaxy. While the 
velocity field in this extended arm appears to follow the general pattern
of rotation, the gas distribution shows numerous irregularities, clumps and 
bifurcations. The 20\arcmin\ long ridge in the northwest ends in a noticable 
\HI\ clump and marks a kink in the outer arm.
The \HI\ velocity field of M\,83 highlights the warped nature of the disc. 
Using 3D FAT Kamphuis et al. (2015) obtained an \HI\ rotation curve indicating 
\vrot\ = 157.0\kms\ at \Rmax\ = 50.0 kpc (for $i$ = 40\fdg3 and $PA$ = 
226\fdg9; see Table~9) and \Mdyn\ = $2.8 \times 10^{11}$\Msun.  

The eastern-most \HI\ emission of M\,83 which forms part of its peculiar, outer
arm lies at $\alpha,\delta$(J2000) = 13:39:40, --29:51:45 (\vhel\ = 536\kms), 
i.e. $\sim$34.5\arcmin\ (45 kpc) away from the centre of M\,83. We note that 
the dwarf irregular galaxy NGC~5264 (HIPASS J1341--29; \vsys\ = 478\kms) lies 
at a projected distance of only 25\farcm5 (33~kpc) from the eastern \HI\ edge 
of M\,83. Given that the independently measured distances to M\,83 and 
NGC~5264 are very similar (see Table~2) both galaxies are likely to be 
interacting.  \\

\begin{figure}  
\mbox{\psfig{file=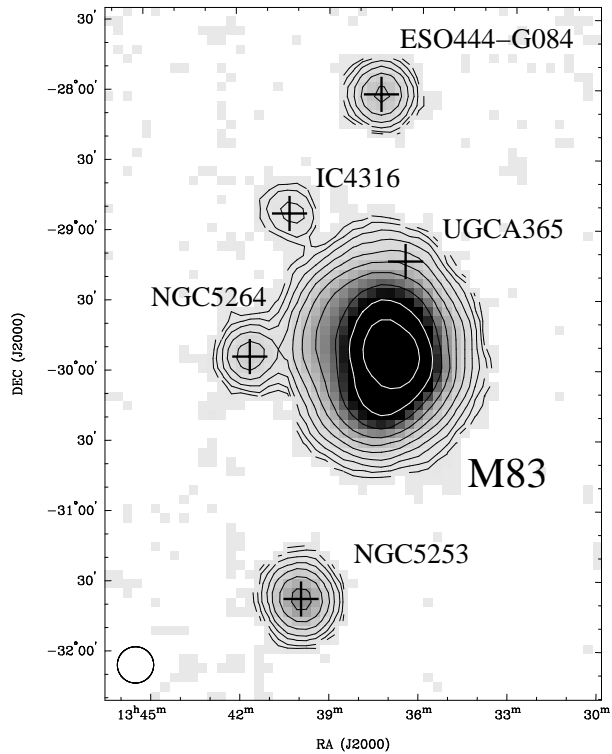,width=8cm}}
\caption{\HI\ distribution of the M\,83 group as obtained from the 
  re-calibrated deep Parkes \HI\ multibeam survey (HIDEEP). The contour levels 
  are 0.5, 1, 2, 4, 8, 16, 32, 64, 128 and 256 Jy\,beam$^{-1}$\kms\ (the first 
  contour corresponds to an \HI\ column density of $\sim7 \times 10^{17}$ 
  atoms\,cm$^{-2}$). The gridded Parkes beam of 15\farcm5 is indicated at the 
  bottom left.}
\end{figure}

{\bf UGCA~365 (HIPASS J1336--29)} is a dwarf irregular galaxy located 38\farcm4
from the spiral galaxy M\,83, just north of its extended \HI\ envelope. Its 
TRGB distance is $5.25 \pm 0.43$ Mpc (Karachentsev et al. 2007), similar to 
that of M\,83 within the given uncertainty. UGCA\,365 has an optical size 
of $1.2\arcmin \times 0.5\arcmin$ ($PA$ = 31\degr) and a $B$-band magnitude of 
15.43 mag. Our low-resolution ATCA \HI\ data of UGCA\,365 show an unresolved 
source; at higher resolution a velocity pattern along the optical major axis
is detected. We measure \FHI\ = 2.9 Jy\kms\ and derive \MHI\ = $1.9 \times 
10^7$\Msun.
Both the stellar and the gas distribution show some extent to the south-east
along the galaxy minor axis, possibly due to the tidal interaction with M\,83.
Using an inclination-corrected ($i$ = 66\degr) rotational velocity of 50\kms\
we calculate a total dynamical mass of $\sim1.2 \times 10^9$\Msun\ for
UGCA\,365. The tidal limit at the position of UGCA\,365, 38\farcm4 (50~kpc)
from the centre of M\,83, is $\sim$5 kpc.
Begum et al. (2008) observed UGCA~365 as part of the FIGGS project. They 
measure an \HI\ diameter of 0\farcm9 ($\sim$1.7 $\times$ the optical diameter) 
and an \HI\ flux density of \FHI\ = $2.3 \pm 0.2$ Jy\kms. Their \HI\ map
hints at an extension towards the north-west. \\

\begin{figure*} 
\begin{tabular}{cc}
  \mbox{\psfig{file=ngc5237.na.mom0.ps,width=8cm,angle=-90}} &
  \mbox{\psfig{file=ngc5237.r0.mom0.ps,width=8cm,angle=-90}} \\
  \mbox{\psfig{file=ngc5237.r06.mom0.ps,width=8cm,angle=-90}} & 
  \mbox{\psfig{file=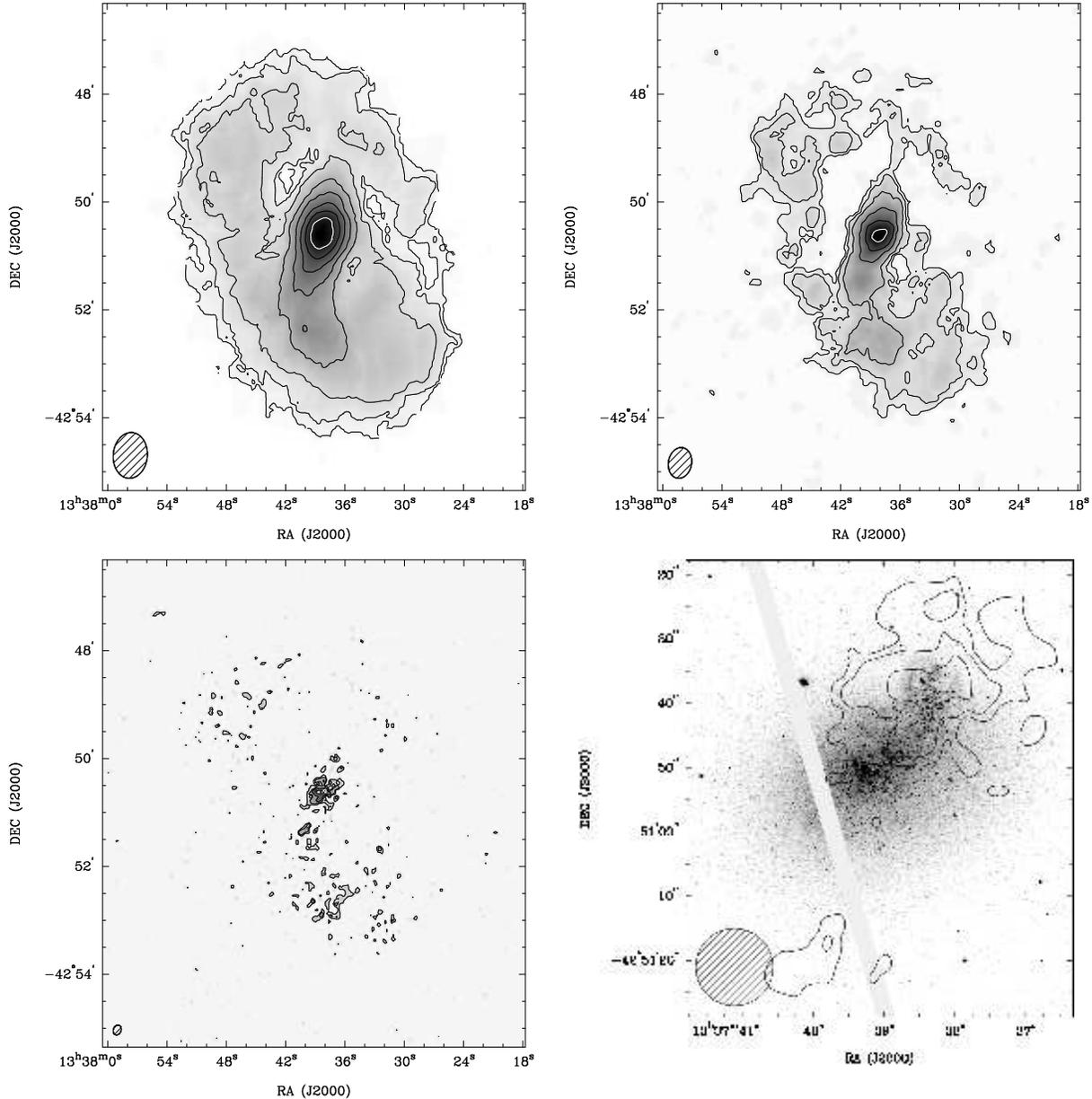,width=7.5cm,angle=-90}} \\
\end{tabular}
\caption{\HI\ moment maps of the dwarf galaxy NGC~5237 (HIPASS J1337--42)
   at different weightings of the ATCA $uv$-data. The resulting synthesized
   beam is displayed in the bottom left corner of each panel and given in 
   brackets here. {\bf Top:} natural weighting (left: beam = $51\farcs1 \times 
   37\farcs9$; contour levels = 0.05, 0.1, 0.2, 0.4, 0.6, 0.8, 1.0 and 1.2
   Jy\,beam$^{-1}$\kms) and robust $r=0$ weighting (right: beam = $34\farcs5 
   \times 25\farcs8$; contours = 0.05, 0.1, 0.2, 0.4, 0.6, and 0.8
   Jy\,beam$^{-1}$\kms) excluding the longest baselines. {\bf Bottom:} robust 
   weighting including the longest baselines (left: beam = $11\farcs2 \times 
   8\farcs6$; contours = 0.025, 0.05, 0.1, and 0.2 Jy\,beam$^{-1}$\kms) with
   a zoom-in to the central area overlaid on to an HST F606-band image (right: 
   beam = 12\arcsec; contours = 0.05, 0.1, and 0.2 Jy\,beam$^{-1}$\kms), 
   showing the high density \HI\ emission associated with a young star-forming 
   region to the north-west of the galaxy's stellar core. The measured \HI\ 
   flux densities decrease with increasing resolution (10.9, 7.5 and 2 Jy\kms\
   from top left to bottom left) due to extended \HI\ emission being filtered
   by the interferometer.}
\end{figure*}

{\bf ESO444-G084 (HIPASS J1337--28)} is a dwarf irregular galaxy located at a 
distance of $D_{\rm TRGB}$ = $4.61 \pm 0.46$~Mpc (Karachentsev et al. 2002). 
Its closest neighbours are IC\,4316 (64\farcm2), UGCA~365 (72\farcm2), M\,83
(109\farcm3) and NGC~5264 (125\farcm4), all members of the M\,83 group. 
ESO444-G084 is located 147~kpc north of the massive spiral galaxy M\,83. Our 
ATCA \HI\ moment maps reveal a symmetric disc, at least twice as large as the 
stellar $B_{25}$ extent, with a significant ($\sim$35\degr) warp of the outer
parts (see also C\^ot\'e et al. 2000). We measure \FHI\ = 16.5 Jy\kms, lower 
than the HIPASS \FHI\ of $21.1 \pm 3.2$ Jy\kms\ (Koribalski et al. 2004), and 
derive \MHI\ = $8.3 \times 10^7$\Msun. We calculate an \HI\ mass to light 
ratio of \MHI/\LB\ = 2.1, well above the typical value for dwarf irregular 
galaxies (Roberts \& Haynes 1994; see also Warren et al. 2004). C\^ot\'e et 
al. (2000) obtain an \HI\ rotation curve which reaches \vrot\ = 63.1\kms\ at 
a radius of 3.2 kpc. C\^ot\'e et al. (2009) estimate an SFR of $0.8 \times 
10^{-3}$\Msun\,yr$^{-1}$. \\

{\bf NGC~5253 (HIPASS J1339--31)} is a peculiar starburst galaxy at a distance 
of $D_{\rm TRGB}$ = 3.56~Mpc (Mould \& Sakai 2008), located $\sim$2\degr\ south
of M\,83 ($D_{\rm TRGB}$ = 4.92~Mpc) but significantly closer. For a detailed, 
multi-wavelength study see L\'opez-S\'anchez et al. (2012). NGC~5253 is one of 
the closest known BCD galaxies; its outer optical isophotes resemble that of a 
dwarf elliptical galaxy but the core is dominated by a young starburst and it 
contains a large amount of gas. Fig.~10 shows a very deep Parkes \HI\ map of 
the M\,83 group, using re-calibrated data from the HIDEEP survey (Koribalski 
2006, Minchin et al. 2003); no diffuse \HI\ gas is detected between NGC~5253 
and M\,83 down to \NHI\ = 10$^{18}$ atoms\,cm$^{-2}$ for gas filling the 
beam. Furthermore, the large separation between NGC~5253 and M\,83 (at least 
1~Mpc based on their independent distances), suggests that no recent tidal 
interactions occured between the two galaxies.
The \HI\ dynamics of NGC~5253 was studied by Kobulnicky \& Skillman (1995)
using very short (45 min.) VLA observations. They found that most of the \HI\
gas appears to rotate around the major axis of the stellar distribution.
We note that the integrated \HI\ flux density measured with the VLA (\FHI\ 
= 24 Jy\kms) is only 54\% of that detected by HIPASS (see Table~2). Kobulnicky 
\& Skillman (2008) explore if gas inflow, outflow or galaxy interactions are 
the cause of NGC~5253's unusual \HI\ gas dynamics. Using ATCA \HI\ data from
the LVHIS project L\'opez-S\'anchez et al. (2008, 2012) further investigate 
NGC~5253 and conclude that NGC~5253 experienced infall of a low-metallicity 
\HI\ cloud along the minor axis, triggering the powerful starburst.
NGC~5253 is one of several dwarf starburst galaxies with highly unusual \HI\ 
kinematics, including NGC~625 (Cannon et al. 2004), M\,82 (Yun, Ho \& Lo 1993),
and IC\,10 (Huchtmeier 1979; Wilcots \& Miller 1998; Manthey \& Oosterloo 
2008). The latter two, M\,82 and IC\,10, show \HI\ streamers most likely due 
to galaxy interactions. \\

{\bf IC\,4316 (HIPASS J1340--28)} is a dwarf irregular galaxy at a distance 
of $D_{\rm TRGB}$ = $4.41 \pm 0.44$ Mpc (Karachentsev et al. 2002), located 
72\farcm5 north-east of M\,83 (see Fig.~10). The large (mostly red) low 
surface brightness stellar body resembles that of an early-type galaxy while
optical and GALEX $UV$ images of the inner region are typical of dIrr galaxies. 
Our ATCA \HI\ data show a marginally resolved source with just a hint of 
gas motions along the N--S axis, well offset from the stellar major axis
(see also L\'opez-S\'anchez et al. 2012). We measure \FHI\ = 2.6 Jy\kms, 
somewhat more than in HIDEEP (Minchin et al. 2003), and derive \MHI\ = $1.2 
\times 10^7$\Msun. Begum et al. (2008) show a barely resolved GMRT \HI\ 
intensity map of IC\,4316. \\

{\bf NGC~5264 (HIPASS J1341--29)} is a Magellanic irregular galaxy of type 
IB(s)m), located very close to the eastern tidal arm of M\,83 (see Figs.~9
and 10). Our ATCA \HI\ maps reveal an extended, slowly rotating, possibly 
warped disc. The independent distance estimates for NGC~5264 ($D_{\rm TRGB}$ 
= $4.53 \pm 0.45$ Mpc, Karachentsev et al. 2002) and M\,83 ($D_{\rm TRGB}$ = 
$4.92 \pm 0.10$~Mpc) suggest they may be physically close and are likely 
tidally interacting. Their respective Local Group velocities are 300\kms\
(NGC~5264) and 332\kms\ (M\,83); see Table~6. NGC~5264 lies $\sim$1\degr\ 
from the centres of M\,83 
(HIPASS 1337--29) and IC\,4316 (HIPASS J1340--28). Our ATCA \HI\ maps show a 
well-resolved \HI\ distribution and rather peculiar velocity field. We measure 
\FHI\ = 10.3 Jy\kms, in agreement with HIPASS and HIDEEP (Koribalski et al. 
2004, Minchin et al. 2003), and derive \MHI\ = $5.0 \times 10^7$\Msun. Very 
short VLA observations by Simpson \& Gottesman (2000) show 
\HI\ emission in an asymmetric ring-like distribution or central depression; 
they detect only $\sim$40\% of the total \HI\ flux. L\'opez-S\'anchez et al. 
(2012) study the nearby starburst dwarf galaxy NGC~5253, south of M\,83, and 
provide a comparison with its dwarf neighbours, incl. NGC~5264. For a 
separation of 80~kpc between NGC~5264 and M\,83 the tidal radius at the 
position of NGC~5264 is $\sim$7 kpc, similar to the size of the ATCA \HI\ 
distribution. GALEX $UV$ and \Ha\ emission is detected in NGC~5264, which is 
dominated by two bright \HII\ regions; SFR $\sim$ 0.02\Msun\,yr$^{-1}$ (Lee 
et al. 2009). ATCA 20-cm radio continuum emission is detected (Shao et al. 
2017). \\

{\bf AM1321--304 (HIPASS J1324--30)} is a Magellanic dwarf irregular (dIm) 
galaxy at a distance of $D_{\rm TRGB}$ = 4.63 Mpc (Karachentsev et al. 2002).
Its nearest large neighbour is the spiral galaxy M\,83 at a projected distance 
of 173\farcm7. Our ATCA \HI\ maps reveal a marginally resolved source without 
a distinct velocity gradient. We measure \FHI\ = 1.7 Jy\kms, in agreement with 
Begum et al. (2008) who observed AM1321--304 as part of the FIGGS project. The 
HIPASS \FHI\ of $3.9 \pm 2.5$ Jy\kms\ is higher but with a large uncertainty 
(Banks et al. 1999). Begum et al. (2008) measure an \HI\ diameter of 1\farcm4 
($\sim$ the optical diameter).  \\

{\bf IC\,4247 (HIPASS J1326--30A)} is a dwarf irregular galaxy in the M\,83
sub-group with $D_{\rm TRGB}$ = $4.97 \pm 0.49$ Mpc (Karachentsev et al. 2007).
Its systemic velocity is 420\kms\ (Banks et al. 1999; Minchin et al. 2003).
IC\,4247's nearest neighbours are AM 1321--304 (45\farcm8), HIPASS J1321--31 
(100\farcm5), KKs\,54 (113\farcm2) and M\,83 (136\farcm6); projected distances
are given in brackets. Our low-resolution ATCA \HI\ data show a marginally 
resolved source with a velocity gradient along the major axis. Higher 
resolution ATCA \HI\ images reveal a tear-drop shaped gas distribution, 
hinting at tidal interactions. We  measure \FHI\ = 3.3 Jy\kms, in agreement 
with HIDEEP (Minchin et al. 2003), and derive an \HI\ mass of $1.9 \times 
10^7$\Msun. Lee et al. (2007) measure the properties of two \HII\ regions 
in IC\,4247. There are luminous AGB stars, helium-burning stars and some 
early type stars detected in this galaxy, indicating ongoing star formation. 
A detailed analysis of its star formation history shows the galaxy was 
constantly forming stars during its lifetime (Crnojevi\'c et al. 2011).  \\

\begin{figure} 
  \mbox{\psfig{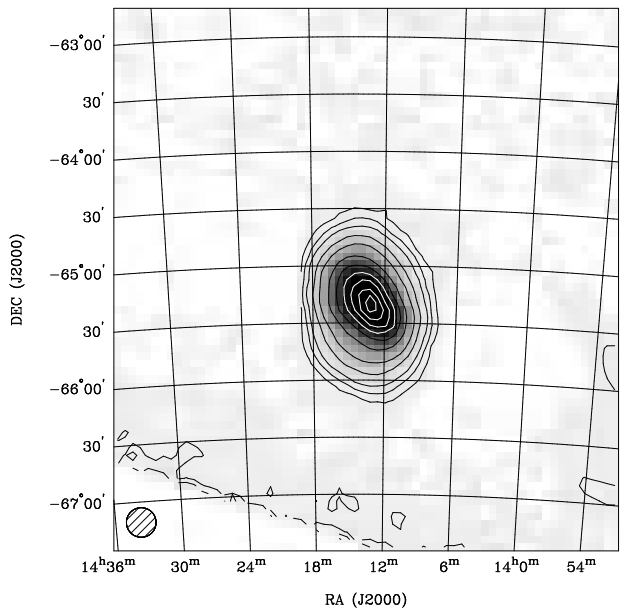}} \\ 
\caption{Low-resolution \HI\ distribution of the Circinus galaxy and 
  surroundings as obtained from the deep Parkes \HI\ multibeam survey of the 
  Zone of Avoidance (Juraszek et al. 2000, Staveley-Smith et al. 2016). The 
  contour levels are 7 $\times$ (0.4, 1, 2, 5, 10, 20, 30, 40, 50, 60 and 66) 
  Jy\,beam$^{-1}$\kms, where 7 Jy\,beam$^{-1}$\kms\ corresponds to an \HI\ 
  column density of $10^{19}$ cm$^{-2}$. The levels were chosen to match 
  those by Freeman et al. (1977; their Fig.~6) above their detection limit 
  of \NHI\ = $5 \times 10^{19}$ cm$^{-2}$; our detection limit is a factor 
  $\sim$10 lower. The gridded beam of 15\farcm5 is shown in the bottom left 
  corner.}
\end{figure}

\begin{figure*} 
\begin{tabular}{cc}
  \mbox{\psfig{file=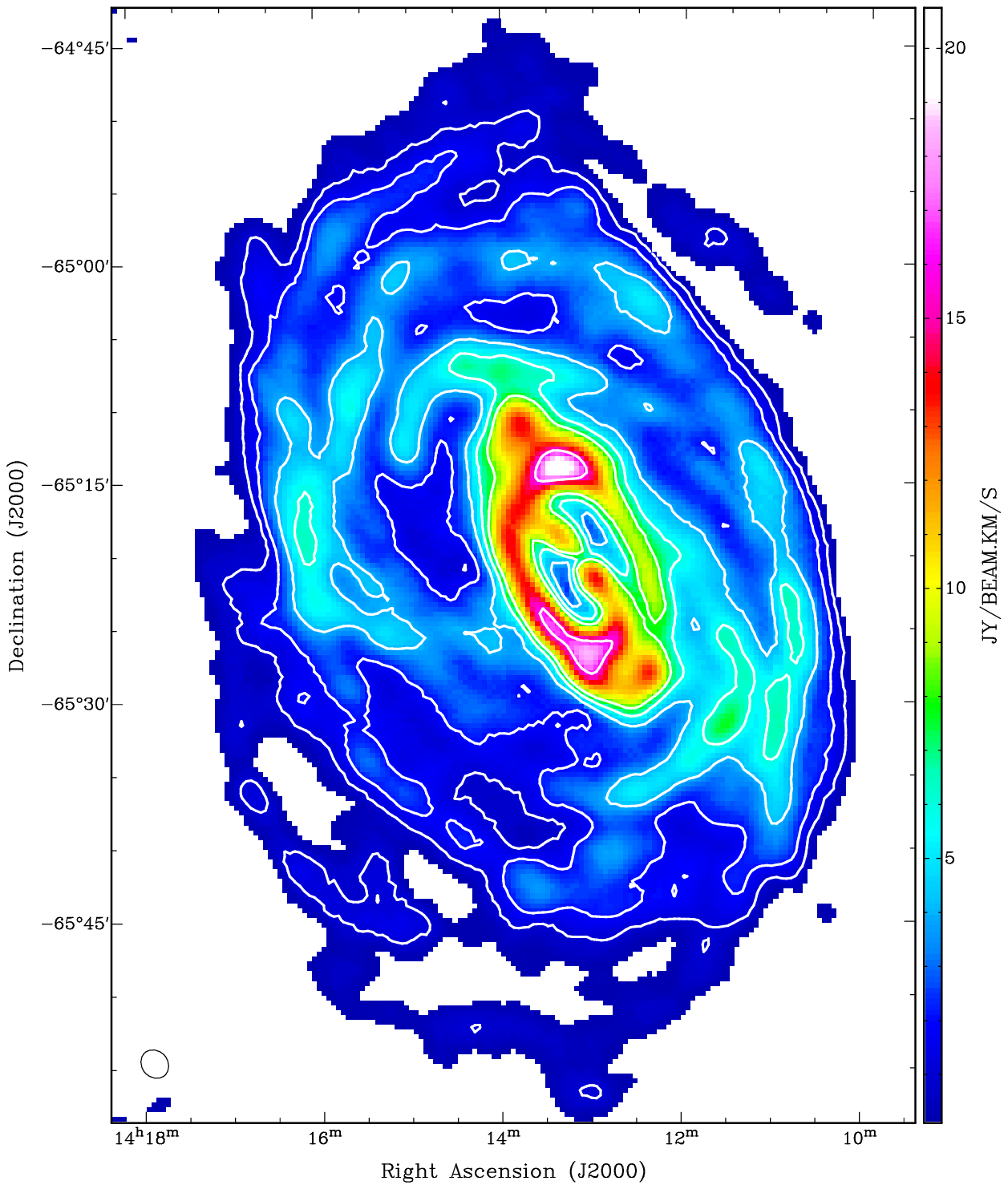,width=9cm}} & 
  \mbox{\psfig{file=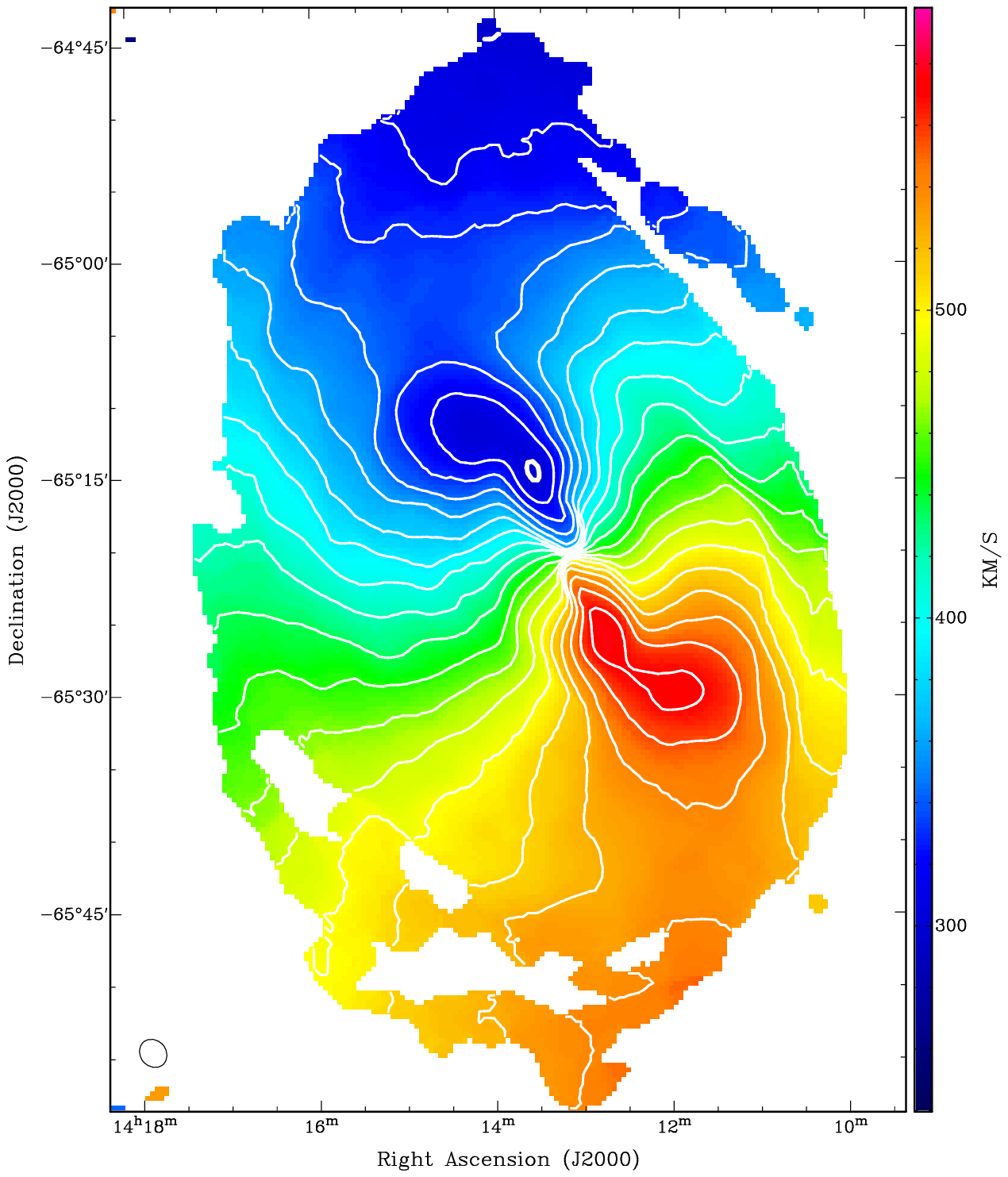,width=9cm}} \\
\end{tabular}
\caption{ATCA \HI\ distribution (left) and mean velocity field (right) of the 
   giant Circinus Galaxy (HIPASS J1413--65, $D_{\rm TF}$ = 4.2~Mpc) from 
   12-pointing ATCA mosaic observations. The contour levels are 1, 2, 4, 8, 
   and 16 Jy\,beam$^{-1}$\kms\ (mom0) and 300 to 570\kms\ in steps of 15\kms\ 
   (mom1). The angular resolution is $124\arcsec \times 107\arcsec$.}
\end{figure*}

\subsubsection{The Cen\,A subgroup}
The Cen\,A subgroup is dominated by the giant elliptical galaxy NGC~5128 
(Cen\,A), an active radio galaxy which lies $\sim$13.3\degr\ ($\sim$1~Mpc) 
south of M\,83. Fig.~14 shows the 20-cm radio continuum emission of Cen\,A 
as mapped by the Parkes 64-m telescope (Calabretta et al. 2014), overlaid 
with HIPASS contours and galaxy labels. Cen\,A's giant radio lobes extend 
over 10\degr\ in the north-south direction. New radio and optical images 
of Cen\,A are presented by McKinley et al. (2018) who investigate its jets,
outflows, and filaments. LVHIS galaxies in the vicinity of Cen\,A are 
visualised in Johnson et al. (2015; their Fig.~1) using our ATCA \HI\ maps 
colour-coded by distance. A multitude of gas-poor dwarf galaxies are also 
known in the vicinity of Cen\,A.  \\

\begin{figure}  
 \mbox{\psfig{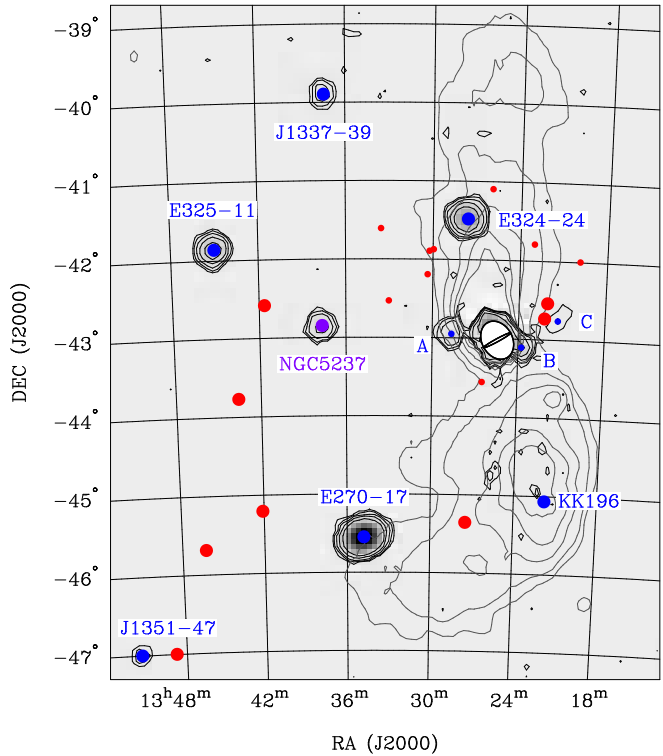}}
\caption{Galaxies in the vicinity of the radio galaxy Centaurus\,A. Three
   newly discovered \HI\ sources near Cen\,A are labelled A, B, and C. Black 
   contours (1, 2, 4, 8, and 16 Jy\,beam$^{-1}$\kms) and greyscales show the 
   \HI\ distribution from HIPASS, while grey contours show the 20-cm radio 
   continuum emission from CHIPASS (Calabretta et al. 2014). All known group 
   members in the displayed area are marked: dIrr galaxies (blue circles; 
   named), dSph galaxies (red circles), and the dwarf transitional galaxy 
   NGC~5237 (purple circle; named). The gridded HIPASS beam is 15\farcm5.}
\end{figure}

Recent deep imaging of NGC~5128 (Cen\,A) by Crnojevi\'c et al. (2016) using 
Megacam on the Magellan telescope reveals new stellar streams, shells, and 
faint dwarf galaxies in its vicinity. For the nine dwarf galaxies with 
distance estimates, they give upper limits of \MHI\ $\sim 5 \times 10^6$\Msun\
from HIPASS. 

Grossi et al. (2007) present an in-depth study of three unusual dwarf galaxies
in the Cen\,A Group: HIPASS J1321--31, HIPASS J1337--39 and HIDEEP J1337--3320.
They obtain deep HST WFPC2 images through the F555W and F814W filters 
($\sim$5000~s each), $R$-band and \Ha\ images with the WIYN 3.5-m telescope as 
well as ATCA \HI\ moment maps. Only HIPASS J1337--39 is detected in \Ha. They 
also note that HIPASS J1328--30 is a distant LSB galaxy with \vhel\ $\sim$
8100\kms\ which lies in the line-of-sight to a Galactic HVC. Crnojevi\'c et 
al. (2012) study five dIrr galaxies in the Cen\,A Group: CEN06, ESO269-G058, 
AM1318--444, HIPASS J1348--37 and ESO384-G016. They analyse archival HST ACS 
data in the F606W and F814W filters to study the stellar populations, 
metallicities and stellar masses. Neither AM1318--44 nor ESO384-G016 were 
detected in HIPASS and therefore not selected as part of the LVHIS sample. 
AM1318--444 (KK\,196) has a distance of $D_{\rm TRGB}$ = $3.98 \pm 0.29$~Mpc 
(Karachentsev et al. 2007), located at $\alpha,\delta$(J2000) = 13:21:47.1, 
--45:03:48.0. It appears to be in or near the southern radio lobe of Cen\,A. 
Lee et al. (2007) detect \Ha\ emission and derive a stellar mass of about 
$2.7 \times 10^7$\Msun. Using HIPASS we determine an upper limit of \FHI\ = 
5~Jy\kms, assuming \vhel\ = 741\kms\ and a velocity width of 100\kms. The 
r.m.s. noise in the area is increased due to the bright radio emission from 
Cen\,A's southern lobe. \\

{\bf ESO376-G016 (HIPASS J1043--37)} is a dwarf irregular galaxy at a distance
of $D_{\rm TF}$ = 7.1~Mpc. Its nearest neighbour is ESO318-G013 (HIPASS 
J1047--38; \vhel\ = 716\kms) about 2\degr\ away. Our ATCA data show a regularly
rotating \HI\ disc extending well beyond the stellar body. We measure \FHI\ = 
10.2 Jy\kms, in agreement with HIPASS (Koribalski et al. 2004), and derive 
\MHI\ = $1.2 \times 10^8$\Msun. \\

{\bf ESO318-G013 (HIPASS J1047--38)} is a small, edge-on spiral galaxy at a 
distance of $D_{\rm TF}$ = 6.5 Mpc. Its nearest neighbour is ESO376-G016 
(HIPASS J1043--37; \vhel\ = 668\kms) about 2\degr\ away. Our ATCA data show 
a well-resolved, but somewhat lopsided \HI\ disc extending well beyond the 
stellar body. While its rotation appears regular, the gas distribution shows 
some peculiar features. A faint \HI\ tail on the western (receding) side may 
hint at tidal interactions. We measure \FHI\ = 9.1 Jy\kms, in agreement with 
HIPASS (Koribalski et al. 2004), and derive \MHI\ = $9.1 \times 10^7$\Msun. \\

{\bf ESO215-G?009 (HIPASS J1057--48)} is a highly obscured ($b$ = 10.5\degr) 
dwarf irregular galaxy at a TRGB distance of $5.25 \pm 0.41$ Mpc (Karachentsev 
et al. 2007). It appears to be rather isolated with a low current star 
formation rate. Its nearest known neighbour is likely the small spiral galaxy
ESO318-G013 (HIPASS J1047--38) nearly 10\degr\ away. Warren, Jerjen \& 
Koribalski (2004) determine an \HI\ mass to light ratio of \MHI\,/\,\LB\ = 
$22 \pm 4$\Msun\,/\Lsun, the highest recorded for a galaxy in the literature. 
Our ATCA data show the \HI\ disc of ESO215-G?009 to extend over six times the 
Holmberg radius of the stellar disc. We measure \FHI\ = 110.1 Jy\kms, in
agreement with HIPASS (Koribalski et al. 2004), and derive \MHI\ = $7.1 \times
10^8$\Msun. Warren et al. (2004) note that its azimuthally averaged surface 
density remains below the critical gas surface density at 
all radii. While its rotation pattern is very regular, the galaxy's near 
circular shape (suggesting a near face-on orientation if the gas rotates in
circular orbits) is somewhat in contrast to the measured rotation amplitude.
Using 3D FAT Kamphuis et al. (2015) obtain an \HI\ rotation curve indicating 
\vrot\ = 93.0\kms\ at \Rmax\ = 9.3 kpc (for $i$ = 18\fdg5 and $PA$ = 119\fdg7; 
see Table~9) and \Mdyn\ = $1.9 \times 10^{10}$\Msun. See also previous \HI\ 
kinematics analysis by Warren et al. (2004) and Kirby et al. (2012). \\

{\bf ESO320-G014 (HIPASS J1137--39)} is a dwarf irregular galaxy at a distance 
of $D_{\rm TRGB}$ = $6.08 \pm 0.65$ Mpc (Karachentsev et al. 2007), located in
the outskirts of the Cen\,A group. Its closest known companion is the dwarf 
galaxy HIPASS J1132--32, and its most massive companions are NGC~3621 (HIPASS 
J1118--32) and Cen\,A (HIPASS J1324--42). Our ATCA data show a barely resolved 
\HI\ source centred on the fuzzy stellar body; a faint GALEX $UV$ source is 
also detected. Deep infrared photometry was obtained by Young et al. (2014). 
A weak gradient is seen in the \HI\ velocity field, but higher resolution is 
needed for further analysis of the gas kinematics. We measure \FHI\ = 2.0 
Jy\kms, in agreement with HIPASS (Meyer et al. 2004), and derive \MHI\ = $1.7 
\times 10^7$\Msun. \\

{\bf ESO379-G007 (HIPASS J1154--33)} is a dwarf irregular galaxy at a distance 
of $D_{\rm TRGB}$ = $5.22 \pm 0.52$ Mpc (Karachentsev et al. 2002), located in
the outskirts of the Cen\,A group. The stellar core appears to be embedded in
a low-surface brightness disc, which is somewhat extended to the east. Its 
nearest neighbour is likely ESO379-G024 (181\farcm6 away), discussed below. 
Our ATCA data show the \HI\ distribution and velocity field to be somewhat 
asymmetric, with the eastern side more diffuse and extended than the western 
side. We measure \FHI\ = 4.8 Jy\kms\ and derive \MHI\ = $3.1 \times 10^7$\Msun.
Begum et al. (2008) observed ESO379-G007 as part of the FIGGS project. They 
measure an \HI\ diameter of 3\farcm6 and an \HI\ flux density of \FHI\ = $5.0 
\pm 0.5$ Jy\kms, in agreement with our HIPASS and ATCA measurements. Bouchard 
et al. (2009) find just one \HII\ region, located $\sim$10\arcsec\ west of the 
optical centre. The galaxy shape and kinematics suggest that ram pressure or 
tidal forces may play a role and could possibly be responsible for the offset 
star-forming region. \\

{\bf ESO379-G024 (HIPASS J1204--35)} is a dwarf irregular galaxy at a Hubble 
distance of 4.9 Mpc, also located in the outskirts of the Cen\,A group. 
Using HIPASS we measure \vsys\ = 631\kms\ and obtain \vlg\ = 356\kms. The 6dF 
optical velocity for ESO379-G024 is $652 \pm 55$\kms. Its nearest neighbour is 
likely the galaxy ESO379-G007 (HIPASS J1154--33), $\sim$3\degr\ away. The ATCA 
\HI\ distribution of ESO379-G024 is noticeably lopsided, with more \HI\ gas 
on the south-western side. The \HI\ velocity field is somewhat irregular, 
suggesting a mixture of low rotation and peculiar velocities. We measure \FHI\ 
= 2.6 Jy\kms, within the uncertainties of the HIPASS \FHI, and derive \MHI\ =
$1.5 \times 10^7$\Msun.  \\

{\bf ESO321-G014 (HIPASS J1214--38)} is a dwarf irregular galaxy (type IABm)
at $D_{\rm TRGB}$ = $3.18 \pm 0.23$ Mpc (Dalcanton et al. 2009), located in
the outskirts of the Cen\,A group. Its nearest known neighbour, the galaxy 
ESO379-G024 (discussed above), lies 3\degr\ away. Using our ATCA \HI\ data we 
measure \FHI\ = 5.1 Jy\kms, more than three times the \HI\ emission detected 
by Begum et al. (2008) with the GMRT. The HIPASS \FHI\ value of $6.4 \pm 1.6$ 
Jy\kms\ (Koribalski et al. 2004) agrees within the uncertainties. We derive an 
\HI\ mass of $1.2 \times 10^7$\Msun. We find the \HI\ distribution to be quite 
symmetric apart from small deviations of the velocity field towards the 
northern end. The stellar distribution has a peculiar arrow shape (very 
prominent in GALEX $UV$ images) with star formation most prominent in the 
southern part. Bouchard et al. (2009) present an \Ha\ image, indicating weak 
emission along the galaxy minor axis. \\

{\bf ESO381-G018 (HIPASS J1244--35)} is a dwarf irregular galaxy in the Cen\,A
group, located at a TRGB distance of $5.32 \pm 0.51$ Mpc (Karachentsev et al. 
2007). Its nearest neighbour is ESO381-G020 (HIPASS J1246--33), just over 
2\degr\ away. Our ATCA \HI\ maps show a barely resolved source with a regular 
rotating disc. We measure \FHI\ = 2.6 Jy\kms, in agreement with HIPASS (Meyer 
et al. 2004), and derive an \HI\ mass of $1.7 \times 10^7$\Msun. A detailed 
study of its star formation history suggests there were two recent star 
formation bursts (Crnojevi\'c et al. 2011). \\

{\bf ESO381-G020 (HIPASS J1246--33)} is a dwarf irregular galaxy in the Cen\,A 
group, located at a TRGB distance of $5.44 \pm 0.37$~Mpc (Karachentsev et al. 
2007). Its outer stellar disc has a slightly triangular shape, appearing much
wider in the SE than the NW. Our ATCA \HI\ data reveal a well-resolved, regular
rotating disc, $\sim$10 times more massive than that of its nearest neighbour 
ESO381-G018 (HIPASS J1244--35). The mean \HI\ velocity field indicates a mild 
warp of the outer \HI\ disc (C\^ot\'e et al. 2000, Kirby et al. 2012). We 
measure \FHI\ = 32.8 Jy\kms, in agreement with HIPASS (Koribalski et al. 2004),
and derive $2.3 \times 10^8$\Msun. In contrast to its neighbour, ESO381-G020 
is clearly detected in \Ha, indicating ongoing star formation (Lee et al. 2007,
Bouchard et al. 2009, C\^ot\'e et al. 2009) at a rate of $\sim$3--6 $\times 
10^{-3}$\Msun\,yr$^{-1}$; see also Crnojevi\'c et al. (2011) . \\

{\bf NGC~4945 (HIPASS J1305--49)} is a well-known starburst spriral galaxy 
(type SBcd) in the Cen\,A group. It is oriented close to edge-on, highly 
obscured in the optical 
and located at a distance of $D_{\rm TRGB}$ = 3.80~Mpc (Mould \& Sakai 2008). 
Its nearest neighbours are ESO269-G058 (157\farcm1), ES0269-G?066 (285\farcm1; 
\vhel\ = 784\kms) and NGC~5206 (290\farcm2); projected distances are given in
brackets. About 7\arcmin\ north-west of the centre of NGC~4945 lies a faint 
galaxy, sometimes known as CEN05, but identified to be a distant spiral galaxy 
by Bouchard et al. (2004). Early ATCA \HI\ maps of NGC~4945 were published by 
Ott et al. (2001). They also map the extended 20-cm radio continuum emission 
of NGC~4945 and estimate a flux density of $4.2 \pm 0.1$ Jy over the central 
source. Consequently, \HI\ absorption dominates in the galaxy core, extending
over $\pm$200\kms\ with respect to the systemic velocity. This is similar to 
the velocity range of the main \HI\ emission. Here we present \HI\ results 
from new ATCA mosaic observations in the EW367 and 750A arrays. The overall
impression is that of a large, symmetric, regularly rotating \HI\ disc, which
does not extend much beyond the bright stellar disc.
Using 3D FAT Kamphuis et al. (2015) obtain an \HI\ rotation curve indicating 
\vrot\ = 173.6\kms\ at \Rmax\ = 16.7 kpc (for $i$ = 82\fdg8 and $PA$ = 44\fdg2;
see Table~9) and \Mdyn\ = $1.2 \times 10^{11}$\Msun. NGC~4945 has the highest
rotational velocity in our sample. A peculiar \HI\ feature is seen towards the
south, possibly extraplanar gas lagging behind. Combining all the available 
ATCA \HI\ data will allow a much more detailed analysis.
Using our low-resolution ATCA mosaic we measure \FHI\ = 405.3 Jy\kms\ and
derive \MHI\ = $1.4 \times 10^9$\Msun. This is $\sim$20\% higher than the 
HIPASS \FHI, which is significantly affected by \HI\ absorption (Koribalski 
et al. 2004). \\

Bouchard et al. (2007) obtained \HI\ observations for 18 dwarf galaxies in 
the Cen\,A Group. They detect five galaxies (ESO269-G037, CEN06, UGCA~365,
ESO384-G016 and ESO272-G025) with the ATCA and quote upper limits for the
remaining galaxies. 
For the dE galaxy ESO269-G?066 (KK\,190; $D_{\rm TRGB}$ = $3.82 \pm 0.26$ Mpc; 
Karachentsev et al. 2007), which was observed with the 64-m Parkes telescope 
for over 10~h, Bouchard et al. (2007) give an upper limit of \FHI\ = 0.026 
Jy\kms. This corresponds to an \MHI\ limit of $\sim$9 $\times 10^4$\Msun.
In the vicinity is also the dwarf galaxy ESO269-G037 (type dSph or dIrr).
Using ATCA \HI\ observation Bouchard et al. (2007) find \vhel\ = 744\kms\
and \FHI\ = $0.14 \pm 0.02$ Jy\kms. The latter corresponds to an \HI\ mass 
of only $4 \times 10^5$\Msun\ for $D_{\rm TRGB}$ = $3.48 \pm 0.35$~Mpc 
(Karachentsev et al. 2002).  \\

{\bf CEN06 (HIPASS J1305--40)} is a dwarf galaxy, also known as KK\,182, at 
a distance of $D_{\rm TRGB}$ = $5.78 \pm 0.42$ Mpc (Karachentsev et al. 2007).
It is located behind M\,83 and Cen\,A, separated by $\sim$1 and 2~Mpc, 
respectively. CEN06's nearest neighbour is likely NGC~5011C, located more than 
3\fdg5 away (\vhel\ = 647\kms, Saviane \& Jerjen 2007). Our ATCA \HI\ maps 
show the emission centred on the optical galaxy and a north-south velocity 
gradient. We measure \FHI\ = 4.5 Jy\kms, in agreement with HIPASS (Koribalski 
et al. 2004), and derive an \HI\ mass of $3.5 \times 10^7$\Msun. CEN06 was 
detected in \Ha\ by C\^ot\'e et al. (2009), and Crnojevi\'c et al. (2012) 
derive a stellar mass close to 10$^6$\Msun. \\

{\bf ESO269-G058 (HIPASS J1310--46A)} is a peculiar I0-type dwarf galaxy at a 
distance of $D_{\rm TRGB}$ = $3.80 \pm 0.29$~Mpc (Karachentsev et al. 2007). 
It was first catalogued in \HI\ by Banks et al. (1999), who established it as 
a member of the Cen\,A Group. Using our ATCA \HI\ data we measure a systemic 
velocity of 400\kms\ and \FHI\ =  5.4 Jy\kms, slightly lower than the HIPASS
\FHI\ (Koribalski et al. 2004), and derive \MHI\ = $1.8 \times 10^7$\Msun.
ESO269-G058 exhibits a regular \HI\ velocity, but we note that its \HI\ 
emission barely extends beyond the optical body. Its nearest neighbour is 
probably NGC~4945, over 2\fdg5 away. \Ha\ emission was detected by Phillips 
et al. (1986), and Crnojevi\'c et al. (2012) derive a stellar mass of $\sim9 
\times 10^8$\Msun. The enhanced star formation observed in ESO269-G058 may 
have been triggered by tidal interactions $\sim$1 Gyr in the past (Davidge 
2007). \\

{\bf HIPASS J1321--31} (KK\,195) is an unusual dwarf galaxy at a distance of 
$D_{\rm TRGB}$ = $5.22 \pm 0.30$~Mpc (Pritzl et al. 2003). Its stellar body 
(diameter $\sim$70\arcsec) is very faint and diffuse; no \Ha\ emission was
detected (Banks et al. 1999, Meurer et al. 2006, Grossi et al. 2007). Using 
our ATCA \HI\ data we determine a center position of $\alpha,\delta$(J2000) =  
13:21:09.4, --31:32:01.2, just east of the stellar body. Faint GALEX $UV$
emission is also detected. The ATCA \HI\ distribution is resolved and shows an
east-west velocity gradient. Grossi et al. (2007) detect only the bright \HI\ 
component, which is located to the south-east of the optical centre, while
we find further \HI\ emission to the west. We measure \FHI\ = 5.2 Jy\kms, in 
agreement with HIPASS (Koribalski et al. 2004), and derive = \MHI\ = $3.3 
\times 10^7$\Msun. HIPASS J1321--31 has a high \MHI/\LB\ ratio of 4.6. Begum 
et al. (2008) observed HIPASS J1321--31 as part of the FIGGS project. They 
measure an \HI\ diameter of 5 arcmin and an \HI\ flux density of \FHI\ = $4.8 
\pm 0.5$ Jy\kms, slightly lower than our HIPASS and ATCA measurements.  \\

{\bf NGC~5102 (HIPASS J1321--36)} is a large lenticular galaxy of type SA0 at 
a distance of $D_{\rm TRGB}$ = $3.40 \pm 0.39$ Mpc (Karachentsev et al. 2002) 
located in the outskirts of the Cen\,A group. Its nearest neighbours are 
HIPASS J1337--39 (267\farcm5), HIPASS J1305--40 (287\farcm2), and ESO324-G024 
(298\farcm4); projected distances are given in brackets. Our ATCA \HI\ moment 
maps reveal a large, somewhat asymmetric \HI\ disc with a central \HI\ 
depression that coincides with the bright stellar body. Most of the galaxy's
\HI\ emission resides in a ring-like structure of 3\farcm5 radius, also seen 
by van Woerden et al. (1993), surrounded by much fainter emission in the 
outskirts. A prominent \HI\ extension to the south-west resembles a tidal arm,
suggesting ongoing gas accretion. NGC~5102's \HI\ velocity field highlights 
the lopsided and mildly warped nature of the disc. We measure \FHI\ = 85.4 
Jy\kms, within the uncertainties of the HIPASS \FHI\ (Koribalski et al. 2004), 
resulting in \MHI\ = $2.3 \times 10^8$\Msun. Using the VLA CnD-array van 
Woerden et al. (1993) only 
detect the \HI\ emission within the inner disc/ring of NGC~5102 and measure 
\FHI\ = 50 Jy\kms. Using 3D FAT Kamphuis et al. (2015) obtain an \HI\ rotation 
curve indicating \vrot\ = 94.3\kms\ at \Rmax\ = 10.5 kpc (for $i$ = 75\fdg3 
and $PA$ = 42\fdg2; see Table~9) and \Mdyn\ = $2 \times 10^{10}$\Msun. 
Based on a stellar population study Davidge (2008) find NGC~5102 to be a 
post-starburst galaxy. 
Mitzkus et al. (2016) provide an excellent literature overview and analyse the 
stellar population and kinematics in the central region of NGC~5102 via MUSE 
data, suggesting two counter-rotating stellar discs. Faint radio continuum 
emission is detected in the galaxy core ($1.7 \pm 0.5$ mJy). \\

{\bf NGC~5128 (HIPASS J1324--42; Cen\,A)} is a nearby radio galaxy ($D_{TRGB}$ 
= $3.77 \pm 0.38$~Mpc; Rejkuba 2004) with two giant lobes spanning $\sim$8 
degrees on the sky (ie., $\sim$0.5 Mpc). It is the dominant elliptical galaxy 
in the Cen\,A/M\,83 galaxy group, which is known to contain $\sim$100 members. 
The bright and extended 20-cm radio continuum emission provides some technical
challenges for accurate \HI\ measurements which are further complicated by the 
presence of both \HI\ emission and absorption over a wide velocity range (see 
Koribalski et al. 2004). In Fig.~14 we show the radio lobes of Cen\,A as 
well as the locations of gas-rich or gas-poor group members.  

Cen\,A's optical appearance is that of a giant elliptical surrounded by a 
prominent band of highly opaque dust, which matches the \HI\ disc detected 
with the ATCA. Struve et al. (2010) measure \FHI\ = 144~Jy\kms, somewhat more 
than obtained by van Gorkom et al. (1990), and derive \MHI\ = $4.9 \times 
10^8$\Msun. For a comparison with Parkes \FHI\ measurements see the discussion
in Koribalski et al. (2004; their Section 3.6). Struve et al. (2010) obtain an 
\HI\ extent of 814\arcsec\ or 15~kpc and find a regular rotating, highly warped
disc. Further \HI\ emission is detected in the outer disc, somewhat aligned 
with the faint optical shells of Cen\,A and likely the result of gas accretion 
in a recent merger event. Schiminovich et al. (1994) also find \HI\ emission 
associated with the diffuse shells of NGC~5128 and measure \FHI\ = 208 Jy\kms.
Parkes \HI\ spectra obtained by Gardner \& Whiteoak (1976) shows emission from 
$\sim$250 to 850\kms. Using HIPASS we estimate an \HI\ mass of at least 
$6 \times 10^8$\Msun. 

Furthermore, we detect \HI\ emission east and west of Cen\,A (sources A, B 
and C in Fig.~14) without stellar counterparts. These could either be very 
faint dwarf galaxies near Cen\,A or another set of \HI\ shells. Their 
$\alpha,\delta$(J2000) positions and \HI\ properties are as follows. Source~A: 
13:28:39.9, --42:56:56.5 ($\sim$349 -- 507\kms, 35\farcm4 east of Cen\,A, 
\FHI\ = $7.9 \pm 1.1$ Jy\kms; slightly elongated N--S); source B: 13:23:46.2, 
--43:06:19.9 ($\sim$257 -- 414\kms, 19\farcm2 west of Cen\,A, unresolved in 
HIPASS), and source C: 13:21:16.9, --42:45:24.1 ($\sim$507 -- 534\kms, 
48\farcm5 east of Cen\,A, elongated NW--SE in HIPASS. GALEX $UV$ images do 
not reveal any counterparts. \\

{\bf ESO324-G024 (HIPASS J1327--41)} is a Magellanic irregular galaxy located
near or within the northern radio lobe of Cen\,A (Johnson et al. 2015). Its 
TRGB distance of $3.73 \pm 0.43$ Mpc (Karachentsev et al. 2002; Jacobs et al. 
2009) places it $\ga$100 kpc from the centre of Cen\,A, making it the closest 
dIrr companion. Numerous dSph galaxies (eg., KKs\,55, AM1318--444, KK\,197) 
are possibly located even closer to Cen\,A than ESO324-G024 (see Fig.~15). Our 
ATCA data show a tadpole-shaped \HI\ distribution, the head of which agrees
with the stellar body. The \HI\ tail points in a north-east direction and
shows a disturbed velociy field. C\^ot\'e et al. (2009) detect just a few \HII\
regions, mostly outside the bright stellar disc. For a detailed analysis of 
ESO324-G024's \HI\ morphology, kinematics, star formation and polarisation 
properties see Johnson et al. (2015). \\

{\bf ESO270-G017 (HIPASS J1334--45)}, also known as the Fourcade-Figueroa
galaxy, is a highly obscured edge-on spiral. Karachentsev et al. (2013) 
assigned a distance of $D_{\rm mem}$ = 3.6~Mpc based on the assumption 
that it is a member of the Cen\,A subgroup. Here we adopt the TRGB distance 
of $6.95 \pm 0.16$ Mpc derived by Jacobs et al. (2009), in agreement with 
the Hubble distance obtained from ESO270-G017's Local Group velocity (\vlg\ 
= 611\kms). Our ATCA \HI\ maps reveal a well-resolved disc with a mostly 
regular rotation pattern (see also Koribalski 2008). Some peculiar motions 
are discernible on the western (approaching) side, which also shows a higher 
velocity dispersion. The observed \HI\ extent of ESO270-G017 is only slightly 
larger than the stellar disc detected in deep optical images by David Malin. 
Mild asymmetries in the stellar disc (typical for SBm type galaxies) match 
those in the \HI\ distribution. We measure \FHI\ = 224.7 Jy\kms, $\sim$10\% 
higher than the HIPASS \FHI\ (Koribalski et al. 2004), and derive \MHI\ = 
$2.6 \times 10^9$\Msun. \\

{\bf NGC~5237 (HIPASS J1337--42)} appears to be a dwarf transitional galaxy,
located in the Cen\,A group at a distance of $D_{\rm TRGB}$ = $3.40 \pm 0.23$ 
Mpc (Karachentsev et al. 2007). Its nearest neighbours are the dSph galaxy
KKs\,57 ($D_{\rm TRGB}$ = $3.93 \pm 0.28$), the IBm galaxy ESO325-G?011 
(HIPASS J1345--41) and the giant elliptical Cen\,A, separated by 46\farcm7, 
100\farcm7 and 134\farcm3 ($\sim$133~kpc), respectively. Our ATCA \HI\ moment 
maps reveal a large \HI\ disc, extending well beyond the bright stellar body, 
and a slightly warped velocity field. In Fig.~11 we show the NGC~5237 \HI\ 
distribution at different angular resolutions, revealing areas of high-density 
\HI\ emission within the extended \HI\ disc. Interestingly, we find the densest
\HI\ peak to coincide with a young star-forming region, well-resolved in HST 
images from the Hubble Legacy Archive. While the overall stellar population 
resembles that of a dwarf elliptical galaxy, the single \HII\ region to the 
north-west and offset \HI\ emission suggest that NGC~5237 is a dwarf 
transitional galaxy. Its vicinity to Cen\,A makes it likely that tidal 
interactions / harassment are responsible for the observed peculiarities. 
Fig.~11 also highlights the difference in position angles between the outer 
and inner \HI\ distribution, which are both significantly offset from the 
stellar distribution. Thomson (1992) presents numerical simulations of galaxy 
interactions in the Cen\,A group and discusses the likely impact on NGC~5237 
and the edge-on galaxy ESO270-G017, also known the Fourcade-Figueroa shred. 
Using our low-resolution ATCA \HI\ maps we measure \FHI\ = 10.9 Jy\kms, 
comparable to the HIPASS \FHI\ of $12.1 \pm 2.6$ Jy\kms\ (Koribalski et al. 
2004), and derive \MHI\ = $3 \times 10^7$\Msun. Using 3D FAT Kamphuis et al. 
(2015) obtain an \HI\ rotation curve indicating \vrot\ = 75.2\kms\ at \Rmax\ 
= 4.9 kpc (for $i$ = 33\fdg8 and $PA$ = 50\fdg2; see Table~9) and \Mdyn\ = 
$6.5 \times 10^9$\Msun. C\^ot\'e et al. (2009) detect \Ha\ emission from the 
bright \HII\ region and derive SFR = $4.6 \times 10^{-3}$\Msun\,yr$^{-1}$. 
Radio continuum emission from that region is detected in our ATCA 20-cm data 
(see Shao et al. 2017). \\

{\bf HIPASS J1337--39} appears to be an old dwarf galaxy at a distance of 
$D_{\rm TRGB}$ = $4.83 \pm 0.20$ Mpc (Grossi et al. 2007). It is located in
the outskirts of the Cen\,A group, $\sim$4\degr\ from NGC~5128. Its closest
known neighbour is ESO325-G?011, $\sim$2\fdg5 to the south. HIPASS J1337--39's
near spherical stellar distribution (diameter $\sim$40\arcsec) resembles that 
of dSph galaxies, while the ongoing star formation activity, as revealed 
by the detected \Ha\ emission (Grossi et al. 2007), is more typical of dIrr 
galaxies. Our low-resolution ATCA \HI\ moment maps show a symmetric, regularly 
rotating disc aligned NE to SW. Its centre position is $\alpha,\delta(J2000) 
= 13^{rm h}\,37^{\rm m}\,25^{\rm s}$, --39\degr\,53\arcmin\,46\farcs7 (see
Table~8), in agreement with the optical centre. In contrast, at high resolution
the \HI\ distribution is elongated is the direction perpendicular to the 
main rotation axis and the velocity field and dispersion becomes noticeably 
more irregular. The ATCA \HI\ moment maps shown by Grossi et al. (2007), at 
a resolution of $31\arcsec \times 24\arcsec$, reveal a NW extension with a 
distinct velocity field (possibly an accreted \HI\ gas cloud or companion that 
triggered the recent SF). The overall impression is that of a dwarf 
transitional galaxy. We find \FHI\ = 6.9 Jy\kms, comparable to Grossi's 
\FHI\ = 7.2 Jy\kms\ and in agreement with the HIPASS \FHI\ of $6.6 \pm 1.8$ 
Jy\kms\ (Koribalski et al. 2004), and derive \MHI\ = $3.8 \times 10^7$\Msun.
HIPASS J1337--39 somewhat resembles the Sagittarius dwarf irregular galaxy 
(SagDIG; HIPASS J1929--17; $D$ = 1.16~Mpc) which is located in the outskirts 
of the Local Group (Hunter et al. 2012; Begum et al. 2006; Higgs et al. 
2016). \\

{\bf HIDEEP J1337--3320} is a small, low luminosity dwarf galaxy, located 
at a projected distance of $\sim$3.5\degr\ ($\la$300~kpc) from M\,83. Grossi 
et al. (2007) derive a distance of $D_{\bf TRGB}$ = $4.4 \pm 0.2$~Mpc. They 
suggest it is a transition type dwarf galaxy, similar to LGS3, Phoenix and 
DDO\,210 (the Aquarius dwarf galaxy), based on its very smooth and regular 
optical morphology (diameter $\sim$30\arcsec) and \HI\ gas content (\MHI\ = 
$5.1 \times 10^6$\Msun; \MHI/\LB\ = 1.4). -- This galaxy is currently not 
part of the LVHIS galaxy atlas. For a detailed discussion and ATCA \HI\ maps
see Grossi et al. (2007). \\

{\bf ESO325-G?011 (HIPASS J1345--41)} is a Magellanic barred irregular galaxy 
at a distance of $D_{\rm TRGB}$ = $3.40 \pm 0.39$ Mpc (Karachentsev et al. 
2002). Its nearest neighbour appears to be NGC~5237 at a projected distance 
of 100\farcm7, while the giant elliptical Cen\,A is nearly 4\degr\ away. 
Our ATCA \HI\ maps show a well-resolved, rather regular rotating disc. 
ESO325-G?011's \HI\ flux density and velocity dispersion are somewhat higher 
at the north-western (receding) side, which also shows minor asymmetries in 
the velocity field. The stellar
brightness is also significantly higher on the \HI-bright side. Using 
tilted-ring modelling Kirby et al. (2012), who provide a detailed description 
of ESO325-G?011, find \vrot\ = 46\kms\ (for $PA$ = 302\degr, $i$ = 42\degr) 
at $R_{\rm max}$ = 3.1 kpc, similar to C\^ot\'e et al. (2000). We measure \FHI\ 
= 26.4 Jy\kms, in agreement with HIPASS (Koribalski et al. 2004). C\^ot\'e et 
al. (2009) identify eight \HII\ regions across the stellar disc and derive 
SFR = $2.5 \times 10^{-3}$\Msun\,yr$^{-1}$. \\

{\bf ESO174-G?001 (HIPASS J1348--53)} is very low surface brightness galaxy
at a distance of $D_{\rm TF}$ = 3.6 Mpc. This is much lower than the Hubble 
distance of 6.2~Mpc derived from the HIPASS \vlg\ of 466\kms\ (Koribalski 
et al. 2004). ESO174-G?001 --- not be confused with the more distant spiral 
galaxy ESO174-G001 --- appears to be rather isolated with the nearest 
neighbours being NGC~5206, a gas-poor lenticular galaxy, and NGC~4945 (HIPASS 
J1305--49), a gas-rich starburst galaxy, both more than 5\degr\ away. 
ESO174-G?001's stellar disc is highly obscured by foreground stars due to its 
low Galactic latitude of 8\fdg6. The galaxy was not detected in deep $H$-band 
observations by Kirby et al. (2008b). Our ATCA \HI\ moment maps show a very
extended \HI\ disc, $\sim$3 times larger than the stellar disc, and a clear 
rotation pattern (see also Koribalski 2015, 2017). We find the 
kinematic major axis of ESO174-G?001 ($PA$ = 218\degr) differs significantly 
from the morphological major axis ($PA \sim 165$\degr) both of the stellar and 
the inner \HI\ disc. Kirby et al. (2012) discuss these misalignments and employ
tilted-ring fitting to show that the position angle decreases with radius from 
233\degr\ to 202\degr. Their rotation curve, which appears to flatten reaching
\Rmax\ = 360\arcsec\ (6.3 kpc), shows \vrot\ = 66\kms\ for $i$ = 40\degr.
Using 3D FAT Kamphuis et al. (2015) obtain a much higher rotational velocity
(\vrot\ = 97.3\kms) assuming $i$ = 22\fdg7 and $PA$ = 218\fdg4 (see Table~9).
Neither model fits the observed kinematics well. Similar misalignments are 
found in NGC~625 (C\^ote\' et al. 2000; Cannon et al. 2004), IC\,4662 (van 
Eymeren
et al. 2010) and GR8 (Begum \& Chengalur 2003), who discuss infall/outflow, 
polar rings and minor mergers. The X-shaped \HI\ distribution in the inner 
region of ESO174-G?001 provides further clues as to the nature of its peculiar 
gas kinematics. Using our ATCA \HI\ data we measure \FHI\ = 52.6 Jy\kms, in 
agreement with the HIPASS \FHI\ (Koribalski et al. 2004), and derive \MHI\ = 
$1.6 \times 10^8$\Msun. A 20-cm radio continuum source ($\sim$7 mJy) is 
detected just offset from the galaxy centre (Shao et al. 2017). \\

{\bf HIPASS J1348--37} is a dwarf irregular galaxy at $\alpha,\delta$(J2000) = 
13:48:34.1, --37:58:08.0 (see Table~7) as measured using our ATCA \HI\ data.
It was discovered by Banks et al. (1999) in a detailed HIPASS study of the 
Cen\,A Group. Its distance of $D_{\rm TRGB}$ = $5.75 \pm 0.66$~Mpc 
(Karachentsev et al. 2007) places it beyond M\,83 and more than 2~Mpc from 
Cen\,A. Its nearest neighbours are the dwarf galaxies ESO383-G087 (HIPASS 
J1349--36), ESO384-G016, ESO325-G?011 (HIPASS J1345--41) and NGC~5408 (HIPASS
J1403--41). Crnojevi\'c et al. (2012) suggest that HIPASS J1348--37 may be a 
transition-type dwarf galaxy, given its absence of recent star formation. They 
derive a stellar mass of $\sim$2 ($\pm$1) $\times 10^7$\Msun. Our ATCA \HI\ 
data show a barely resolved source, centred on the faint optical body. There 
is a hint of a north-south velocity gradient. We measure \FHI\ = 1.6 Jy\kms, 
in agreement with HIPASS (Meyer et al. 2004), and derive an \HI\ mass of at 
least $1.2 \times 10^7$\Msun.  \\

{\bf ESO383-G087 (HIPASS J1349--36)} is a barred Magellanic dwarf galaxy at a 
TRGB distance of $3.45 \pm 0.27$ Mpc (Karachentsev et al. 2007), located 
roughly between M\,83 and Cen\,A. Its nearest neighbours are the dwarf galaxies
ESO384-G016 (see Bouchard et al. 2009, Crnojev\'c et al. 2012) and HIPASS 
J1348--37, at projected distances of 103\farcm9 and 114\farcm8, respectively, 
while the more massive S0 galaxy NGC~5102 (HIPASS J1321--36) is 5\fdg5 away. 
Our ATCA data show an extended, but highly asymmetric \HI\ distribution of 
ESO383-G087 and an unusual \HI\ velocity field. The latter is similar to the
\HI\ velocity dispersion map and shows no clear sign of rotation (Koribalski 
2007), which is likely due to the galaxy's face-on orientation. This 
interpretation is supported by the narrow velocity widths of the integrated 
\HI\ profile. The very faint spheroidal halo seen in co-added optical images 
by Kemp \& Meaburn (1994) is similar in size to the detected ATCA \HI\ 
envelope. The somewhat asymmetric \HI\ extensions to the north and south are 
likely spiral arms. We measure \FHI\ = 25.4 Jy\kms, in agreement with HIPASS
(Koribalski et al. 2004), and derive \MHI\ = $7.1 \times 10^7$\Msun. Only a 
partial GALEX $UV$ image is currently available. \\

ESO384-G016 is a lenticular or transition-type dwarf galaxy, located at 
$\alpha,\delta$(J2000) = 13:57:01.6, --35:20:02.0 and $D_{\rm TRGB}$ = $4.53 
\pm 0.31$~Mpc (Karachentsev et al. 2007). It is not detected in HIPASS.  
Beaulieu et al. (2006) report \vhel\ = 504\kms and \FHI\ = 1.3 Jy\kms, ie. 
\MHI\ = $6.5 \times 10^6$\Msun. They show a marginally resolved ATCA \HI\ 
distribution and suggest the galaxy may be experiencing ram pressure stripping 
from its passage through the intragroup medium. Jerjen et al. (2000) measure 
a stellar velocity of $561 \pm 32$\kms. ESO384-G016 is detected in \Ha\ by 
Bouchard et al. (2009), is clearly visible in GALEX $UV$ images and looks like
blue compact galaxy in optical images.  \\

{\bf HIPASS J1351--47} is a dwarf irregular galaxy at $\alpha,\delta$(J2000) = 
13:51:21.2, --46:59:53.0 (see Table~7) as measured using our ATCA \HI\ data.
It was discovered by Banks et al. (1999) who measured \vsys\ = 530\kms. Its 
distance of $D_{\rm TRGB}$ = $5.73 \pm 0.73$~Mpc (Karachentsev et al. 2007) 
is similar to that of HIPASS J1348--37; both are located in the outskirts of 
the Cen\,A Group. Its nearest neighbours appear to be several dSph galaxies,
for example LEDA\,166179 (see Fig.~15). The low Local Group velocity (\vlg\ 
= 291\kms, Table~2) of HIPASS J1351--47 indicates an infall velocity of 
$\sim$140\kms\ towards the group. Our ATCA \HI\ data show a marginally 
resolved source and hint at a kinematic $PA$ of $\sim$45\degr. We measure 
\FHI\ = 3.6 Jy\kms, in agreement with HIPASS (Meyer et al. 2004), and derive 
an \HI\ mass of $2.8 \times 10^7$\Msun.  \\

{\bf NGC~5408 (HIPASS J1403--41)} is a barred Magellanic irregular galaxy 
(type IB(s)m) in the Cen\,A group, located at a distance of $4.81 \pm 0.48$ 
Mpc (Karachentsev et al. 2002). Its closest neighbours are ESO325-G?011
(HIPASS J1345--41), HIPASS J1348--37, and NGC~5237 (HIPASS J1337--42) at 
projected distances of 207\farcm7, 264\farcm5, and 299\farcm1, respectively.
Our ATCA data show a well-resolved, fairly asymmetric \HI\ distribution with
a reasonably regular velocity field. The galaxy's south-western side shows
peculiar motions and relatively high velocity dispersion. The disturbed 
appearance, studied in detail by van Eymeren et al. (2010), could be due to 
gas outflows and/or accretion. The total \HI\ extent of NGC~5408 is about 
$10\farcm0 \times 6\farcm3$ ($PA$ = 302\degr), four times larger than its 
optical size. Van Eymeren et al. (2010) also studied the distribution and 
kinematics of the ionized gas using \Ha\ images and spectra, revealing a 
diffuse filamentary structure similar to IC\,4662. NGC~5408 is a well-studied 
galaxy, mainly because of its ultra-luminous X-ray source, possibly an
intermediate mass black hole. Strong 20-cm radio continuum emission is 
associated with the bright \HII\ region detected by van Eymeren et al. 
(2010). \\

{\bf Circinus (HIPASS J1413--65)} is a highly obscured, late-type spiral 
galaxy with a very large, warped \HI\ disc (Freeman et al. 1977; Jones et al. 
1999; Curran et al. 2008; For, Koribalski \& Jarrett 2012).
It is located behind the Galactic Plane (at $l,b$ = 311.3\degr, --3.8\degr),
which makes it difficult to study its optical properties and large-scale 
environment due to high stellar density and dust extinction (\AB\ = 5.3 mag, 
Schlafly \& Finkbeiner 2011). For the same reason, the distance to Circinus
is still rather uncertain. Based on a variety of methods, Freeman et al. 
(1977) derived an approximate distance of $4.2 \pm 0.8$ Mpc, which we adopt 
here. Karachentsev et al. (2007) estimated $D_{\rm TF}$ = $2.82 \pm 0.28$~Mpc
using 2MASS infrared magnitudes, while K13 give $D_{\rm TF}$ = 4.2~Mpc. 
Freeman et al. (1977) determine de Vaucouleurs and Holmberg optical diameters 
for the Circinus Galaxy of 11\farcm9 and 17\farcm2, respectively, after 
extinction correction (their Table~3) as well as $PA$ = 210\degr\ and $i$ = 
65\degr. They give \LB\ = $7.2 \times 10^9$\Lsun. Using our ATCA \HI\ mosaic 
we find an enormous \HI\ envelope for the Circinus Galaxy, extending over 
60\arcmin\ (at the outermost \HI\ contour), well beyond the optical Holmberg 
radius. We measure an \DHI\ = 48\farcm7 at 1\Msun\,pc$^{-2}$. Its large \HI\ 
extent was also noted in the \HI\ surveys of the Zone of Avoidance (HIZOA, 
Juraszek et al. 2000, Staveley-Smith et al. 2016) and in HIPASS (Koribalski 
et al. 2004); see Fig.~13. The ATCA \HI\ distribution and 
mean \HI\ velocity field of the Circinus galaxy, obtained with a 12-pointing 
mosaic in the 375-m configuration, show its large-scale warped \HI\ disc. 
For, Koribalski \& Jarrett (2012) analyse high-resolution Spitzer mid-infrared
images of the Circinus Galaxy and compare with Parkes and ATCA \HI\ data. They
derive star formation rates and analyse the local star formation conditions. 
The ionized gas in Circinus was studied by Elmouttie et al. (1998a) via 
Fabry-Perot \Ha\ imaging. No GALEX $UV$ images are available yet. Circinus
also hosts bipolar radio lobes extending perpendicular to the inner disc
(Elmouttie et al. 1998b, Wilson et al. 2011). 
The large \HI\ discs of Circinus and M\,83 are typical for their \MHI\ (see
Wang et al. 2016) and allow us to obtain accurate rotation curves out to radii
of $\sim$50 kpc. Using 3D FAT Kamphuis et al. (2015) obtain an \HI\ rotation 
curve indicating \vrot\ = 161.4\kms\ at \Rmax\ = 47.2 kpc (for $i$ = 62\fdg2 
and $PA$ = 199\fdg6; see Table~9) and \Mdyn\ = $2.9 \times 10^{11}$\Msun. \\

{\bf ESO222-G010 (HIPASS J1434--49)} is a dwarf irregular galaxy at low 
Galactic latitude, located in the outskirts of the Cen\,A group. Its systemic 
velocity of \vhel\ = $622 \pm 5$\kms\ (Koribalski et al. 2004) and \vlg\ = 
431\kms\ suggest a Hubble distance of 5.8 Mpc. No TRGB distance is currently 
available. Our ATCA \HI\ data show a resolved distribution and a regular 
velocity field ($PA \sim 20\degr$), aligned with the faint stellar body. Higher 
resolution maps are needed to study its \HI\ morphology and velocity
field in detail. We measure \FHI\ = 6.1 Jy\kms, in agreement with HIPASS 
(Koribalski et al. 2004), and derive \MHI\ = $4.8 \times 10^7$\Msun. Bright 
\Ha\ emission detected by Kaisin et al. (2007) agrees with the centre of the 
ATCA \HI\ distribution presented here.  \\

{\bf HIPASS J1441--62} is a dwarf irregular galaxy at $\alpha,\delta$(J2000) 
= 14:41:42.2, --62:46:04.2 (see Table~7) as measured from our ATCA \HI\ data.
The galaxy is located at low Galactic latitude ($l,b$ = 315\fdg2, --2\fdg55, 
\AB\ = 4.9). Its systemic velocity of \vhel\ = $672 \pm 8$\kms\ (Koribalski 
et al. 2004) and \vlg\ = 461\kms\ suggest an approximate Hubble distance of 
6.0 Mpc. No stellar counterpart is identified.  Its nearest known neighbour is
the Circinus galaxy, $\sim$4\degr\ away. Our ATCA data show a resolved \HI\ 
distribution and a regular velocity field ($PA \sim 260\degr$). We measure 
\FHI\ = 2.4 Jy\kms, within the uncertainty of the HIPASS \FHI\ (Koribalski 
et al. 2004), and derive \MHI\ = $2.0 \times 10^7$\Msun.  \\

{\bf ESO223-G009 (HIPASS J1501--48)} is a Magellanic irregular galaxy at a
TRGB distance of $6.49 \pm 0.51$~Mpc (Karachentsev et al. 2007). Due to its
low Galactic latitude ($b$ = 9\fdg2) it is obscured by dust and foreground 
stars. While the $B$-band Galactic extinction in this direction is moderate 
(\AB\ = 0.94), the density of Galactic foreground stars is quite high, making 
it difficult to estimate the galaxy's optical dimensions and orientation. No 
optical velocity has yet been measured. Using HST ACS images we find that the 
main stellar body of ESO223-G009 is best described by an ellipse of size 
$160\arcsec \times 130\arcsec$ (5~kpc $\times$ 4~kpc), $PA$ = 135\degr, 
centered at $\alpha,\delta$(J2000) = 15:01:08, --48:17:30. ESO223-G009 appears 
to be a relativly isolated galaxy, with only one known neighbour, the late-type
spiral ESO274-G001 (HIPASS 1514--46 at $D$ = 3.1 Mpc, i.e. in the foreground),
within 3\degr. It may form a loose association with the dwarf irregular 
galaxies ESO222-G010, ESO272-G025 and HIPASS J1526--51 (at projected distances 
of 266\arcmin, 282\arcmin, and 300\arcmin, respectively), about 2~Mpc behind 
the Cen\,A Group. Our ATCA \HI\ maps of ESO223-G009 reveal a large, nearly 
circular gas distribution and a very peculiar velocity field (see Koribalski 
2010). We measure \FHI\ = 97.0 Jy\kms, in agreement with HIPASS (Koribalski 
et al. 2004), and derived \MHI\ = $9.6 \times 10^8$\Msun. 
The highly twisted and warped \HI\ velocity field is indicative of a 
face-on warp where the gaseous disc wobbles in and out of the plane. Using 3D 
FAT Kamphuis et al. (2015) obtain an \HI\ rotation curve indicating \vrot\ = 
85.6\kms\ at \Rmax\ = 15.3 kpc (for $i$ = 20\fdg3 and $PA$ = 256\fdg1; see 
Table~9) and \Mdyn\ = $2.6 \times 10^{10}$\Msun. ESO223-G009 shows similarities
to the spiral galaxies NGC~628 (Kamphuis \& Briggs 1992), NGC~3642 
(Verdes-Montenegro et al. 2002), IC~10 (Manthey \& Oosterloo 2008) and the 
lenticular galaxy ESO381-G047 (Donovan et al. 2009). The common feature of 
these rather different galaxies is their warped, nearly face-on \HI\ discs. \\

{\bf ESO274-G001 (HIPASS J1514--46)} is a late-type spiral galaxy in the 
outskirts of the Cen\,A group, located at a distance of $D_{\rm TRGB}$ = $3.09 
\pm 0.23$ Mpc (Karachentsev et al. 2007). Its Local Group velocity of 360\kms\ 
suggests peculiar motions of $\sim$130\kms\ towards the Cen\,A group. Its 
nearest known neighbours are ESO223-G009 (HIPASS J1501--48) and HIPASS 
J1526--51 at projected distances of 160\arcmin\ and 287\arcmin, respectively.
The edge-on stellar disc of ESO274-G001 is obscured by Galactic foreground 
dust and stars. Our ATCA data show an extended, regular rotating \HI\ disc 
with signs of a minor warp in the outskirts (see also Koribalski 2008). We 
measure \FHI\ = 138.4 Jy\kms, $\sim$10\% more than the published HIPASS 
\FHI\ (Koribalski et al. 2004), and derive \MHI\ = $3.1 \times 10^8$\Msun. 
We measure \HI\ dimensions of about $17\arcmin \times 4\arcmin$, corresponding 
to 15~kpc $\times$ 3~kpc. Using high-resolution (10\arcsec) ATCA \HI\ maps 
O'Brien et al. (2010) measure \FHI\ = 152.6 Jy\kms, which seems rather high 
compared to our measurements. They derive a maximum rotational velocity of 
89.4\kms. \tiri\ modeling shows that ESO274-G001 is slightly warped. The 
scale height obtained in the best-fit model is $\sim$250~pc. \Ha\ imaging 
by Rossa \& Dettmar (2003) reveals prominent extraplanar ionised gas. Lee 
et al. (2007) and C\^ot\'e et al. (2009) also obtain \Ha\ images and find most 
\HII\ regions located in the southern (approaching) side of the galaxy; they 
derive an SFR of $1.5 \times 10^{-2}$\Msun\,yr$^{-1}$. Our ATCA 20-cm radio 
continuum maps reveal extended emission in the disc as well as a bright 
nucleus. \\

{\bf UKS1424--460 (HIPASS J1428--46)} is a Magellanic irregular galaxy (type 
IB(s)m) of low surface brightness, located close to a foreground star. Its 
location and TRGB distance of $3.58 \pm 0.33$ Mpc (Karachentsev et al. 2004) 
places it in the far outskirts of the Cen\,A group. Its nearest neighbours are 
ESO272-G025 (HIPASS J1443--44) and ESO222-G010 (HIPASS J1434--49) at projected 
distances of 188\arcmin\ and 200\arcmin, respectively. Both are located well
behind UKS1424--460. Our ATCA data show an extended, regularly rotating \HI\ 
disc. We measure an \HI\ flux density of 16.7 Jy\kms, in agreement with HIPASS 
(Koribalski et al. 2004), corresponding to an \HI\ mass of \MHI\ = $5.0 \times 
10^7$\Msun. Begum et al. (2008) observed UKS1424--460 as part of the FIGGS 
project. They measure an \HI\ diameter of 6\farcm5 ($\sim$2.7 $\times$ the 
optical diameter) and an \HI\ flux density of \FHI\ = $17.4 \pm 1.7$ Jy\kms, 
in agreement with our measurements. Using 3D FAT Kamphuis et al. (2015) obtain 
an \HI\ rotation curve indicating \vrot\ = 22.0\kms\ at \Rmax\ = 3.2 kpc (for 
$i$ = 74\fdg6 and $PA$ = 122\fdg8; see Table~9) and \Mdyn\ = $3.6 \times 
10^8$\Msun; note these values are highly uncertain. Optical magnitudes are 
hard to obtain due to significant foreground dust and stars. Kaisin et al. 
(2007) and C\^ot\'e et al. (2009) show a marginal detection of \Ha\ emission 
on the north-western (approaching) side of the galaxy. They derive an SFR of 
$1.4 \times 10^{-4}$\Msun\,yr$^{-1}$. \\

{\bf ESO272-G025 (HIPASS J1443--44)} is a dwarf irregular galaxy located in 
the outskirts of the Cen\,A group at a Hubble distance of 5.9~Mpc. Its 
nearest neighbour is most likely the galaxy ESO223-G009 (HIPASS J1501--48).
Our ATCA data show a barely resolved \HI\ source centred on the bright and 
compact optical body; no rotation is discernible in our velocity field (see 
also Bouchard et al. 2007). Galactic cirrus filaments are visible in the 
optical image. We measure \FHI\ = 1.6 Jy\kms, in agreement with HIPASS (Meyer 
et al. 2004), and derive \MHI\ = $1.3 \times 10^7$\Msun.
Reduzzi \& Rampazzo (1995) suggest ESO272-G025 has an extended low-surface 
brightness disc or is possibly a galaxy pair. 
ESO272-G025 shows clumpy and diffuse \Ha\ emission (Kaisin et al. 2007; 
C\^ot\'e et al. 2009), with an estimated star formation rate of $1.5 \times 
10^{-3}$\Msun\,yr$^{-1}$.

\subsection{Other Galaxies (in RA order)}

{\bf ESO115-G021 (HIPASS J0237--61)} is a Magellanic barred spiral galaxy at 
a distance of $D_{\rm TRGB}$ = $4.99 \pm 0.10$~Mpc (Tully et al. 2006). Apart
from its small \HI\ companion noted below, ESO115-G021 has no known neighbours
within 5\degr. Our ATCA maps show an extended, regular rotating, nearly edge-on
\HI\ disc with a diameter of $\sim$15\arcmin, corresponding to 22~kpc. We 
measure \FHI\ = 110.7 Jy\kms, $\sim$10\% higher than quoted in the HIPASS BGC 
using a point-source fit (Koribalski et al. 2004), and derive \MHI\ = $6.5 
\times 10^8$\Msun. The \HI\ velocity field of ESO115-G021 shows some peculiar 
motions: (1) a disturbance in the disc area closest to the companion, 
suggesting mild interactions, and asserting the close proximity of the two 
galaxies, and (2) a significant twist in the iso-velocity contours towards 
the south-western (approaching) end of the disc, indicating a warp. 
Using high-resolution (9\arcsec) ATCA \HI\ maps O'Brien et al. (2010) measure 
a projected scale height of 26\farcs7 (646~pc) with vertical \HI\ structures 
extending to nearly 2~kpc. They measure a rotational velocity of 63\kms.
Furthermore, our deep ATCA \HI\ observations reveal a dwarf companion, ATCA 
J023658--611838 at $\alpha,\delta$(J2000) = 02:36:58.8, --61:18:38.5, with a 
systemic velocity of 508\kms\ at a projected distance of 6\arcmin\ (8.6~kpc) 
from ESO115-G021's centre (\vsys\ = $515 \pm 2$\kms). The companion's \HI\ 
flux density is 0.11 Jy\kms, corresponding to an \HI\ mass of $6.5 \times 
10^5$\Msun\ (assuming $D$ = 5~Mpc). A faint GALEX $UV$ counterpart to the \HI\ 
source is visible at $\alpha,\delta$ = $02^{\rm h}\,36^{\rm m}\,59.5^{\rm s}$, 
--61\degr\,18\arcmin\,38\farcs7; we measure FUV and NUV magnitudes of 21.02 
and 21.17, corrected for Galactic extinction (NUV--FUV = 0.15). The detected 
\HI\ source is therefore not an HVC complex. For further discussions see
Koribalski (2008) and O'Brien et al. (2010). \\

{\bf ESO154-G023 (HIPASS J0256--54)} is a Magellanic barred spiral galaxy at
a distance of $D_{\rm TRGB}$ = $5.76 \pm 0.10$ Mpc (Tully et al. 2006). Its 
closest neighbour is NGC~1311 ($D_{\rm TRGB}$ = 5.20 Mpc), more than 4\degr\ 
away. ESO154-G023 is similar to the galaxy ESO115-G021 in size and orientation
(Koribalski 2008). When comparing the optical discs, ESO154-G023 appears to 
have either a thicker disc or is slightly less inclined than ESO115-G021. Our
ATCA \HI\ maps show an extended, regularly rotating \HI\ disc with a pronounced
warp on both sides. The latter appears to start at the edge of the bright 
stellar disc and is likely contributing to the apparent disc thickness. We 
measure ATCA \FHI\ = 130.6 Jy\kms, within the $\sim$10\% uncertainty of the 
HIPASS \FHI\ (Koribalski et al. 2004), and derive \MHI\ = $10^9$\Msun. Using 
3D FAT Kamphuis et al. (2015) obtain an \HI\ rotation curve indicating \vrot\ 
= 63.6\kms\ at \Rmax\ = 15.7 kpc (for $i$ = 79\fdg7 and $PA$ = 218\fdg3; see 
Table~9) and \Mdyn\ = $1.5 \times 10^{10}$\Msun. SINGG \Ha\ images reveal star 
formation throughout the stellar disc (Meurer et al. 2006). Furthermore, our 
deep ATCA \HI\ observations reveal a dwarf galaxy at $\alpha,\delta$(J2000) = 
02:56:40.3, --54:35:38.8 (see Table~7) behind the disc of ESO154-G023 at 
velocities of $\sim$1100 to 1140\kms. We measure \FHI\ = 0.31 Jy\kms\ and 
derive \MHI\ = $1.2 \times 10^7$\Msun\ for $D$ = 12.7 Mpc. The newly discovered
galaxy is also detected in the AAT $H$-band (see Kirby et al. 2008b), SINGG 
\Ha\ (Meurer et al. 2006) and GALEX $UV$ images. \\

{\bf NGC~1313 (HIPASS J0317--66)} is a late-type barred spiral galaxy at a
TRGB distance of $4.07 \pm 0.22$~Mpc (Grise et al. 2008). For a detailed study
see Ryder et al. (1995) who find a large \HI\ disc, extending $\sim 18\arcmin 
\times 12\arcmin$, and measure \FHI\ = 455 Jy\kms. By combining the ATCA \HI\ 
data of NGC~1313 with our ATCA \HI\ data for the dwarf companion AM0319--662, 
discussed below, we create an even more sensitive mosaic of the area. We 
measure \FHI\ = 491.9 Jy\kms\ for NGC~1313, in agreement with our recent 
H75-array \HI\ mosaic (\FHI\ = 496.2 Jy\kms) and Parkes \HI\ mosaics (Ryder 
et al. 1995, Barnes \& de Blok 2004, Koribalski et al. 2004), and derive 
\MHI\ = $1.9 \times 10^9$\Msun. Our high-resolution ATCA \HI\ data show an 
extended, somewhat asymmetric \HI\ distribution and mildly disturbed velocity 
field. Using 3D FAT Wang et al. (2017) obtain an \HI\ rotation curve 
indicating \vrot\ = 220\kms\ at \Rmax\ = 10.3 kpc (see Table~9) and \Mdyn\ 
= $1.1 \times 10^{11}$\Msun. \Ha\ imaging by Ryder \& Dopita (1993) reveals 
the brightest \HII\ regions in the bar and inner spiral arms of NGC~1313, 
surrounded by low surface brightness emission in the form of filaments, arcs 
and shells. \\

{\bf AM0319--662 (HIPASS J0321--66)} is a dwarf irregular galaxy at a TRGB 
distance of $3.98 \pm 0.36$~Mpc (Karachentsev et al. 2003). It is a close 
companion of the large spiral galaxy NGC~1313 (HIPASS J0317--66), separated 
by only 20\arcmin\ or 23 kpc. Our ATCA data show a barely resolved \HI\ source
and no discernible rotation in the velocity field ($\sim$524 -- 544\kms). 
The peak of the ATCA \HI\ distribution is slightly offset to the south of 
AM0319--662's stellar body, suggesting it may be a transition dwarf galaxy 
(see also Makarova et al. 2005). We measure \FHI\ $\sim$ 0.3 Jy\kms\ and 
derive \MHI\ $\sim$ 10$^6$\Msun. \\

{\bf NGC~1311 (HIPASS J0320--52)} is a Magellanic barred spiral galaxy at
distance of $D_{\rm TRGB}$ = 5.22~Mpc (Tully et al. 2013). Its closest 
neighbours are IC\,1959 (HIPASS J0333--50) and ESO154-G023 (HIPASS J0256--54) 
and ESO199-G007 (HIPASS J0258--49), all separated by $\ga$1~Mpc. Eskridge et 
al. (2008) conduct a detailed study of ultraviolet, optical and near-infrared
HST images of NGC~1311 revealing 13 candidate star clusters. NGC~1311 and 
IC\,1959, both low-luminosity star-forming galaxies, are located in the 
foreground to the galaxy group LGG\,93 (Pisano et al. 2011). Our ATCA 
\HI\ maps show a box-shaped \HI\ distribution, somewhat truncated on the 
north-eastern side where faint stellar emission is prominent. The fat inner 
disc suggests the presence of extra-planar \HI\ gas; some deviations of the 
velocity field are observed to the north-eastern side. We measure \FHI\ = 
13.6 Jy\kms, in agreement with HIPASS (Koribalski et al. 2004), and derive 
\MHI\ = $8.7 \times 10^7$\Msun. \\

{\bf HIPASS J0457--42} is a nearby dwarf irregular galaxy at $D_{\rm TF}$ = 
7.2~Mpc (Karachentsev et al. 2013). It is the western galaxy ($PA$ = 56\degr)
of the apparent galaxy pair ESO252-IG001 (see Koribalski et al. 2004; their
Appendix~A). From the ATCA \HI\ intensity map we measure a position of
$\alpha,\delta$(J2000) = 04:56:59.1, --42:47:58.3 (see Table~7). The galaxy
was also detected in \Ha, where Meurer et al. (2006) find a {\em curious
near linear \Ha\ arc through centre along the minor axis}. Our ATCA data show 
an extended, regular rotating \HI\ disc, aligned with optical appearance. We 
note a small \HI\ extension or tail on the north-eastern (approaching) side. 
The optical properties, listed in Table~2, are of the \HI-detected galaxy
(ESO252-IG001 NED01). The eastern, nearly edge-on galaxy (ESO252-IG001 NED02,
$PA \sim 130\degr$) is much more distant and not a companion. The closest 
neighbour to HIPASS J0457--42, ESO305-G002 (\vopt\ = 259\kms) is more than 
3\degr\ away. If the optical velocity (da Costa et al. 1991) is correct, 
ESO305-G002 would be a nearby dwarf 
spiral galaxy at a projected distance of $\sim$200\arcmin. Inspecting HIPASS 
data we find \HI\ emission at the position/velocity of ESO305-G002, but it 
is situated close to an HVC complex (Putman et al. 2002). At the position of 
05:01:21.8, --39:36:13 we measure $S_{\rm peak}$ = 120 mJy, \FHI\ = 4.3 
Jy\kms, $w_{\rm 20}$ = 55\kms\ and $w_{\rm 50}$ = 30\kms. \\

{\bf ESO199-G007 (HIPASS J0258--49)} is a dwarf spiral galaxy at a Hubble 
distance of $D_{\rm Ho}$ = 6.6~Mpc. Using HIPASS we measure \vsys\ = 631\kms\ 
and obtain \vlg\ = 479\kms, showing this galaxy to be a member of the Local 
Volume. No optical velocity measurement has been reported yet. ESO199-G007's 
closest neighbour is NGC~1311 (HIPASS J0320--52), separated by $\sim$1.5 Mpc. 
Our ATCA \HI\ data show a marginally resolved, regularly rotating disc. We
measure \FHI\ = 1.6 Jy\kms, in agreement with HIPASS (Meyer et al. 2004), 
and derive \MHI\ = $1.6 \times 10^7$\Msun. Karachentsev \& Kaisina (2013) 
derive a star formation rate of $\sim$2.3 $\times 10^{-3}$\Msun\,yr$^{-1}$, 
and Young et al. (2014) analyse a deep $H$-band image of ESO199-G007.   \\

{\bf IC\,1959 (HIPASS J0333--50)} is a Magellanic barred spiral galaxy at a 
distance of $D_{\rm TRGB}$ = $6.05 \pm 0.14$~Mpc (Tully et al. 2006). Its 
nearest neighbour is NGC~1311 (HIPASS J0320--52), separated by $>1$~Mpc. Deep 
$H$-band images of IC\,1959 and NGC~1311 are analysed by Kirby et al. (2008b). 
Using 11HUGS Lee et al. (2009) report recent star formation with a rate of 
$4-5 \times 10^{-2}$\Msun\,yr$^{-1}$ for IC\,1959. The galaxy is 
also detected in LVHIS 20-cm radio continuum maps (Shao et al. 2017). Our 
ATCA \HI\ data show a well-resolved, symmetric and regularly rotating disc 
galaxy. Minor deviations in the mean \HI\ velocity field are seen towards the 
north-western (approaching) side.  We measure \FHI\ = 26.0 Jy\kms, in agreement 
with HIPASS (Koribalski et al. 2004), and derive \MHI\ = $2.2 \times 
10^8$\Msun. Using 3D FAT Kamphuis et al. (2015) obtain an \HI\ rotation curve 
indicating \vrot\ = 65.9\kms\ at \Rmax\ = 5.3 kpc (for $i$ = 78\fdg8 and 
$PA$ = 149\fdg0; see Table~9) and \Mdyn\ = $5.3 \times 10^9$\Msun. \\

{\bf NGC~1705 (HIPASS J0454--53)} is a well-studied BCD galaxy (type S0) at 
a distance of $D_{\rm TRGB}$ = $5.1 \pm 0.6$~Mpc (Tosi et al. 2001). It has 
no known neighbours within 5\degr. Our high-resolution ATCA data show an 
extended \HI\ distribution elongated nearly north-south ($PA \sim 20\degr$) 
with a mostly regular rotation pattern. Small peculiar velocities are seen 
towards the north-west. We measure \FHI\ = 12.2 Jy\kms, $\sim$10\% lower than 
the published HIPASS \FHI\ (Koribalski et al. 2004), and derive \MHI\ = $7.5 
\times 10^7$\Msun. For a more detailed \HI\ study see Elson et al. (2013), who 
improve on previous ATCA \HI\ results by Meurer et al. (1996, 1998) and Bureau 
et al. (1999). The young stellar population and star clusters of NGC~1705 are 
studied by Annibali et al. (2009) using HST images. Extended stellar emission
is seen in GALEX $UV$ images. \\

{\bf ESO364-G?029 (HIPASS J0605--33)} is an irregular looking barred Magellanic
dwarf galaxy at a Hubble distance of $D_{\rm Ho}$ = 7.6~Mpc. Its nearest
neighbour is the dwarf galaxy ATCA J060511--332534 (\vsys\ $\sim$ 830\kms; 
\FHI\ $\sim$ 1 Jy\kms), 21\farcm9 away, detected here (see Table~7). Other 
known
neighbours are AM0605--341 (HIPASS J0607--34) and NGC~2188 (HIPASS J0610--34) 
at projected distances of 70\farcm2 and 82\farcm5, respectively. Our ATCA maps 
show a regularly rotating, well-resolved \HI\ disc, encompassing the very 
irregular-shaped low-surface brightness stellar body. We measure \FHI\ = 22.3 
Jy\kms, somewhat larger than the HIPASS \FHI\ (Koribalski et al. 2004), and 
derive \MHI\ = $1.2 \times 10^8$\Msun. For a detailed study of ESO364-G?029, 
including a comparison with the LMC and other dwarf irregular galaxies see 
Kouwenhoven et al. (2007).\\

{\bf AM0605--341 (HIPASS J0607--34)} is a rather compact, barred Magellanic 
dwarf galaxy at a distance of $D_{\rm mem}$ = 7.4~Mpc. Its nearest neighbours 
are NGC~2188 (HIPASS J0610--34) and ESO364-G?029 ((HIPASS J0605--33) at 
projected distances of 35\farcm6 and 70\farcm2. Our ATCA \HI\ maps show a 
resolved source centred on the optical galaxy plus extended \HI\ emission on 
the western (receding) side. Kirby et al. (2012) model the \HI\ kinematics 
of the main component and fit a rising rotation curve. They provide a detailed 
discussion, including on likely tidal interactions between AM0605--341 and its 
eastern neighbour NGC~2188 which show opposing \HI\ extensions. We measure 
\FHI\ = 9.3 Jy\kms, in agreement with HIPASS (Koribalski et al. 2004), and 
derive \MHI\ = $1.2 \times 10^8$\Msun. \\

{\bf NGC~2188 (HIPASS J0610--34)} is a Magellanic barred spiral galaxy seen 
nearly edge-on, at a distance of $D_{\rm TF}$ = 7.4~Mpc (Karachentsev et al. 
2013). Its nearest neighbours are AM0605--341 ((HIPASS J0607--34) and 
ESO364-G?029 (HIPASS J0605--33) at projected distances of 35\farcm6 and 
82\farcm5 (see Kirby et al. 2012). Our ATCA data of NGC~2188 show an extended, 
but rather asymmetric \HI\ distribution and somewhat peculiar velocity field. 
Most notably, there is significant extraplanar \HI\ emission extending towards 
the east, while the western side of NGC~2188 appears compressed, resembling 
discs affected by ram pressure forces. This asymmetry is also, to a lesser
degree, evident in the stellar disc. The mean \HI\ velocity fields appears to 
show two components, a rotating disc plus peculiar motions associated with the 
extraplanar gas. These features were also noticed by Domg\"orgen et al. (1996) 
who analyse high-resolution VLA \HI\ and ESO \Ha\ images. Kirby et al. (2012) 
model the ATCA \HI\ velocity fields of NGC~2188's two neighbours. They suggest
that tidal interactions between NGC~2188 and AM0605--341, which are separated 
by $\sim$77 kpc, may be responsible for the peculiar features. Using our ATCA 
\HI\ data we measure \FHI\ = 34.3 Jy\kms, 
in agreement with HIPASS (Koribalski et al. 2004), and derive \MHI\ = $4.4 
\times 10^8$\Msun. For comparison, Domg\"orgen et al. (1996) measure \FHI\ = 
$20.4 \pm 0.6$ Jy\kms, losing much of the diffuse \HI\ emission. In a 
follow-up study Domg\"orgen \& Dettmar (1997) find spectacular filaments of 
ionised gas extending several hundred parsecs perpendicular to the plane of 
the galaxy, mostly from the massive \HII\ region in the southern (approaching) 
part of NGC~2188. They note that at least 24\% of the \Ha\ emitting is gas 
is diffuse. Deep $H$-band images are presented by Young et al. (2014). \\

{\bf ESO121-G020 and ATCA J061608--574552 (HIPASS J0615--57)} are a close 
dwarf galaxy pair, separated by $\sim$3\arcmin. Karachentsev et al. (2006)
obtain a distance of $D_{\rm TRGB}$ = $6.05 \pm 0.49$ Mpc for ESO121-G020. The 
HIPASS detection includes \HI\ emission from ESO121-G020 (\vsys\ = 584\kms) 
and its dwarf galaxy companion ATCA~J061608--574552 (\vsys\ = 540\kms), which 
was discovered by Warren et al. (2006). No other companions are known within
at least 5\degr. Our ATCA \HI\ maps show two regularly rotating galaxies. We 
measure \FHI\ = 7.3 Jy\kms\ for ESO121-G020 and 2.1 Jy\kms\ for the companion. 
Assuming the above distance for both we derive \MHI\ = 6.3 and $1.8 \times 
10^7$\Msun, respectively. Warren et al. (2006) obtain \HI\ mass to light ratios
of $\sim$1.5 and $\sim$2.2\Msun,/\Lsun. Both galaxies are clearly detected in
GALEX $UV$ images. For a detailed analysis of ESO121-G020's \HI\ kinematics 
see Kirby et al. (2012); their results are summarised in Table~9. \\

{\bf ESO308-G022 (HIPASS J0639--40)} is a rather isolated dwarf irregular 
galaxy at a Hubble distance of 7.7~Mpc. Optical and infrared H-band imaging 
by Parodi et al. (2002) and Kirby et al. (2008b), respectively, reveal its 
stellar structure. Our ATCA \HI\ maps show a resolved, regularly rotating \HI\ 
disc. We measure \FHI\ = 4.4 Jy\kms, in agreement with HIPASS (Meyer et al. 
2004), and derive an \HI\ mass of $6.1 \times 10^7$\Msun. A kinematic analysis 
of the LVHIS data by Kirby et al. (2012) results in a rising rotating curve 
with \vmax\ = 40\kms\ at a radius of 100\arcsec\ (4.1 kpc). \\

{\bf AM0704--582 (HIPASS J0705--58)} is also known as the Argo Dwarf Irregular.
It is a rather isolated, low surface brightness galaxy at a distance of 
$D_{\rm TRGB}$ = $4.90 \pm 0.45$ Mpc (Karachentsev et al. 2003). Parodi et al. 
(2002) and Kirby et al. (2008b) measure basic optical and infrared properties, 
respectively. The low extremely surface brightness of AM0704--582 makes it 
hard to determine its shape. Using HIPASS, Koribalski et al. (2004) fit an 
\HI\ systemic velocity of $564 \pm 2$\kms; no optical velocity measurement 
is available yet. Our ATCA \HI\ maps show a well-resolved, regularly rotating 
disc extending well beyond the faint stellar body. We measure \FHI\ = 33 
Jy\kms, in agreement with HIPASS (Koribalski et al. 2004), and derive \MHI\ 
= $1.9 \times 10^8$\Msun. Kirby et al. (2012) analyse the \HI\ kinematics 
and discuss the galaxy in detail. Using 3D FAT Kamphuis et al. (2015) obtain 
an \HI\ rotation curve indicating \vrot\ = 38.5\kms\ at \Rmax\ = 6.1 kpc 
(for $i$ = 53\fdg6 and $PA$ = 275\fdg9; see Table~9) and \Mdyn\ = $2.1 
\times 10^9$\Msun. \\

{\bf ESO059-G001 (HIPASS J0731--68)} is a Magellanic barred irregular galaxy 
at a TRGB distance of $4.57 \pm 0.36$ Mpc (Karachentsev et al. 2006). It 
appears very isolated with no known neighbours within at least 5\degr. Our 
ATCA \HI\ maps reveal a well-resolved, regularly rotating \HI\ disc extending 
well beyond the irregular stellar disc. We measure \FHI\ = 17.9 Jy\kms, in 
agreement with HIPASS (Koribalski et al. 2004), and derive \MHI\ = $8.8 \times 
10^7$\Msun. Using 3D FAT Kamphuis et al. (2015) obtain an \HI\ rotation curve 
indicating \vrot\ = 61.0\kms\ at \Rmax\ = 5.4 kpc (for $i$ = 50\fdg2 and $PA$ 
= 323\fdg4; see Table~9) and \Mdyn\ = $4.7 \times 10^9$\Msun; see also van 
Eymeren et al. (2009c) and Kirby et al. (2012). \\

{\bf NGC~2915 (HIPASS J0926--76)} is a rather isolated BCD galaxy at a TRGB 
distance of $3.78 \pm 0.43$ Mpc (Karachentsev et al. 2003), much studied in
\HI\ by Meurer et al. (1996) and more recently Elson et al. (2010). No 
neighbouring galaxies are known within 5\degr. The \HI\ cloud HIPASS J0851--75 
(\vsys\ = 482 km/s) west of NGC~2915 appears to be the highest velocity cloud
complex near the start of the Magellanic Leading Arm. 
Our ATCA maps of NGC~2915 show an enormous \HI\ disc -- compared to the 
compact stellar body -- with a mostly regular rotation pattern. Meurer et 
al. (1996) measured an \HI\ extent of $\sim$20\arcmin, i.e., $5\times$ its 
Holmberg radius. They note a short central bar and extended spiral arms. We 
measure \FHI\ = 108.7 Jy\kms, in agreement with HIPASS (Koribalski et al. 
2004), and derive \MHI\ = $3.7 \times 10^8$\Msun. For recent \HI\ studies of 
NGC~2915 see Elson et al. (2010, 2011a,b), who improve on previous ATCA \HI\ 
results by Meurer et al. (1998, 1996) and Bureau et al. (1999). Elson et al. 
(2012) also investigate the star formation in NGC~1705 and NGC~2915. \\

{\bf NGC~3621 (HIPASS J1118--32)} is a late-type spiral galaxy at a cepheid 
distance of $6.70 \pm 0.47$ Mpc (Ferrarese et al. 2000).  
To date NED shows a diverse collection of $\sim$50 independent distances. Its 
stellar disc has a diameter of about $12\arcmin \times 7\arcmin$, while the 
GALEX $UV$ emission extends about $20\arcmin \times 10\arcmin$ (Thilker et al. 
2007). NGC~3621 appears to be rather isolated; its nearest neighbours are the 
newly discovered dwarf galaxies HIPASS J1131--31 and HIPASS J1132--32, both 
$\sim$3\degr\ away. Our 3-pointing ATCA \HI\ mosaic reveals the extended \HI\ 
emission of NGC~3621, spanning over 40\arcmin\ (Koribalski 2017), a factor 
two beyond its remarkable $XUV$ disc. The ATCA \HI\ velocity field hints at a 
strong warp of the outer disc. Peculiar motions are also evident in the \HI\ 
velocity dispersion map. Our ATCA \FHI\ estimate of 856.8 Jy\kms\ is in good 
agreement with previous single-dish estimates (see Koribalski et al. 2004). 
We derive \MHI\ = $9.1 \times 10^9$\Msun. 
Single-pointing VLA \HI\ maps, taken as part of the THINGS project (Walter et 
al. 2008), reveal only the inner part of the galaxy (\FHI\ = 679 Jy\kms),
missing the peculiar extended emission towards the north and south.  
De Blok et al. (2008) find a slightly rising rotation curve out to a radius of 
13.5 arcmin (26 kpc) with \vrot\ = 150\kms, $i \sim 65\degr$, and $PA \approx 
345\degr$. Beyond that radius the \HI\ distribution and kinematics change 
significantly, as shown in our ATCA \HI\ maps, possibly due to the accretion 
of a companion. The ATCA \HI\ maps were first analysed and discussed by Walsh 
(1997). \\

{\bf HIPASS J1131--31} is a dwarf irregular galaxy with an \HI\ systemic 
velocity of 716\kms, originally discovered in HIPASS by us. Our ATCA \HI\ maps 
reveal an unresolved source, centered at $\alpha,\delta$(J2000) = 
11:31:34.6, --31:40:28.3 (see Table~7). The star TYC 7215-199-1 (10.4 mag) 
nearly fully obscures the stellar body of this galaxy.
GALEX $UV$ images reveal a compact blue dwarf galaxy at the above position.
Its closest neighbours are HIPASS J1132--32 (80\arcmin) and NGC~3621 
(182\arcmin); projected distances are given in brackets. Based on its likely 
association with NGC~3621 (HIPASS J1118--32) we assign the same distance, ie.  
$D_{\rm mem}$ = 6.7~Mpc. We measure \FHI\ = 1.13 Jy\kms, which corresponds 
to \MHI\ = $1.2 \times 10^7$\Msun.  \\

{\bf HIPASS J1132--32} is a dwarf irregular galaxy with an \HI\ systemic 
velocity of $\sim$680\kms, originally discovered in HIPASS (Meyer et al. 2004). 
Our low-resolution ATCA data show an unresolved \HI\ source, centered at 
$\alpha,\delta$(J2000) = 11:33:10.6, --32:57:45.2 (see Table~7). It coincides  
with a compact blue dwarf galaxy (PGC\,683190) that shows a faint extension 
to the west. No GALEX $UV$ images are available. Higher resolution ATCA data 
reveal a marginally resolved \HI\ source, matching the shape of the stellar 
body.
Doyle et al. (2005) give optical magnitudes 
of $B_{\rm J}$, $R$, and $I$ = 16.28, 16.17, and 17.22 mag. Its nearest 
neighbours are HIPASS J1131--31 (80\arcmin), NGC~3621 (188\arcmin), and 
ESO379-G007 (273\arcmin); projected distances are given in brackets. Based on 
its likely association with NGC~3621 (HIPASS J1118--32) we assign the same 
distance, ie. $D_{\rm mem}$ = 6.7~Mpc. We measure ATCA \FHI\ = 1.4 Jy\kms, 
which corresponds to \MHI\ = $1.5 \times 10^7$\Msun.  \\

{\bf IC\,3104 (HIPASS J1219--79)} is a Magellanic barred irregular galaxy at 
a TRGB distance of $2.27 \pm 0.19$ Mpc (Karachentsev et al. 2002). Its nearest 
neighbour is probably HIPASS J1247--77 (\vhel\ = 413\kms) at a projected 
distance of 154\farcm3. Our ATCA \HI\ images show a well-resolved, regularly 
rotating galaxy. We measure \FHI\ = 8.1 Jy\kms, consistent with the HIPASS 
\FHI\ (Koribalski et al. 2004), and derive \MHI\ = $9.8 \times 10^6$\Msun. \\

{\bf HIPASS J1247--77} is a highly obscured (\AB\ = 2.75) dwarf irregular 
galaxy located at $\alpha,\delta$(J2000) = 12:47:32.4, --77:34:53.9 (see 
Table~7) as measured using our ATCA \HI\ distribution (see also Kilborn et 
al. 2002). HST ACS images show a young stellar population and one bright star 
cluster; the galaxy's TRGB distance is $3.16 \pm 0.25$ Mpc (Karachentsev et 
al. 2006). Our ATCA \HI\ data reveal a resolved source, centred on the faint 
optical counterpart, with a regular velocity field ($PA \sim 55\degr$). We 
measure \vhel\ = 413\kms\ and \FHI\ = 4.2 Jy\kms, in agreement with HIPASS 
(Koribalski et al. 2004), and derive \MHI\ = $10^7$\Msun. Furthermore, our 
ATCA data reveal a neighbouring \HI\ source (not shown here) at 
$\alpha,\delta$(J2000) = 12:48:50.3, --77:49:30.6 (see Table~7) without 
any obvious optical counterpart. It has a systemic velocity of $\sim$402\kms\ 
and \FHI\ = 0.3 Jy\kms. We derive \MHI\ = $7 \times 10^5$\Msun, assuming the 
same distance as above. GALEX $UV$ images are not available at either position. 
The nearest neighbour to HIPASS J1247--77 is IC\,3104. \\

{\bf HIPASS J1526--51}, also known as HIZOA J1526--51, appears to be a dwarf 
irregular galaxy located in the Zone-of-Avoidance. It was first discovered 
in the Parkes \HI\ multibeam survey of the Zone of Avoidance (Juraszek et al. 
2000). No optical counterpart has been identified due to the high optical 
extinction (\AB\ = 2.30 mag) and high stellar density (Ryan-Weber et al. 2002).
Using our ATCA \HI\ maps we measure a centre position of $\alpha,\delta$(J2000)
= 15:26:22.4, --51:10:30.2 (see Table~7). The Local Group velocity of HIPASS 
J1526--51 is 438\kms\ which corresponds to a Hubble distance of $D_{\rm Ho}$ 
= 5.7~Mpc. The ATCA data reveal an extended \HI\ distribution and a peculiar 
velocity field, hinting at two distinct components. Its nearest known 
neighbours are ESO274-G001 (HIPASS J1514--46) and ESO223-G009 (HIPASS 
J1501--48), both close to 5\degr away. We measure \FHI\ = 5.0 Jy\kms\ for 
HIPASS J1526--51, in agreement with HIPASS (Koribalski et al. 2004), and 
derive \MHI\ = $3.8 \times 10^7$\Msun. \\


{\bf ESO137-G018 (HIPASS J1620--60)} appears to be a rather isolated, 
Magellanic type dwarf irregular galaxy located at low Galactic latitude. Its 
TRGB distance is $6.40 \pm 0.48$ Mpc (Karachentsev et al. 2007); there are no 
known neigbours within 3\degr. Our ATCA data reveal a well-resolved, regularly 
rotating \HI\ disc that extends a factor two beyond the stellar body. We 
measure \FHI\ = 43.5 Jy\kms, somewhat higher than the HIPASS \FHI\ (Koribalski 
et al. 2004), and derive \MHI\ = $4.2 \times 10^8$\Msun. Bonne (2008), Kirby 
et al. (2012) and most recently Kamphuis et al. (2015) obtained \HI\ rotation 
curves. Using 3D FAT Kamphuis et al. (2015) determine \vrot\ = 71.2\kms\ at 
\Rmax\ = 8.6 kpc (for $i$ = 71\fdg5 and $PA$ = 29\fdg2; see Table~9) and 
\Mdyn\ = $1.0 \times 10^{10}$\Msun. \Ha\ imaging by Kaisin et al. (2007) 
reveals an extended disc of clumpy and diffuse ionised gas.  \\

{\bf IC\,4662 (HIPASS J1747--64)} is a barred Magellanic irregular galaxy at 
a distance of $2.44 \pm 0.19$ Mpc (Karachentsev et al. 2006). Our ATCA \HI\ 
data reveal extented emission and a 
rather peculiar velocity field. The \HI\ emission is brightest in the galaxy 
centre and along the direction of the optical minor axis. At large radii the 
\HI\ envelope becomes nearly circular with a faint \HI\ extension towards the 
east. Apart from irregular motions in and near the eastern extension, the \HI\ 
velocity field clearly indicates rotation. The twisted iso-velocity contours 
highlight the disturbed nature of the kinematics in the outer disc. Van 
Eymeren et al. (2010) measure a total \HI\ extent of $15\farcm0 \times 
11\farcm5$, six times larger than the bright stellar body. Their deep \Ha\ 
images reveal a diffuse filamentary structure surrounding the box-shaped 
body of IC\,4662 plus a detached \HII\ region (a companion galaxy ?) 1\farcm5 
towards the south-east (see also Kaisin et al. 2007). 
Surprisingly, IC\,4662 appears to be a rather isolated galaxy, with no known 
neighbours within 5\degr. Its disturbed appearance, partly due to the bar and 
gas outflows, could be caused by accretion of or merging with close companions. 
For a more detailed discussion of IC\,4662 and comparison with other dwarf 
irregular galaxies see Janine van Eymeren's PhD Thesis (2008) and van Eymeren
et al. (2009a,b,c,d). \\

{\bf ESO461-G036 (HIPASS J2003--31)} is a dwarf irregular galaxy at a distance 
of $D_{\rm TRGB}$ = $7.83 \pm 0.63$ Mpc (Karachentsev et al. 2006). It appears
very isolated, having no known neighbours within at least 5\degr. Our ATCA \HI\
data show a remarkable, regularly rotating \HI\ disc extending well beyond the 
stellar body. The HIPASS and ATCA \HI\ observations both indicate a total \HI\ 
flux density of 7.5 Jy\kms, corresponding to an \HI\ mass of $1.1 \times 
10^8$\Msun. The velocity field suggests a mild warp of the outer disc, which 
has a diameter of $\sim$6\arcmin. Our \HI\ maps show a significant swing of the
position angle. While the \HI\ emission in the inner most region is aligned 
with the stellar extent ($PA \sim 20\degr$, the kinematic major axis suggests 
rotation along a different axis ($PA \sim 340\degr$). Begum et al. (2008) 
observed ESO461-G036 as part of the FIGGS project. Their estimates of \FHI\ 
and \DHI\ are underestimates due to the limited observed velocity range as 
evident by the truncated \HI\ spectrum in their Fig.~5. Kreckel et al. (2011) 
analyse VLA \HI\ data and obtain a rotation curve showing \vrot\ $\sim$ 51\kms\
(for $i \sim 65\degr$), and \Rmax\ = 6.8 kpc. Without the influence of 
neighbouring galaxies, the kinematics of the large \HI\ disc of ESO461-G036 
reflect its intrinsic evolution. Similar to NGC~2915, the galaxy is dominated 
by a large dark matter halo. \\

{\bf IC\,5052 (HIPASS J2052--69)} is a barred late-type spiral galaxy at a 
distance of $D_{\rm TRGB}$ = $6.03 \pm 0.10$~Mpc (Seth et al. 2005). It is 
a relatively isolated galaxy with star-formation throughout the edge-on 
stellar disc (Meurer et al. 2006) but no known neighbours within 5\degr.
Our low-resolution \HI\ maps show a large gas disc ($\sim 16\arcmin \times 
8\arcmin$, corresponding to 28~kpc $\times$ 14~kpc), i.e. about 2--3 times the 
size of the stellar disc. Both the \HI\ morphology and kinematics indicate 
significant warping of the outer \HI\ disc which bends symmetrically by 
$\sim$30\degr. As usual, the warp starts beyond the stellar disc; the $PA$ of 
the outermost \HI\ disc is $\sim$290\degr\ compared to $\sim$322\degr\ for 
the inner disc (see also Wang et al. 2017). We measure \FHI\ = 90.0 Jy\kms, 
$\sim$10\% less than the HIPASS \FHI\ (Koribalski et al. 2004), and derive 
\MHI\ = $7.7 \times 10^8$\Msun. Using high-resolution (9\arcsec) ATCA \HI\ 
maps O'Brien et al. (2010) study the structure of IC\,5052's inner disc. 
They estimate the exponential scale height of the projected minor axis to be 
20\farcs5 (600~pc), which is confirmed by \tiri\ modeling. 
O'Brien et al. (2010) give a rotational velocity of 90\kms\ for IC\,5052.
Rossa \& Dettmar (2003) study the extraplanar diffuse ionised gas in IC\,5052
as well as prominent \Ha\ filaments and shells. HST ACS images by Seth et al. 
(2005) reveal an intricate web of dust filaments in the edge-on, stellar disc 
as well as two relatively bright star forming regions. ATCA 20-cm radio 
continuum emission is clearly detected (see Shao et al. 2017). \\

\section{Summary \& Outlook} 

The Local Volume \HI\ Survey (LVHIS) consists of all galaxies with Local Group 
velocities \vlg\ $< 550$\kms\ or distances $D < 10$~Mpc that are detected in 
the \HI\ Parkes All Sky Survey (HIPASS) at declinations $\delta \la -30\degr$.
We present the results of deep ATCA \HI\ spectral line observations of a 
complete sample of 82 nearby, gas-rich galaxies, including a comprehensive 
\HI\ atlas (Appendix~A) and on-line database\footnote{LVHIS project page: 
{\em www.atnf.csiro.au/research/LVHIS}}. Furthermore, we list the optical and 
HIPASS properties of all LVHIS galaxies and provide information on their 
multi-wavelength coverage (see also Wang et al. 2017). We discuss the \HI\ 
properties of each galaxy together with a brief literature overview. Members 
of galaxy groups, such as the Local Group, the Sculptor Group and the Cen\,A 
Group, are organised together, while the remaining LVHIS galaxies are sorted 
in RA order. We provide accurate ATCA \HI\ positions for nine dwarf galaxies 
discovered in HIPASS and a further six dwarf galaxies discovered in our \HI\ 
data cubes (either companion or background galaxies). 

The LVHIS galaxy atlas includes \HI\ moment maps and $pv$-diagrmas for the 
majority of LVHIS galaxies. For each LVHIS galaxy we determine their \HI\ flux 
densities and diameters, analyse their structure and kinematics, search for 
companions and neighbouring \HI\ clouds and compare these with the stellar 
distribution and optical properties. The ATCA \HI\ cubes and moment maps are 
made available as FITS files via our on-line LVHIS database; requests for 
calibrated $uv$-data will be considered. Scientific analysis of the LVHIS 
data is on-going and will be presented in subsequent papers. Recent highlights
include the kinematic analysis of LVHIS galaxies (Kamphuis et al. 2015, Oh et
al. 2018), investigations of the \DHI-\MHI\ relation (Wang et al. 2016) and a 
comprehensive multi-wavelength study (Wang et al. 2017). \\

Within the LVHIS sample we identify several transition type dwarf galaxies
based on their optical and \HI\ morphologies; these are Phoenix (ESO245-G007),
ESO294-G010, ESO410-G005, NGC~5237, HIPASS J1337--39, HIPASS J1348--37 and 
AM0319--662. Other transition type LV galaxies discussed in this paper include 
ESO540-G030, ESO540-G032, NGC~59, and HIDEEP J1337--3320. Their \HI\ masses 
range from $\sim10^5$ to $4 \times 10^7$\Msun, and their \HI\ distributions 
are characterised by being offset and/or misaligned from the stellar body. We
note that these dwarf transitional galaxies all reside in groups, apart from 
AM0319--662 close to NGC~1313. 

The LVHIS sample contains two starburst spirals (NGC~253 and NGC~4945), a 
dwarf starburst galaxy (NGC~5253), an early-type post-starburst galaxy 
(NGC~5102) and two active galaxies with radio lobes (Cen\,A and Circinus). 
The largest \HI\ distributions ($\sim$100~kpc) are found for the spiral 
galaxies M\,83, Circinus and NGC~3621. While \HI\ warps are detected in 
numerous LVHIS galaxies, only one LVHIS galaxy (M\,83) shows strong signs 
of tidal interactions. 
We note that the most isolated galaxies in the LVHIS sample (AM0704--582, 
ESO215-G?009, NGC~2915, NGC~3621, ESO174-G?001, ESO308-G022, ESO461-G036,
ESO149-G003, ESO223-G009, IC\,5052) have typically much larger \HI\ discs 
compared to their optical diameters than galaxies with many neighbours. 
This trend needs to be explored further. \\

We plan to obtain more sensitive high-resolution \HI\ observations with the 
Compact Array Broadband Backend (CABB), which was installed in 2009 (Wilson 
et al. 2011). By using several 6-km ATCA configurations we aim to reach 
10\arcsec\ angular resolution and 1--4\kms\ velocity resolution. The 2~GHz of 
bandwidth available in the new ATCA 1--3~GHz frequency band as well as 
high-bit sampling will allow us to also make very deep radio continuum and 
polarisation maps for all LVHIS galaxies and their surroundings. \\

Major \HI\ surveys are planned with three new radio interferometers, which are 
precursors/pathfinders for the Square Kilometre Array (SKA). The Australian 
SKA Pathfinder (ASKAP; Johnston et al. 2008), $36 \times 12$-m dishes forming
a 6-km diameter synthesis array, is currently being equipped with chequerboard 
Phased Array Feeds (PAFs; Chippendale et al. 2015) operating from 0.7 to 1.8 
GHz and providing 30 sq degr field of view\footnote{ASKAP Early Science started
in Oct 2016 with an array of twelve PAF-equipped antennas.}. First ASKAP 
science results were published by Serra et al. (2015). At the same time, the old 
WSRT is being outfitted with Vivaldi PAFs (Apertif; Oosterloo
et al. 2010), operating from 1 to 1.8 GHz and providing 8 sq degr field of 
view. The MeerKAT radio telescope array (Blyth et al. 2015), $64 \times 13.5$-m
dishes forming an 8-km synthesis array, located in the Northern Cape of South 
Africa, is currently being equipped with a range of traditional, single-horn 
receivers. The fast 21-cm survey speed of ASKAP and Apertif will allow a full 
census of nearby gas-rich galaxies, while MeerKAT is ideal for deep follow-up 
\HI\ observations. In particular, the {\em ASKAP HI All-Sky Survey} (known as 
WALLABY) and its sister survey, the {\em Westerbork Northern Sky HI Survey} 
(WNSHS), together aim to cover the whole sky out to a redshift of $z$ = 0.26 
(Koribalski 2012, Duffy et al. 2012). For galaxies at $D$ = 10~Mpc, WALLABY 
will have a 5$\sigma$ \HI\ mass limit of $5 \times 10^6$\Msun\ and a spatial 
resolution of 1.5~kpc (30\arcsec). \\ 

LVHIS is a significant step towards the ASKAP \HI\ All Sky Survey (WALLABY;
Koribalski 2012), as it delivers similar spectral line sensitivity and 
resolution. With WALLABY Early Science observations and data processing 
under way, the LVHIS data are essential for ASKAP data verification (e.g., 
flux, velocity and position accuracy). Four southern fields, each 30 square 
degress in size, have been observed with an array of twelve and later sixteen
PAF-equipped ASKAP antennas. One of the fields currently under investigation
includes the M\,83 galaxy group for which plenty of LVHIS data exist. Once
successfully calibrated we aim to combine ASKAP and ATCA \HI\ data.

\section*{Acknowledgements}
\begin{itemize}
\item This research has made extensive use of the NASA/IPAC Extragalactic 
      Database (NED) which is operated by the Jet Propulsion Laboratory, 
      Caltech, under contract with the National Aeronautics and Space 
      Administration. 
\item The Digitised Sky Survey was produced by the Space Telescope Science
      Institute (STScI) and is based on photographic data from the UK Schmidt 
      Telescope, the Royal Observatory Edinburgh, the UK Science and 
      Engineering Research Council, and the Anglo-Australian Observatory.
\item We thank Janine van Eymeren, Emma Kirby, Nic Bonne, Helmut Jerjen and 
      Igor Karachentsev for their contributions in the early stages of the 
      LVHIS project.
\item We thank the referee for suggesting the addition of major and minor
      axes position-velocity ($pv$) diagrams which enrich the LVHIS galaxy
      atlas. 
\end{itemize}

%
\appendix

\section{The LVHIS Galaxy Atlas}

The LVHIS Galaxy Atlas is published by MNRAS as online supplementary material.
It contains the ATCA \HI\ moment maps and position-velocity diagrams of all 
LVHIS galaxies (apart from ESO294-G010, ESO245-G007, and NGC~5128) arranged 
on a single page per galaxy. Each page has six figure panels, displaying --- 
from the top left to the bottom right --- the galaxy's integrated \HI\ 
distribution typically overlaid onto its DSS $B$-band image, the mean \HI\ 
velocity field (1.\,moment), the integrated \HI\ distribution (0.\,moment), 
the \HI\ velocity dispersion (2.\,moment), the major-axis $pv$-diagram and 
the minor-axis $pv$-diagram. Some galaxy properties as well as the contour 
levels are given in the captions. The synthesized beam is shown in the bottom 
left of each panel. North is up and east to the left. Unless otherwise stated,
we show the low-resolution maps obtained using `natural' weighting to 
emphasize the large-scale \HI\ distribution, including diffuse emission in 
the outer disc. The LVHIS Galaxy Atlas (incl. FITS files) is also available 
on the LVHIS webpages.

\clearpage



\begin{thebibliography}{}
\bibitem{annibali:etal} 
Annibali, F., Tosi, M., Monelli, M., Sirianni, M., Montegriffo, P., Aloisi, 
  A., Greggio, L. 2009, AJ 138, 169
\bibitem{banks:etal} 
Banks, G.D., Disney, M.J., Knezek, P.M., and HIPASS Team 1999, 
  ApJ 524, 612  (B99) 
\bibitem{barnes:etal} 
Barnes, D.G., et al. 2001, MNRAS 322, 486 
\bibitem{barnes:deblok} 
Barnes, D.G., de Blok, W.J.G. 2004, MNRAS 351, 333
\bibitem{beaulieu:etal2006} 
Beaulieu, S.F., Freeman, K.C., Carignan, C., Lockman, F.J., Jerjen, H.,
  2006 AJ 131, 325
\bibitem{begum:etal} 
Begum, A., Chengalur, J.N. 2003, A\&A 409, 879
\bibitem{begum:etal} 
Begum, A., Chengalur, J.N., Karachentsev, I.D., Kaisin, S.S., Sharina, M.E.
   2006, MNRAS 365, 1220
\bibitem{begum:etal} 
Begum, A., Chengalur, J.N., Karachentsev, I.D., Sharina, M.E., Kaisin, S.S.
   2008, MNRAS 386, 1667
\bibitem{bell:etal} 
Bell, E.F., McIntosh, D.H., Katz, N., Weinberg, M.D. 2003, ApJS 149, 289
\bibitem{blanton:etal} 
Blanton, M.R., et al. 2003, ApJ 592, 818 
\bibitem{bonne} 
Bonne, N.J. 2008, in "Galaxies in the Local Volume", Sydney, July 2007,
  eds. B.S. Koribalski \& H. Jerjen, Springer, p. \,45
\bibitem{boomsma:etal} 
Boomsma, R., Oosterloo, T.A., Fraternali, F., van der Hulst, J.M., Sancisi,
  R. 2005, A\&A 431, 65
\bibitem{bouchard:etal2004}
Bouchard, A., Da Costa, G.S., Jerjen, H. 2004, PASP 116, 1032
\bibitem{bouchard:etal.2005} 
Bouchard, A., Jerjen, H., Da Costa, G.S., Ott, J. 2005, AJ 130, 205832
\bibitem{bouchard:etal} 
Bouchard, A., Jerjen, H., Da Costa, G.S., Ott, J. 2007, AJ 133, 261
\bibitem{bouchard:etal} 
Bouchard, A., Da Costa, G.S., Jerjen, H. 2009, AJ 137, 3038 (B09)
\bibitem{broeils} 
Broeils, A.H. 1992, PhD Thesis, University of Groningen
\bibitem{broeils:rhee} 
Broeils, A.H., Rhee, M.-H. 1997, A\&A 324, 877
\bibitem{bureau:etal} 
Bureau, M., Freeman, K.C., Pfeitzber, D.W. 1999, AJ 118, 2158
\bibitem{buyle:etal} 
Buyle, P., Michielsen, D., de Ricke, S., Ott, J., Dejonghe, H. 2006,
  MNRAS 373, 793 
\bibitem{calabretta:etal} 
Calabretta, M.R., Staveley-Smith, L., Barnes, D.G. 2014, PASA 31, 7
\bibitem{cannon:etal} 
Cannon, J.M., Dohm-Palmer, R.C., Skillman, E.D., Bomans, D.J.,
  C\^ot\'e, S., Miller, B.W. 2003, AJ 126, 2806
\bibitem{cannon:etal} 
Cannon, J.M., McClure-Griffiths, N.M., Skillman, E.D. C\^ot\'e, S. 2004, 
  ApJ 607, 274 
\bibitem{carignan:puche1990a} 
Carignan, C., Puche D. 1990a, AJ 100, 394
\bibitem{carignan:puche1990b} 
Carignan, C., Puche D. 1990b, AJ 100, 641
\bibitem{carignan:etal1990} 
Carignan, C., Beaulieu, S., Freeman, K.C. 1990, AJ 99, 178 
\bibitem{chemin:etal} 
Chemin, L., Carignan, C., Drouin, N., Freeman, K.C. 2006, AJ 132, 2527
\bibitem{chippendale:etal} 
Chippendale, A.P., et al. 2015, International Symposium on Antennas and 
   Propagation (ISAP), IEEE (astro-ph/1509.05489) 
\bibitem{chung:etal} 
Chung, A., van Gorkom, J.H., Kenney, J.D.P., Crowl, H., Vollmer, B. 2009, 
  AJ 138, 1741
\bibitem{cote2000} 
C\^ot\'e, S., Carignan, C., Freeman, K. 2000, AJ 120, 3027
\bibitem{cote2009} 
C\^ot\'e, S., Draginda, A., Skillman E.D., Miller, B.W. 2009, 
   AJ 138, 1037 (C09)
\bibitem{crnojecic:etal2011}
Crnojevi\'c, D., Grebel, E.K., Cole, A.A. 2011, A\&A 530, 59
\bibitem{crnojecic:etal2012} 
Crnojevi\'c, D., Grebel, E.K., Cole, A.A. 2012, A\&A 541, 131
\bibitem{curran:etal} 
Curran, S.J., Koribalski, B.S., Bains, I. 2008, MNRAS 389, 63
\bibitem{dacosta:etal2008}
Da Costa, G.S., Jerjen, H. Bouchard, A. 2008, in "Galaxies in the Local 
  Volume", Sydney, July 2007, eds. B.S. Koribalski \& H. Jerjen, Springer, 
   p.\,123
\bibitem{dalcanton:etal} 
Dalcanton, J.J., et al. 2009, ApJS 183, 67 
\bibitem{dale:etal} 
Dale, D.A., et al. 2009, ApJ 703, 517
\bibitem{davidge2007} 
Davidge, T.J. 2007, ApJ 664, 820 
\bibitem{davidge2008} 
Davidge, T.J. 2008, AJ 135, 1636
\bibitem{deblok:etal} 
de Blok, W.J.G., Walter, F., Brinks, E., Trachternach, C., Oh, S.-H., 
  Kennicutt, R.C. 2008, AJ 136, 2648
\bibitem{dicaire:etal} 
D\'enes, H., Kilborn, V.A., Koribalski, B.S. 2014, MNRAS 444, 667
\bibitem{domgoergen:dettmar} 
Domg\"orgen, H., Dettmar, R.-J. 1997, A\&A 322, 391
\bibitem{domgoergen:etal} 
Domg\"orgen, H., Dahlem, M., Dettmar, R.-J. 1996, A\&A 313, 96 
\bibitem{donovan:etal} 
Donovan, J.L., et al. 2009, AJ 137, 5037 
\bibitem{duffy:etal} 
Duffy, A.R., Meyer, M.J., Staveley-Smith, L., Bernyk, M., Croton, D.J.,
  Koribalski, B.S., Gerstmann, D., Westerlund, S. 2012, MNRAS 426, 3385 
\bibitem{doyle:etal} 
Doyle, M.T., et al. 2005, MNRAS 361, 34 
\bibitem{elmouttie:etal} 
Elmouttie, M., Koribalski, B.S., Gordon, S., Taylor, K., Houghton, S., 
  Lavezzi, T., Haynes, R.F. Jones, K.L. 1998a, MNRAS 297, 49
\bibitem{elmouttie:etal} 
Elmouttie, M., Haynes, R.F. Jones, K.L., Sadler, E.M., Ehle, M. 1998b, 
  MNRAS 297, 1202
\bibitem{elson:etal} 
Elson, E.C., de Blok, W.J.G., Kraan-Korteweg, R.C. 2013, MNRAS 429, 2550
\bibitem{elson:etal} 
Elson, E.C., de Blok, W.J.G., Kraan-Korteweg, R.C. 2012, AJ 143, 1
\bibitem{elson:etal} 
Elson, E.C., de Blok, W.J.G., Kraan-Korteweg, R.C. 2011b, MNRAS 415, 323 
\bibitem{elson:etal} 
Elson, E.C., de Blok, W.J.G., Kraan-Korteweg, R.C. 2011a, MNRAS 411, 200 
\bibitem{elson:etal} 
Elson, E.C., de Blok, W.J.G., Kraan-Korteweg, R.C. 2010, MNRAS 404, 2061
\bibitem{eskridge:etal} 
Eskridge, P.B., de Grijs, R., Anders, P., Windhorst, R.A., Mager, V.A., 
  Jansen, R.A. 2008, AJ 135, 120
\bibitem{ferguson:etal} 
Ferguson, A.M.N.., Wyse, R.F.G., Gallagher, J.S. 1996, AJ 112, 2567
\bibitem{for:etal} 
For, B.-Q., Koribalski, B.S., Jarrett, T.H. 2012, MNRAS 425, 1934
\bibitem{fouque:etal} 
Fouque, P., Gourgoulhon, E., Chamaraux, P., Paturel, G. 1992, A\&AS 93, 211
\bibitem{freeman:etal} 
Freeman, K.C., Karlsson, B., Lynga, G., Burrell, J.F., van Woerden, H., 
  Goss, W.M., Mebold, U. 1977, A\&A 55, 445
\bibitem{gallart:etal} 
Gallart, C., Martin\'ez-Delgado, D., G\'omez-Flechoso, M.A., Mateo, M. 2001,
  AJ 121, 2572
\bibitem{gardner:whiteoak} 
Gardner, F.F., Whiteoak , J.B. 1976, PASA 3, 63
\bibitem{gieren:etal} 
Gieren, W., Pietrzy\'nski, G., Soszy\'nski, I., Bresolin, F., Kudritzki,
   R.-P. 2008, ApJ 672, 266
\bibitem{grise:etal} 
Grise, F., Pakull, M.W., Soria, R., Motch, C., Smith, I.A., Ryder, S.D., 
  B\"ottcher, M. 2008, A\&A 486, 151
\bibitem{grossi:etal} 
Grossi, M., et al. 2007, MNRAS 374, 107 
\bibitem{heald:etal} 
Heald, G., et al. 1022, A\&A 526, 118 
\bibitem{heisler:etal} 
Heisler, C.A., Hill, T.L., McCall, M.L., Hunstead, R.W. 1997, MNRAS 285, 374
\bibitem{herrmann:etal} 
Herrmann, K.A., Ciardullo, R., Feldmeier, J.J., Vinciguerra, M. 2008,
  ApJ 683, 630
\bibitem{higgs:etal} 
Higgs, C.R., et al. 2016, MNRAS 458, 1678 
\bibitem{huchtmeier} 
Huchtmeier, W.K. 1979, A\&A 75, 170
\bibitem{huchtmeier:bohnenstengel} 
Huchtmeier, W.K., Bohnenstengel, H.-D. 1981, A\&A 100, 72
\bibitem{hunter:etal} 
Hunter, D.A., et al. 2012, AJ 144, 134 
\bibitem{irwin:tolstoy} 
Irwin, M., Tolstoy, E. 2002, MNRAS 336, 643
\bibitem{jacobs:etal} 
Jacobs, B.A., Rizzi, L., Tully, R.B., Shaya, E.J., Makarov, D.I., Makarova, 
   L. 2009, AJ 138, 332
\bibitem{jarrett:etal} 
Jarrett, T.H. et al. 2013, AJ 145, 6 
\bibitem{jerjen:etal1998} %
Jerjen, H., Freeman, K.C., Binggeli, B. 1998, AJ 116, 2873
\bibitem{jerjen:etal2000} 
Jerjen, H., Freeman, K.C., Binggeli, B. 2000, AJ 119, 166
\bibitem{johnson:etal} 
Johnson, M., Kamphuis, P., Koribalski, B.S., Wang, J., Oh, S.-H., Hill, A.,
  O'Sullivan, S., Haan, S., Serra, P. 2015, MNRAS 451, 3192
\bibitem{johnston:etal} 
Johnston, S., et al. 2008, ExA 22, 151 
\bibitem{jones:etal} 
Jones, K.L., Koribalski, B.S., Elmouttie, M., Haynes, R.F. 1999, 
  MNRAS 302, 649
\bibitem{jozsa:etal} 
J\'ozsa, G.I.G., Kenn, F., Klein, U., Oosterloo, T.A. 2007, A\&A 468, 731
\bibitem{juraszek:etal} 
Juraszek, S., et al. 2000, AJ 119, 1627 
\bibitem{kaisin:karachentsev} 
Kaisin, S.S., Karachentsev, I.D. 2006, Astrophysics 49, 287
\bibitem{kaisin:etal} 
Kaisin, S.S., Kasparova, A.V., Knyazev, A.Yu., Karachentsev, I.D. 2007, 
  Astronomy Letters 33, 283
\bibitem{kamphuis:etal}
Kamphuis, P., J\'ozsa, G.I.G., Oh, S.-H., Spekkens, K., Urbanic, N., Serra, 
  P., Koribalski, B.S., Dettmar, R.-J. 2015, MNRAS 452, 3139 
\bibitem{karachentsev:etal}
Karachentsev, I.D., et al. 2000, ApJ 542, 128 
\bibitem{karachentsev:etal} 
Karachentsev, I.D., Sharina, M.E., Dolphin, A.E., Grebel, E.K., Geisler, D., 
  Guhathakurta, P., Hodge, P.W., Karachentseva, V.E., Sarajedini, A.,
  Seitzer, P.  2002, A\&A 385, 21
\bibitem{karachentsev:etal} 
Karachentsev, I.D., Makarov, M.E., Sharina, M.E., Dolphin, A.E., Grebel, E.K.,
  Geisler, D., Guhathakurta, P., Hodge, P.W., Karachentseva, V.E., 
  Sarajedini, A., Seitzer, P. 2003, A\&A 398, 479
\bibitem{karachentsev:etal} 
Karachentsev, I.D., Grebel, E.K., Sharina, M.E., Dolphin, A.E., Geisler, D., 
  Guhathakurta, P., Hodge, P.W., Karachentseva, V.E., Sarajedini, A., 
  Seitzer, P. 2003, A\&A 404, 93 
\bibitem{karachentsev:etal} 
Karachentsev, I.D., Karachentseva, V.E., Huchtmeier, W.K., Makarov, D.I.
  2004, AJ 127, 2031 
\bibitem{karachentsev:etal2005}
Karachentsev, I.D., Kaisin, S.S., Tsvetanov, Z., Ford, H. 2005, A\&A 434, 935
\bibitem{karachentsev:etal2006}
Karachentsev, I.D., et al. 2006, AJ 131, 1361 
\bibitem{karachentsev:etal}
Karachentsev, I.D., et al. 2007, AJ 133, 504
\bibitem{karachentsev:etal} 
Karachentsev, I.D., Makarov, D.I., Kaisina, E.I. 2013, AJ 145, 101 (K13)
\bibitem{karachentsev:etal} 
Karachentsev, I.D., Kaisina, E.I. 2013, AJ 146, 46
\bibitem{karachentsev:etal} 
Karachentsev, I.D., Makarov, D.I., Kaisina, E. 2013, AJ 145, 101
\bibitem{kennicutt:etal} 
Kennicutt, R., et al. 2003, PASP 115, 928  
\bibitem{kennicutt:etal} 
Kennicutt, R.C. Jr., Lee, J.C., Funes, S.J., Jos\'e, G., Sakai, S., Akiyama, S.
  2008, ApJS 178, 247 
\bibitem{kennicutt:etal} 
Kennicutt, R.C., et al. 2011, PASP 123, 1347 
\bibitem{karachentsev}
Karachentsev, I.D., et al. 2008, in "Galaxies in the Local Volume", Sydney, 
  July 2007, eds. B.S. Koribalski \& H. Jerjen, Springer, p.\,21
\bibitem{kilborn:etal} 
Kilborn, V.A., et al. 2002, AJ 124, 690 
\bibitem{kirby:etal} 
Kirby, E., Jerjen, H., Ryder, S., Driver, S. 2008a, in "Galaxies in the Local 
  Volume", Sydney, July 2007, eds. B.S. Koribalski \& H. Jerjen, 
  Springer, p.\,49
\bibitem{kirby:etal} 
Kirby, E., Jerjen, H., Ryder, S., Driver, S. 2008b, AJ 136, 1866
\bibitem{kirby:etal} 
Kirby, E.M., Koribalski, B.S., Jerjen, H., L\'opez-S\'anchez, \'A.R. 2012, 
  MNRAS 420, 2924
\bibitem{kobulnicky:skillman} 
Kobulnicky, H.A., Skillman, E.D. 1995, ApJ 454, L121
\bibitem{koribalski:etal} 
Koribalski, B.S., Whiteoak, J.B., Houghton, S. 1995, PASA 12, 20 
\bibitem{koribalski:etal} 
Koribalski, B.S., Staveley-Smith, L., + HIPASS/ZOA Teams 2004, AJ 128, 16
\bibitem{koribalski2007}
Koribalski, B.S. 2007, in "Groups of Galaxies in the Nearby Universe", 
  ESO workshop, eds. Savine, L., Ivanov, V.D., Borissova, J., p.\,27
\bibitem{koribalski2008} 
Koribalski, B.S., and the LVHIS team 2008, in "Galaxies in the Local Volume", 
  Sydney, July 2007, eds. B.S. Koribalski \& H. Jerjen, Springer, p.\,41
\bibitem{koribalski:lopezsanches} 
Koribalski, B.S., L\'opez-S\'anchez, \'A.R. 2009, MNRAS 400, 1749 
\bibitem{koribalski2010} 
Koribalski, B.S. 2010, in "Galaxies in Isolation: Exploring Nature vs
  Nurture", eds. L. Verdes-Montenegro, A. del Olmo, and J.  Sulentic, 
  ASPC 421, 137
\bibitem{koribalski2012} 
Koribalski, B.S. 2012, PASA 29, 359 
\bibitem{koribalski2015} 
Koribalski, B.S. 2015, Proc. IAU Symposium 309, Cambridge University Press, 
  eds.  B. L. Ziegler, F. Combes, H. Dannerbauer, M. Verdugo, p.\,39
\bibitem{koribalski2017} 
Koribalski, B.S. 2017, Proc. IAU Symposium 321, eds. A. Gil de Paz, J. Knapen 
  \& J. Lee, Cambridge University Press, p.\,232 
\bibitem{kouwenhoven:etal} 
Kouwenhoven, M.B.N., Bureau, M., Kim, S., de Zeeuw, P.T. 2007, A\&A 470, 123
\bibitem{kraan:tammann}
Kraan-Korteweg, R.C., Tammann, G.A. 1979, AN 300, 181
\bibitem{kreckel:etal2011} 
Kreckel, K., Peebles, P.J.E., van Gorkom, J.H., van de Weygaert, R., van der
  Hulst, J.M. 2011, AJ 141, 204
\bibitem{larsen:richtler} 
Larsen, S.S., Richtler, T. 1999, A\&A 345, 59 
\bibitem{lauberts1982}
Lauberts, A. 1982, ESO/Uppsala survey of the ESO(B) atlas, 
  Garching: European Southern Observatory 
\bibitem{lauberts:valentijn1989}
Lauberts, A., Valentijn, E.A. 1989, The surface photometry catalogue of 
  the ESO-Uppsala galaxies, Garching: European Southern Observatory
\bibitem{lee:byunA} 
Lee, M.G., Byun, Y.-Ik. 1999, AJ 118, 817
\bibitem{lee2003:etal} 
Lee, H., Grebel, E.K., Hodge, P.W. 2003, A\&A 401, 141
\bibitem{lee2009:etal} 
Lee, J.C., Kennicutt, R., Funes, S.J.J.G. Funes, Sakai, S., Akiyama, S.
  2009, ApJ 692, 1305
\bibitem{lee2011:etal} 
Lee, J.C., Gil de Paz, A., Kennicutt, R., et al. 2011, ApJS 192, 6 
\bibitem{leroy:etal} 
Leroy, A.K., Walter, F., Bigiel, F., et al. 2009, AJ 137, 4670 
\bibitem{lianou:cole} 
Lianou, S., Cole, A.A. 2013, A\&A 549, 47 
\bibitem{lopez-sanchez:etal} 
L\'opez-S\'anchez, \'A.R., Koribalski, B.S., van Eymeren, J., Esteban, C.,
  Kirby, E., Jerjen, H., Lonsdale, N. 2012, MNRAS 419, 1051 
\bibitem{lopez-sanchez:etal} 
L\'opez-S\'anchez, \'A.R., Koribalski, B.S., Esteban, C., Garcia-Rojas, J. 
  2008, in "Galaxies in the Local Volume", Sydney, July 2007, eds. B.S. 
  Koribalski \& H. Jerjen, Springer, p.\,53 
\bibitem{lucero:etal2015}
Lucero, D., Carignan, C., Elson, E.C., Randriamampandry, T.H., Jarrett, T.H., 
  Oosterloo, T.A., Heald, G.H. 2015, MNRAS 450, 3935
\bibitem{madore:etal}
Madore, B.F., Freedman, W.L., Catanzarite, J., Navarrete, M. 2009, 
  ApJ 694, 1237
\bibitem{malin:hadley} 
Malin, D., Hadley, B. 1997, PASA 14, 52
\bibitem{manthey:oosterloo} 
Manthey, E., Oosterloo, T. 2008, in `Galaxies in the Local Volume', 
  Astrophysics and Space Science Proc., Springer Netherland, eds.
  Koribalski \& Jerjen, p.\,303
\bibitem{marlowe:etal} 
Marlowe, A.T., Meurer, G.R., Heckmann T.M. 1999, ApJ 522, 183
\bibitem{martinez:etal} 
Martin\'ez-Delgado, D., Gallart, C., Aparicio, A. 1999, AJ 118, 862
\bibitem{mcconnachie} 
McConnachie, A.W. 2012, AJ 144, 4
\bibitem{mckinley:etal2017} 
McKinley, B., et al. 2018, MNRAS 474, 4056 
\bibitem{meurer:etal} 
Meurer, G.R., Carignan, C., Beaulieu, S.F., Freeman, K.C. 1996, AJ 111, 1551
\bibitem{meurer:etal} 
Meurer, G.R., Staveley-Smith, L., Killeen, N.E.B. 1998, MNRAS 300, 705 
\bibitem{meurer:etal} 
Meurer, G.R., et al. 2006, ApJS 165, 307 
\bibitem{meyer:etal} 
Meyer, M., + HIPASS Team 2004, MNRAS 350, 1197 
\bibitem{millern:etal} 
Miller, B.W. 1996, AJ 112, 991
\bibitem{minchin:etal} 
Minchin, R., et al. 2003, MNRAS 346, 787 
\bibitem{mould:etal} 
Mould, J., Sakai, S. 2008, ApJ 686, L75
\bibitem{muellern:etal} 
M\"uller, O., Jerjen, H., Binggeli, B. 2015, A\&A 583, 79
\bibitem{obrien:etal} 
O'Brien, J.C., Freeman, K.C., van der Kruit, P.C., Bosma, A. 
   2010, A\&A 515, A60
\bibitem{oh:etal} 
Oh, S.-H., de Blok, W.J.G., Brinks, E., Walter, F., Kennicutt, R.C.Jr. 
   2011, AJ 141, 193
\bibitem{oh:etal} 
Oh, S.-H., Staveley-Smith, L., Spekkens, K., Kamphuis, P., Koribalski, B.S. 
   2018, MNRAS  473, 3256
\bibitem{oosterloo:etal} 
Oosterloo, T., Verheijen, M., van Cappellen, W. 2010, Proc. ISKAF2010 
   (astro-ph/1007.5141)
\bibitem{ott:etal}
Ott, M., Whiteoak, J.B., Henkel, C., Wielebinski, R. 2001, A\&A 372, 463 
\bibitem{ott:etal} 
Ott, J., Stilp, A.M., Warren, S.R., Skillman, E., Dalcanton, J., Walter, F., 
  de Blok, W.J.G., Koribalski, B.S., West, A. 2012, AJ 144, 123
  Weisz, D. 2008, in "Galaxies in the Local Volume", Sydney, July 2007, eds. 
\bibitem{parodi}
Parodi, B.R., Barazza, F.D., Binggeli, B. et al. 2002, A\&A 388, 29  
\bibitem{parodi:etal} 
Parodi, B.R., Binggeli, B. 2003, A\&A 398, 501 
\bibitem{phillips:etal}
Phillips, M.M., Jenkins, C.R., Dopita, M.A., Sadler, E.M., Binette, L. 1986, 
  AJ 91, 1062
\bibitem{pritzl:etal} 
Pritzl, B.J., et al. 2003, ApJ 596, L47
\bibitem{puche:etal} 
Puche, D., et al. 1991, AJ 101, 456 
\bibitem{radburn:etal} 
Radburn-Smith, D.J., et al. 2012, ApJ 753, 138 
\bibitem{radburn:etal} 
Radburn-Smith, D.J., et al. 2014, ApJ 780, 105 
\bibitem{rejkuba} 
Rejkuba, M. 2004, A\&A 413, 903
\bibitem{roberts:haynes} 
Roberts, M.S., Haynes, M.P. 1994, ARA\&A 32, 115
\bibitem{rogstad:etal} 
Rogstad, D.H., Lockhart, I.A., Wright, M.C.H. 1974, ApJ 193, 309
\bibitem{rossa:dettmar} 
Rossa, J., Dettmar, R.-J. 2003, A\&A 406, 505
\bibitem{ryan-weber:etal2002} 
Ryan-Weber, E., et al. 2002, AJ 124, 1954 
\bibitem{ryan-weber:etal2004} 
Ryan-Weber, E., et al. 2004, AJ 127, 1431 
\bibitem{ryan-weber:etal2008} 
Ryan-Weber, E., et al. 2008, MNRAS 384, 535 
\bibitem{ryder:etal} 
Ryder, S.D., Staveley-Smith, L., Malin, D.F., Walsh, W. 1995, AJ 109, 1592
\bibitem{ryder:dopita} 
Ryder, S.D., Dopita, M. 1993, ApJS 88, 415
\bibitem{sault:etal} 
Sault, R.J., Teuben, P.J., Wright, M.C.H. 1995, ASPC 77, 433
\bibitem{saviane:jerjen} 
Saviane, I. \& Jerjen, H. 2007, AJ 133, 1756
\bibitem{schiminovich:etal} 
Schiminovich, D., van Gorkom, J.H., van der Hulst, J.M., Kasow, S. 1994,
  ApJ 423, L101 
\bibitem{schlegel:etal} 
Schlafly, E.F., Finkbeiner, D. P. 2011, ApJ 737, 103
\bibitem{serra2012:etal} 
Serra, P., et al. 2012, MNRAS 422, 1835
\bibitem{serra2015:etal} 
Serra, P., Koribalski, B.S., Kilborn, V., et al. 2015, MNRAS 452, 2680
\bibitem{seth:etal} 
Seth, A.C., Dalcanton, J.J., de Jong, R.C. 2005, AJ 129, 1331
\bibitem{shao2017:etal} 
Shao, L., Koribalski, B.S., Wang, J., et al. 2017, MNRAS, submitted 
\bibitem{simpson:gottesman}
Simpson, C.E., Gottesman, S.T. 2000, AJ 120, 2975
\bibitem{skillman:etal} 
Skillman, E.D., Terlevich, R., Teuben P.J., van Woerden, H. 1988, A\&A 198, 33
\bibitem{stanimirovic:etal} 
Stanimirovic, S. Staveley-Smith, L., Dickey, J.M., Sault, R.J., 
  Snowden, S.L. 1999, MNRAS 302, 417
\bibitem{lss:etal2003} 
Staveley-Smith, L., Kim, S., Calabretta, M.R., Haynes, R.F., Kesteven, M.J.
  2003, MNRAS 339, 87
\bibitem{lss:etal2016} 
Staveley-Smith, L., Kraan-Korteweg, R.C., Schr\"oder, A.C., Henning, P.A.,
  Koribalski, B.S., Stewart, I.M., Heald, G. 2016, AJ 151, 52
\bibitem{stgermain:etal} 
St. Germain, J., Carignan, C., C\^ot\'e, S., Oosterloo, T. 1999, AJ 118, 1235
\bibitem{struve:etal} 
Struve, C., Oosterloo, T.A.,  Morganti, R., Saripalli, L. 2010, A\&A 515, 67
\bibitem{thim:etal} 
Thim, F., Tammann, G.A., Saha, A., Dolphin, A., Sandage, A., Tolstoy, E.,
  Labhardt, L. 2003, ApJ 590, 256
\bibitem{thilker:etal} 
Thilker, D., et al. 2007, ApJS 173, 538 
\bibitem{thomson} 
Thomson, R.C. 1992, MNRAS 257, 689
\bibitem{tosi:etal} 
Tosi, M., et al. 2001, AJ 122, 1271  
\bibitem{tuellmann:etal}
T\"ullmann, R., Rosa, M.R., Elwert, T., Bomans, D.J., Ferguson, A.M.N.,
  Dettmar, R.-J. 2003, ESO Messenger 114, 39
\bibitem{tully:etal2006}
Tully, B.R., et al. 2006, AJ 132, 729 
\bibitem{tully:etal2013}
Tully, B.R., et al. 2013, AJ 146, 86 
\bibitem{vanderhulst:etal} 
van der Hulst, J.M., van Albada, T.S., Sancisi, R. 2001, ASP Conf. Proc, eds.
  J.E. Hibbard, M. Rupen and J.H. van Gorkom, Vol. 240, p. 451
\bibitem{vaneymeren:etal}
van Eymeren, J., Marcelin, M., Bomans, D.J. 2008, in "Galaxies in the Local 
  Volume", Sydney, July 2007, eds. B.S. Koribalski \& H. Jerjen, 
  Springer, p.\,341
\bibitem{vaneymeren} 
van Eymeren, J. 2008, PhD Thesis, University of Bochum
\bibitem{vaneymeren:etal} 
van Eymeren, J., Marcelin, M., Koribalski, B.S., Dettmar, R.-J., Bomans, D.J.,
  Gach, J.-L., Balard, P. 2009a, A\&A 493, 511
\bibitem{vaneymeren:etal} 
van Eymeren, J., Marcelin, M., Koribalski, B., Dettmar, R.-J., Bomans, D.J.,
  Gach, J.-L., Balard, P. 2009b, A\&A 505, 105
\bibitem{vaneymeren:etal} 
van Eymeren, J., Trachternach, C., Koribalski, B., Dettmar, R.-J. 2009c, 
  A\&A 505, 1
\bibitem{vaneymeren:etal} 
van Eymeren, J., Koribalski, B.S., L\'opez-S\'anchez, \'A.R., Dettmar, R.-J.,
  Bomans, D.J. 2010, MNRAS 407, 113
\bibitem{vangorkom:etal} 
van Gorkom, J.H., van der Hulst, J.M., Haschick, A.S., Tubbs, A.D. 1990,
  AJ 99, 1781
\bibitem{woerden:etal} 
van Woerden, H., van Driel, W., Braun, R., Rots, A.H. 1993, A\&A 269, 15
\bibitem{verdes-montenegro:etal} 
Verdes-Montenegro, L., Bosma A., Athanassoula E. 2002, A\&A 389, 825
\bibitem{vaucouleurs:etal}
de Vaucouleurs, G., de Vaucouleurs, A., Corwin, H.G., Jr., Buta, R.J., 
  Paturel, G., Fouque, P. 1991, Third Reference Catalogue of Bright Galaxies, 
  Springer, New York [RC3]
\bibitem{wang2013:etal}
Wang, J., et al. 2013, MNRAS 433, 270
\bibitem{wang2015:etal}
Wang, J., Serra, P., J\'ozsa, G.I.G., Koribalski, B.S., van der Hulst, T.,
  Kamphuis, P., Cheng, L., Fu, J., Xiao, T., Overzier, R., Wieringa, M., 
  Wang, E. 2015, MNRAS 453, 2399  
\bibitem{wang2016:etal} 
Wang, J., Koribalski, B.S., Serra, P., van der Hulst T., Roychowdhury, S.,
Kamphuis, P., Chengalur, J.N. 2016, MNRAS 460, 2143  
\bibitem{wang2017:etal} 
Wang, J., Koribalski, B.S., et al. 2017, MNRAS 472, 3029 
\bibitem{walsh}
Walsh, W. 1997, PhD Thesis, University of NSW 
\bibitem{walter:etal} 
Walter, F., Brinks, E., de Blok, E., Bigiel, F., Kennicutt, R. C., Thornley,
  M.D., Leroy, A.K. 2008, AJ 136, 2563
\bibitem{warren2004:etal} 
Warren, B., Jerjen, H., Koribalski, B. 2004, AJ 128, 1152
\bibitem{warren2006:etal}
Warren, B., Jerjen, H., Koribalski, B. 2006, AJ 131, 2056
\bibitem{warren2007:etal}
Warren, B., Jerjen, H., Koribalski, B. 2007, AJ 134, 1849
\bibitem{westmeier2011:etal} 
Westmeier, T., Braun, R., Koribalski, B.S. 2011, MNRAS 410, 2217
\bibitem{westmeier2013:etal} 
Westmeier, T., Koribalski, B.S., Braun, R. 2013, MNRAS 434, 3511
\bibitem{westmeier2015:etal} 
Westmeier, T., Staveley-Smith, L., Calabretta, M., Jurek, R., Koribalski, 
  B.S., Meyer, M., Popping, A., Wong, O.I. 2015, MNRAS 453, 338 
\bibitem{westmeier2017:etal} 
Westmeier, T., et al. 2017, MNRAS 472, 4832 
\bibitem{wilcots:miller} 
Wilcots, E.M., Miller, B.W. 1998, AJ 116, 2363
\bibitem{wilson:etal} 
Wilson, W.E., et al. 2011, MNRAS 416, 832 
\bibitem{wong:etal2006} 
Wong, O.I., et al. 2006, MNRAS 371, 1855 
\bibitem{wong:etal2016} 
Wong, O.I., Meurer, G.R., Zheng, Z., Heckman, T.M., Thilker, D.A., 
  Zwaan, M.A. 2016, MNRAS 460, 1106
\bibitem{woodley} 
Woodley, K.A. 2006, AJ 132, 2424
\bibitem{young:etal} 
Young, L.M., Lo, K.Y. 1997, ApJ 490, 710
\bibitem{young:etal} 
Young, L.M., Skillman, E.D., Weisz, D.R., Dolphin, A.E. 2007, ApJ 659, 331
\bibitem{young:etal} 
Young, T., Jerjen, H., L\'opez-S\'anchez, \'A.,R., Koribalski, B.S. 2014, 
   MNRAS 444, 3052
\bibitem{yun:etal} 
Yun, M.S., Ho, P.T.P., Lo, K.Y. 1993, ApJ 411, L17
\bibitem{zwaan:etal} 
Zwaan, M., Staveley-Smith, L., Koribalski, B.S., et al. 2003, AJ 125, 2842
\end{thebibliography}
\end{document}